\documentclass[12p]{elsarticle}

\renewcommand{\footnotesize}{\scriptsize}
\usepackage{textcomp}

\usepackage{graphicx}
\usepackage{caption}
\usepackage{subcaption}

\newtheorem{hypothesis}{Hypothesis}
\newtheorem{nullhypothesis}{Null Hypothesis}

\usepackage{csquotes}

\usepackage{tikz}
\usetikzlibrary{tikzmark}
\usetikzlibrary{fit} 
\usetikzlibrary{positioning}
\usetikzlibrary{arrows}
\usetikzlibrary{shapes.multipart}

\usepackage[hidelinks,bookmarks=false]{hyperref}
\usepackage[numbered]{bookmark}
\usepackage{color,soul}
\usepackage{booktabs}
\usepackage{multirow}
\usepackage{url}
\usepackage{tcolorbox}
\usepackage{amsmath,amssymb,amsfonts}
\usepackage{algorithmic}
\usepackage{xcolor}
\usepackage{url}
\usepackage{tabularx} 
\usepackage{array}
\usepackage{lineno,hyperref}
\usepackage{longtable}
\modulolinenumbers[5]
\journal{Journal of \LaTeX\ Templates}

\usepackage{pgfplots}
\pgfplotsset{compat=1.14}

\usepackage{adjustbox}
\usepackage{float}
\usepackage{rotating}
\usepackage{tablefootnote}
\usepackage{nth}
\DeclareCaptionType{TextBox}
\newcolumntype{L}{>{\centering\arraybackslash}m{3cm}}

\tcbset{tab1/.style={fonttitle=\bfseries\large,fontupper=\normalsize\sffamily,
colback=yellow!6!white,colframe=red!75!black,colbacktitle=black!75!black,
coltitle=white,center title,freelance,frame code={
\foreach \n in {north east,north west,south east,south west}
{\path [fill=red!75!black] (interior.\n) circle (3mm); };},}}

\tcbset{tab2/.style={enhanced,fonttitle=\bfseries,fontupper=\tiny\sffamily,
colback=yellow!6!white,colframe=red!50!black,colbacktitle=black!75!black,
coltitle=white,center title}}
\usepackage[T1]{fontenc}
\usepackage{array, booktabs, makecell}
\usepackage[utf8]{inputenc}
\usepackage{dirtytalk}
\usepackage{tcolorbox}

\usepackage[british]{babel}
\usepackage{enumitem}

\newlist{SubItemList}{itemize}{1}
\setlist[SubItemList]{label={$-$}}

\let\OldItem\item
\newcommand{\SubItemStart}[1]{%
    \let\item\SubItemEnd
    \begin{SubItemList}[resume]%
        \OldItem #1%
}
\newcommand{\SubItemMiddle}[1]{%
    \OldItem #1%
}
\newcommand{\SubItemEnd}[1]{%
    \end{SubItemList}%
    \let\item\OldItem
    \item #1%
}
\newcommand*{\SubItem}[1]{%
    \let\SubItem\SubItemMiddle%
    \SubItemStart{#1}%
}

\newcommand{\christian}[1]{\textcolor{red}{{\it [Christian: #1]}}}

\newcommand{\projectsNumber}{800\xspace}
\newcommand{\commitsNumber}{111,884\xspace}

\newcommand{\sarInitialNumber}{513\xspace}

\newcommand{\numberSAR}{230\xspace}

\usepackage{framed}
\usepackage{mdframed}
\usepackage{pstricks}
\usepackage{xspace}
\usepackage{pgf-pie}
\usepackage{comment}
\usepackage{colortbl}
\tcbuselibrary{skins}
\tcbuselibrary{listings}
\usepackage[normalem]{ulem}
\usetikzlibrary{patterns,}
\usepgfplotslibrary{colorbrewer}
\usepackage{verbatim}
\usepackage{tikz}
\usetikzlibrary{tikzmark}
\usetikzlibrary{fit} 
\usetikzlibrary{positioning}
\usetikzlibrary{arrows}
\usetikzlibrary{shapes.multipart}
\usepackage{lscape}
\usepackage{placeins}
\usepackage{afterpage}
\newlength\mylength

\biboptions{authoryear}
\begin{document}

\begin{frontmatter}

\title{How We Refactor and How We Document it? On the Use of Supervised Machine Learning Algorithms to Classify Refactoring Documentation}

\author[RIT]{Eman Abdullah AlOmar\corref{mycorrespondingauthor}}
\cortext[mycorrespondingauthor]{Corresponding author}
\ead{eman.alomar@mail.rit.edu}

\author[RIT]{Anthony Peruma}
\ead{anthony.peruma@mail.rit.edu}

\author[RIT]{Mohamed Wiem Mkaouer}
\ead{mwmvse@rit.edu}

\author[RIT]{Christian Newman}
\ead{cdnvse@rit.edu}

\author[ETS]{Ali Ouni}
\ead{ali.ouni@etsmtl.ca}

\author[UoM]{Marouane Kessentini}
\ead{marouane@umich.edu}


\address[RIT]{Rochester Institute of Technology, Rochester, NY, USA}
\address[ETS]{ETS Montreal, University of Quebec, Montreal, QC, Canada}
\address[UoM]{University of Michigan, Dearborn, MI, USA}

\begin{abstract}
Refactoring is the art of improving the structural design of a software system without altering its external behavior. Today, refactoring has become a well established and disciplined software engineering practice that has attracted a significant amount of research presuming that refactoring is primarily motivated by the need to improve system structures. However, recent studies have shown that developers may incorporate refactoring strategies in other development-related activities that go beyond improving the design especially with the emerging challenges in contemporary software engineering. Unfortunately, these studies are limited to developer interviews and a reduced set of projects.

To cope with the above-mentioned limitations, we aim to better understand what motivates developers to apply a refactoring by mining and automatically classifying a large set of \commitsNumber commits containing refactoring activities, extracted from 800 open source Java projects. We trained a multi-class classifier to categorize these commits into three categories, namely, Internal Quality Attribute, External Quality Attribute, and Code Smell Resolution, along with the traditional Bug Fix and Functional categories. This classification challenges the original definition of refactoring, being exclusive to improving software design and fixing code smells. Furthermore, to better understand our classification results, we qualitatively analyzed commit messages to extract textual patterns that developers regularly use to describe their refactoring activities.

The results of our empirical investigation show that (1) fixing code smells is not the main driver for developers to refactoring their code bases. Refactoring is solicited for a wide variety of reasons, going beyond its traditional definition; (2) the distribution of refactoring
operations differ between production and test files; (3) developers use a variety of patterns to purposefully target refactoring-related activities; (4) the textual patterns, extracted from commit messages, provide a better coverage for how developers document their refactorings.

\end{abstract}

\end{frontmatter}


\section{Introduction}
\label{sec:Introduction}

The success of a software system depends on its ability to retain high quality of design in the face of continuous change. However, managing the growth of the software while continuously developing its functionalities is challenging, and can account for up to 75\% of the total development \citep{erlikh2000leveraging,barry1981software}. 
 One key practice to cope with this challenge is refactoring. Refactoring is the art of remodeling the software design without altering its functionalities \citep{Fowler:1999:RID:311424,7833023}. It was popularized by \citep{Fowler:1999:RID:311424}, who identified 72 refactoring types and provided examples of how to apply them in his catalog.

Refactoring is a critical software maintenance activity that is  performed by developers for an amalgamation of reasons \citep{Tsantalis:2013:MES:2555523.2555539,Silva:2016:WWR:2950290.2950305,palomba2017exploratory}. Refactoring activities in the source code can be automatically detected \citep{Dig2006,Tsantalis:2013:MES:2555523.2555539} providing a unique opportunity to practitioners and researchers to analyze how developers maintain their code during different phases of the development life-cycle and over large periods of time. Such valuable knowledge is vital for understanding more about the maintenance phase; the most costly phase in software development \citep{Boehm:2002:SEE:944331.944370,erlikh2000leveraging}. To detect refactorings, the state-of-the-art techniques \citep{Dig2006,Tsantalis:2013:MES:2555523.2555539} typically search at the level of commits. As a result, these techniques are also able to group commit messages with their corresponding refactorings.

Commit messages are the description, in natural language, of the code-level changes. To understand the nature of the change, recent studies have been using natural language processing to process commit messages for multiple reasons, such as classification of code changes \citep{Hindle:2008:LCT:1370750.1370773}, change summarization \citep{mcburney2017towards}, change bug-proneness \citep{xia2016collective}, and developer's rationale behind their coding decisions \citep{alkadhi2018developers}. That is, commit messages are a common way for researchers to study developer rationale behind different types of changes to the code. There are two primarily challenges to using commit messages to understand refactorings: 1) the commit message does not have to refer to the refactoring that took place at all, 2) developers have many ways of describing the same activity. For example, instead of explicitly stating that they are refactoring, a developer may instead state that they are \textit{performing code clean-up} or \textit{simplifying a method}. Developers are inconsistent in the way they discuss refactorings in commit messages. This makes it difficult to perform analysis on commit messages, since researchers may find it challenging to determine whether a commit message discusses the refactoring(s) being performed or not. Thus, it is hard to determine when the commit message is discussing a refactoring at all and it is hard to determine how a commit message is discussing the refactoring.

To cope with the above-mentioned challenges, the purpose of this study is to augment our understanding of the development contexts that trigger refactoring activities and enable future research to take development contexts into account more effectively when studying refactorings. Thus, the advantages of analyzing the textual description of the code change that was intended to describe refactoring activities are three-fold:  1) it improves our ability to study commit message content and relate this content to refactorings; a challenging task which posed a significant hurdle in recent work on contextualizing rename refactorings \citep{peruma2018empirical, peruma2019contextualizing}, 2) it gives us a stronger understanding of commit message practices and could help us improve commit message generation by making it clear how developers prefer to express their refactoring activities, 3) it provides us with a way of relating common words and phrases used to describe refactorings with one another. Typically frameworks like WordNet, which does not recognize refactoring phrases and terminology, are used for this task. Our dataset and methodology reduces the need to rely on frameworks which are not trained for natural language found in software projects.


In this paper, we present a way to partially-automatically detect how developers document their refactorings in commit messages, and classify these into categories that reflect the type of activity that refactoring was co-located with. The goal of this work is to create a data set of terms and phrases, used by developers, to describe refactorings. Further, we group these words and phrases by maintenance-type (e.g., bug fix, external, code smell) to obtain a fine-grained and maintenance-type-specific dataset of terms and phrases. Recent studies have shown the feasibility of extracting insights of software quality from developers inline documentation. For instance, mining developer's comments has unveiled how developers knowingly commit code that is either incomplete, temporary, or faulty. Such phenomenon is known as \textit{Self-Admitted Technical Debt} (SATD) \citep{potdar2014exploratory}. Similarly, our previous study has introduced \textit{Self-Affirmed Refactoring} (SAR) \citep{alomar2019can,alomar2020toward}, defined as developer's explicit documentation of refactoring operations intentionally introduced during a code change.

To perform this analysis, we formulate the following research questions:

\begin{itemize}

\item \textbf{RQ1.} To what purposes developers refactor their code? 

While previous surveys studied how developers apply refactorings in varying development contexts, none of them have measured the ubiquity of these varying contexts in practice. Therefore, it is important to \textit{quantify} the distribution of refactoring activities performed in varying development contexts to augment our understanding of refactoring in theory versus in practice.


\item \textbf{RQ1.1} Do software developers perform different types of refactoring operations on test code and production code between categories? 

This question further explores the findings of the classification to see to what extent developers refactor production files differently from test files.


\item \textbf{RQ2.} What patterns do developers use to describe their refactoring activities?

Since there is no consensus on how to formally document refactoring changes, we intend to extract (from commit messages) words and phrases commonly used by developers in practice to document their refactorings. Such information is useful from many perspectives. First, it allows to understand the rationale behind the applied refactorings, e.g., fixing code smells or improving specific quality attributes. Moreover, it may reveal what specific refactoring operations are being documented, and whether developers explicitly mention it as part of their documentation. Such details are of crucial importance especially in modern code review the help code reviewers understand the rationale behind such refactorings. 
Little is known about how developers document refactoring as previous studies mainly rely on the keyword \textit{refactor} to annotate such documentation.   


\item \textbf{RQ2.1} Do commits containing the label \textit{Refactor} indicate more refactoring activity than those without the label? 

We revisit the hypothesis raised by \citep{murphy2008gathering} about whether developers use a specific pattern, i.e., \textit{\say{refactor}} when describing their refactoring activities.








\end{itemize}

The dataset of classified refactorings along with textual patterns are available online \citep{SAR2020WEB} for replication and extension purposes.

The remainder of this paper is organized as follows. Section \ref{section:self} discusses the notion of refactoring related documentation or Self-Affirmed Refactoring. Section \ref{sec:RelatedWork} enumerates the previous related studies, and shows how we extracted the categories used for the classification. In Section \ref{sec:Methodology}, we give the design of our empirical study, mainly with regard to the construction of the dataset and classification. Section \ref{sec:Experimental Results} presents the study results while further discussing our findings in Section \ref{sec:Discussion}. The next Section \ref{sec:Threats} reports threats to the validity of our experiments, before concluding the paper in Section \ref{sec:Conclusion}. 

\section{Self-Affirmed Refactoring}
\label{section:self}

Commit messages are the description, in natural language, of the code-level changes. In this paper, we want to automatically detect how refactoring is documented in the commit message, and classify it into categories that reflect the type of activity that refactoring was co-located with. Earlier studies were relying on developer surveys for extracting such information. But multiple studies have been detecting the performed refactoring operations, e.g., rename class, move method etc. within committed changes to better understand how developers cope with bad design decisions, also known as design antipatterns, and to extract their removal strategy through the selection of the appropriate set of refactoring operations \citep{tsantalis2018accurate}. As the accuracy of refactoring detectors has reached a relatively high rate, mined commits' messages and their issues descriptions constitute a rich space to understand how developers describe, in natural language, their refactoring activities. Yet, such information retrieval can be challenging since there are no common standards on how developers should be formally documenting their refactorings, besides inheriting all the challenges related to natural language processing \citep{tan1999text}.

However, recent studies have shown the feasibility of extracting insights of software quality from developer's inline documentation. For instance, mining developers' comments has unveiled how developers knowingly commit code that is either incomplete, temporary, or faulty. Such phenomenon is known as \textit{Self-Admitted Technical Debt} (SATD) \citep{potdar2014exploratory}. Similarly, our previous study has introduced \textit{Self-Affirmed Refactoring} (SAR) \citep{alomar2019can,alomar2020toward}, defined as developers' explicit documentation of refactoring operations intentionally introduced during a code change.

As explained later in the related work section, existing studies locate refactoring documentation through the localization of the keyword \textit{\say{refactor}}, being the most intuitive and widely known. However, a recent study has also shown that the \textit{\say{refactor}} can also be misused, and such information becomes misleading \citep{zhangpreliminary18}. Yet, such findings are mainly taken from interviews. In this study, we leverage the existence of a large set of refactorings, extracted form a wide variety of projects, to design an empirical study to classify the context in which it was performed, for that, we start with the traditional categorization of Swanson \citep{Swanson:1976:DM:800253.807723}, and we extend its \say{Perfective} category to cover what has been known by existing studies as drivers to recommend refactorings. This study also further explores how developers document refactorings, and extracts a new terminology that was found to be consistently used in refactoring-related commit messages.

\section{Related Work}
\label{sec:RelatedWork}

This paper focuses on mining commits to initially detect refactorings and then to classify them. Thus, in this section, we are interested in exploring refactoring documentation, along with the research on refactoring motivations.

\subsection{Refactoring Documentation}

\begin{table*}[h!]
\begin{center}
\caption{Existing works on refactoring identification.}
\label{Table:Related Work-Refactoring Identification}
\begin{adjustbox}{width=1.0\textwidth,center}
 \begin{tabular}{ llllll } \hline
  \toprule
  \bfseries Study  & \bfseries Year & \bfseries Purpose & \bfseries Approach  & \bfseries Source of Info. & \bfseries Ref. Patterns\\
  \midrule
    \citep{stroggylos2007refactoring} & 2007 & Identify refactoring commits &  Mining commit logs & General commits & 1 keyword \\
   \citep{Ratzinger:2008:RRS:1370750.1370759,citeulike:2881658} & 2007 \& 2008 & Identify refactoring commits &  Mining commit logs & General commits & 13 keywords  \\
    \citep{6112738} & 2012 & Identify refactoring commits & Ratzinger's approach & General commits & 13 keywords \\
   \citep{soares2013comparing} & 2013 & Analyze refactoring activity & Ratzinger's approach & General commits & 13 keywords \\
    & & & Manual analysis & & \\
    & & & Dynamic analysis & & \\
    \citep{6802406} & 2014 & Identify refactoring commits & Identifying refactoring branches &  Refactoring branch & Top 10 keywords\\
    & &  &   Mining commit logs \\
     \citep{zhangpreliminary18} & 2018 & Identify refactoring commits & Mining commit logs & General commits & 22 keywords \\
     \citep{alomar2019can} & 2019 & Identify refactoring patterns & Detecting refactorings & Refactoring commits &    87 keywords \& phrases   \\
    & & & Extracting commit messages & & \\
  \bottomrule
  \end{tabular}
 \end{adjustbox}
  \end{center}
\end{table*}

A number of studies have focused recently on the identification and detection of refactoring activities during the software life-cycle. One of the common approaches to identify refactoring activities is to analyze the commit messages in versioned repositories. \citep{stroggylos2007refactoring} opted for searching words stemming from the verb \textit{\say{refactor}} such as \say{refactoring} or \say{refactored} to identify refactoring-related commits. \citep{citeulike:2881658,Ratzinger:2008:RRS:1370750.1370759} also used a similar keyword-based approach to detect refactoring activity between a pair of program versions to identify whether a transformation contains refactoring. The authors identified refactorings based on a set of keywords detected in commit messages, and focusing, in particular, on the following 13 terms in their search approach: \textit{refactor, restruct, clean, not used, unused, reformat, import, remove, replace, split, reorg, rename, and move}. 

Later, \citep{6112738} replicated Ratzinger's experiment in two open source systems using Ratzinger's 13 keywords. They conclude that commit messages in version histories are unreliable indicators of refactoring activities. This is due to the fact that developers do not consistently report/document refactoring activities in the commit messages. In another study, \citep{soares2013comparing} compared and evaluated three approaches, namely,  manual analysis, commit message (Ratzinger et al.'s approach), and dynamic analysis (SafeRefactor approach \citep{Soares2009safetytool}) to analyze refactorings in open source repositories, in terms of behavioral preservation. The authors found, in their experiment, that manual analysis achieves the best results in this comparative study and is considered as the most reliable approach in detecting behavior-preserving transformations. 

In another study, \citep{6802406} surveyed 328 professional software engineers at Microsoft to investigate when and how they do refactoring. They first identified refactoring branches and then asked developers about the keywords that are usually used to mark refactoring events in change commit messages. When surveyed, the developers mentioned several keywords to mark refactoring activities. Kim et al. matched the top ten refactoring-related keywords identified from the survey (\textit{refactor, clean-up, rewrite, restructure, redesign, move, extract, improve, split, reorganize, rename}) against the commit messages to identify refactoring commits from version histories. Using this approach, they found 94.29\% of commits do not have any of the keywords, and only 5.76\% of commits included refactoring-related keywords. 

\citep{zhangpreliminary18} performed a preliminary investigation of Self-Admitted Refactoring (SAR) in three open source systems. They first extracted 22 keywords from a list of refactoring operations defined in the Fowler's book \citep{Fowler:1999:RID:311424} as a basis for SAR identification. After identifying candidate SARs, they used Ref-Finder \citep{kim2010ref} to validate whether refactorings have been applied. In their work, they used code smells to assess the impact of SAR on the structural quality of the source code. Their main findings are the following (1) SAR tends to enhance the software quality although there is a small percentage of SAR that have introduced code smells, and (2) the most frequent code smells that are introduced or reduced depend highly on the nature of the studied projects. They concluded that SAR is a signal that helps to locate refactoring events, but it does not guarantee the application of refactorings. We summarize these state-of-the-art approaches in Table~\ref{Table:Related Work-Refactoring Identification}.

\subsection{Refactoring Motivation}
\citep{Silva:2016:WWR:2950290.2950305} investigate what motivates developers when applying specific refactoring operations by surveying GitHub contributors of 124 software projects. They observe that refactoring activities are mainly caused by changes in the project requirements and much less by code smells. \citep{palomba2017exploratory} verify the relationship between the application of refactoring operations and different types of code changes (i.e.,  \textit{Fault Repairing Modification, Feature Introduction Modification}, and \textit{General Maintenance Modification}) over the change history of three open source systems. Their main findings are that developers apply refactoring to: 1)  improve comprehensibility and maintainability when fixing bugs, 2) improve code cohesion when adding new features, and 3) improve the comprehensibility when performing general maintenance activities. On the other hand,  \citep{6802406} do not differentiate the motivations between different refactoring types. They surveyed 328 professional software engineers at Microsoft to investigate when and how they do refactoring. When surveyed, the developers cited the main benefits of refactoring to be: improved readability (43\%), improved maintainability (30\%), improved extensibility (27\%) and fewer bugs (27\%). When asked what provokes them to refactor, the main reason provided was poor readability (22\%). Only one code smell (i.e, code duplication) was mentioned (13\%). 

\citep{6112738} examine how programmers perform refactoring in practice by monitoring their activity and recording all their refactorings. They distinguished between high, medium and low-level refactorings. High-level refactorings tend to change code elements signatures without changing their implementation e.g., \textit{Move Class/Method, Rename Package/Class}. Medium-level refactorings change both signatures and code blocks, e.g., \textit{Extract Method, Inline Method}. Low level refactorings only change code blocks, e.g., \textit{Extract Local Variable, Rename Local Variable}. Some of the key findings of this study are that 1) most of the refactoring is floss, i.e., applied to reach other goals such as adding new features or fixing bugs, 2) almost all the refactoring operations are done manually by developers without the help of any tool, and 3) commit messages in version histories are unreliable indicators of refactoring activity because developers tend to not explicitly state refactoring activities when writing commit messages. It is due to this observation that, in this study, we do not rely on commits messages to identify refactorings. Instead, we use them to identify the motivation behind the refactoring. 

\citep{moser2006does} study the impact of refactoring on reusability. They showed that refactoring increases the reusability of classes in an industrial, agile environment. In a subsequent study, \citep{moser2008case} question the effectiveness of refactoring on increasing the productivity in agile environments. They performed a comparative study of developers coding effort before and after refactoring their code. They measured the developer's effort in terms of added lines of code and time. Their findings show that not only does the refactored system improve in terms of coupling and complexity, but also that the coding effort was reduced and the difference is statistically significant.

\citep{szHoke2014case} conduct 5 large-scale industrial case studies on the application of refactoring while fixing coding issues, they have shown that developers tend to apply refactorings manually at the expense of a large time overhead.  \citep{szHoke2017empirical} extend their study by investigating whether the refactorings applied when fixing issues did improve the system's nonfunctional requirements with regard to maintainability. They noticed that refactorings performed manually by developers do not significantly improve the system's maintainability like those generated using fully automated tools. They concluded that refactoring cannot be cornered only in the context of design improvement.      

\citep{Tsantalis:2013:MES:2555523.2555539} manually inspect the source code for each detected refactoring with a text diff tool to reveal the main drivers that motivated the developers for the applied refactoring. Besides code smell resolution, they found that introduction of extension points and the resolution of backward compatibility issues are also reasons behind the application of a given refactoring type. In another study, \citep{5306290} generally focuses on the human and social factors affecting the refactoring practice rather than on the technical motivations. He interviewed 10 industrial developers and found a list of \textit{intrinsic} (e.g., responsibility of code authorship) and \textit{external} (e.g., recognitions from others) factors motivating refactoring activity.

\begin{sidewaysfigure}[htbp]
\centering 
\includegraphics[width=\textwidth]{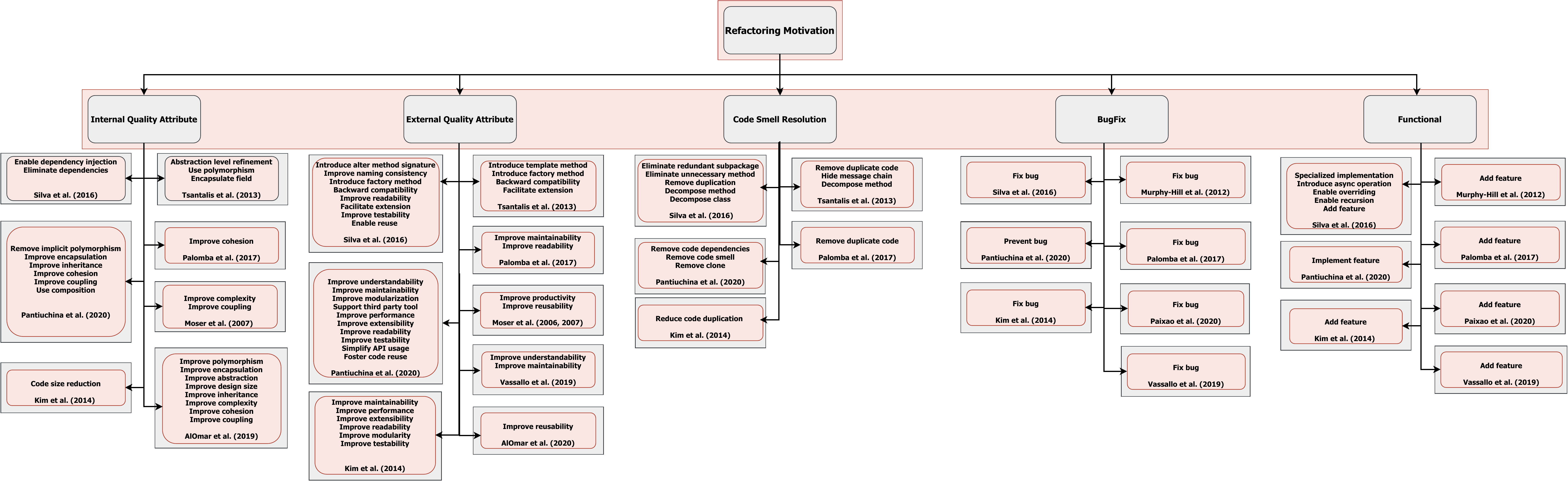}
\caption{Refactoring motivation.}
\label{fig:motivation}
\end{sidewaysfigure}

Another study relevant to our work is by \citep{vassallo2019large}. They performed an exploratory study on refactoring activities in 200 projects, by mining their performed refactoring operations. Their findings show the need for better understanding the rationale behind these operations, and so our study focuses on contextualizing refactoring activities within typical software engineering activities and questions whether such difference in developers’ intentions would infer different refactorings strategies. Such investigation has not been investigated before in the literature.    More recently, \citep{pantiuchina2018developers} present a mining-based study to investigate why developers are performing refactoring in the history of 150 open source systems. Particularly, they analyzed 551 pull requests implemented refactoring operations and reported a refactoring taxonomy that generalizes the ones existing in the literature. \citep{paixao2020behind} perform an empirical study on refactoring activities in code review in which they captured Bug Fix and Feature refactoring categories. \citep{alomar2020how} studied how developers refactor their code to improve its reuse by analyzing the impact of reusability refactorings on the state-of-the-art reusability metrics. Figure \ref{fig:motivation} depicts how our classification clusters the existing refactoring taxonomy reported in the literature \citep{moser2006does,moser2008case,Tsantalis:2013:MES:2555523.2555539,6802406,Silva:2016:WWR:2950290.2950305,palomba2017exploratory,vassallo2019large,alomar2019impact,pantiuchina2018developers,paixao2020behind, alomar2020how}. As can be seen, our classification covers these categories. Furthermore, previous studies have shown that refactoring can be used outside of the \textit{design box}, e.g., correction flaky tests, code naturalness, etc., therefore, our study is the first to engage the automated classification of commit messages in order to cluster the refactoring effort that has been performed in non-design circumstances. 


All the above-mentioned studies have agreed on the existence of motivations that go beyond the basic need for improving the system's design. Refactoring activities have been solicited in scenarios that have been coined by the previous studies as follows: Functional, Bug Fix, Internal Quality Attribute, Code Smell Resolution, and External Quality Attribute. Since these categories are the main drivers for refactoring activities, we decided to cluster our mined refactoring operations according to these groups. 

Our proposal differs from commit classification-related studies, as their classification targeted general maintenance activities (perfective, adaptive) and was not specific to commits containing messages describing refactoring activities. In this study, we subdivide what would have been considered \say{perfective} in previous studies, into three separate categories, namely, Internal Quality Attribute, External Quality Attribute and Code Smell Resolution. This division is inherited from the analysis of previous papers whose detection of refactoring opportunities rely on the optimization of high-level design principles, structural metrics, and reduction of code smells. Thus, this is not a typical commit classification since refactoring related commit messages contain a strong overlap in their terminology and so their classification is challenging. Moreover, as we previously stated, existing studies in recommending refactoring are based on (i) Internal Quality Attribute (ii) External Quality Attribute, and (iii) Code Smell Resolution. The classification of commits according to these categories, will be an empirical evidence of whether and to what extent these factors are being used in practice. 
 To perform the classification, we use existing classifiers (e.g., Random Forest, Naive Bayes Multinominal, etc) that have been used by several studies (e.g., \citep{Hindle:2011:ATN:1985441.1985466,kochhar2014automatic,Levin:2017:BAC:3127005.3127016,honel2019importance,alomar2020toward}) in the context of commit classification and challenge them using our defined set of classes. Although several studies \citep{4686322,Mauczka2012,5090025,article,Levin:2017:BAC:3127005.3127016,Hindle:2008:LCT:1370750.1370773,7180125,YAN2016296} have discussed how to automatically classify change messages into Swanson’s general maintenance categories (i.e., Corrective, Adaptive, Perfective), refactoring, in general, has been classified as a sub-type of \say{Perfective} in these maintenance categories. While we are motivated by the above-mentioned studies, our work is still different from them since we apply the machine learning technique to automatically classify commit messages into five main refactoring motivations defined in this study, i.e., `Functional', `Bug Fix', `Internal Quality Attribute', `Code Smell Resolution', and `External Quality Attribute'.

\section{Empirical Study Setup}
\label{sec:Methodology}
\begin{figure*}[htbp]
\centering 
\includegraphics[width=\textwidth]{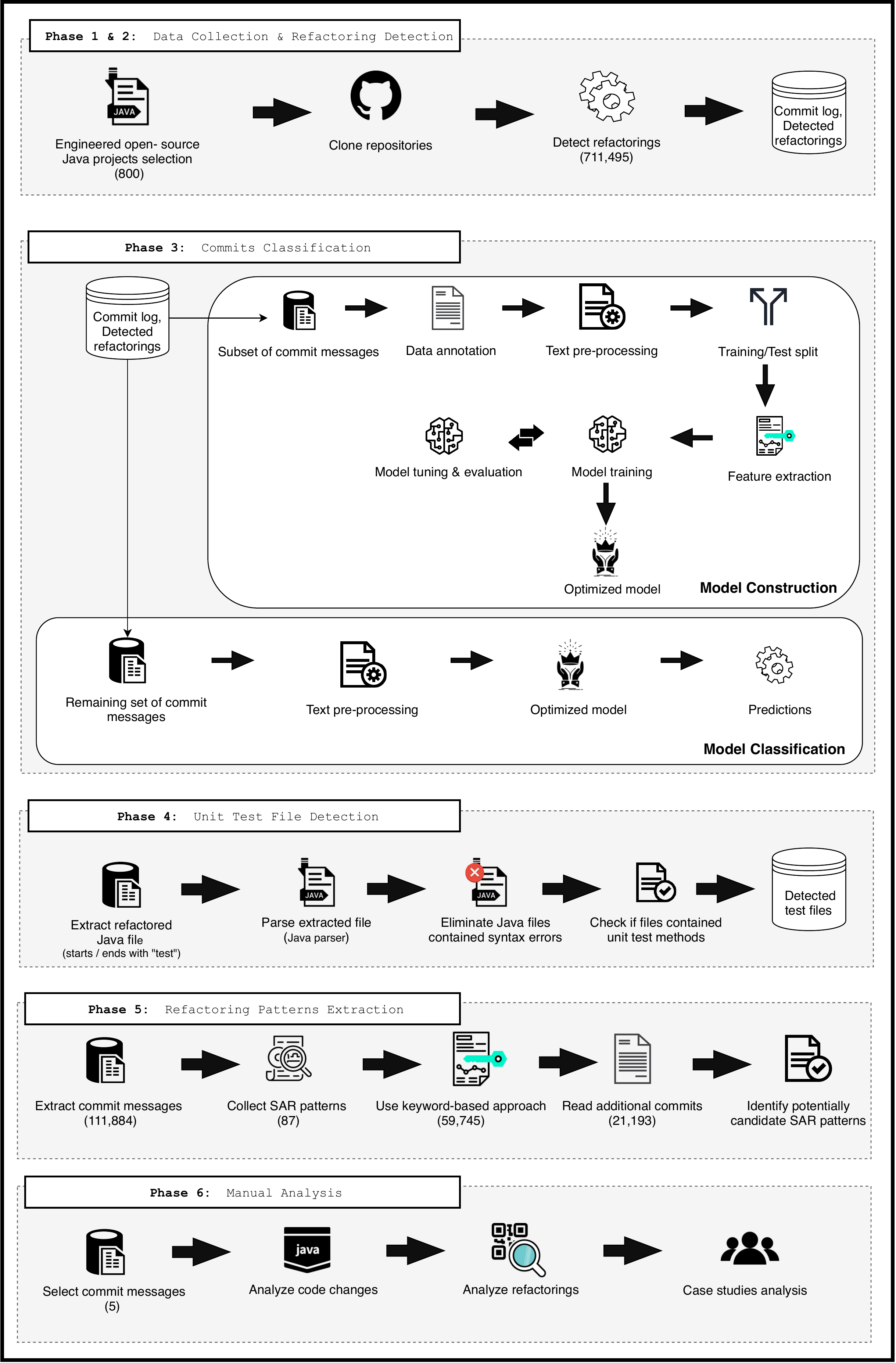}
\caption{Empirical study design overview.} 
\label{fig:approach_overview }
\end{figure*}

To answer our research questions defined in Section \ref{sec:Introduction}, we design a five-steps approach as shown in Figure \ref{fig:approach_overview }. Our approach consists of: (1) data collection, (2) refactoring detection, (3) automatic refactoring classification, (4) unit test files detection, and (5) refactoring 
documentation extraction.

\subsection{Phase 1: Data Collection}
Our first step consists of randomly selecting 800 projects, which were curated open-source Java projects hosted on GitHub. These curated projects were selected from a dataset made available by  \citep{munaiah2017curating}, while verifying that they were Java-based, the only language supported by Refactoring Miner \citep{tsantalis2018accurate}. 
 The authors of this dataset selected \say{well-engineered software projects} based on the projects’ use of software engineering best practices such as documentation, testing, and project management. Additionally, these projects are non-forked (i.e., not cloned from other projects) as forked projects may impact our conclusions by introducing duplicate code and data. 
We cloned the 800 selected projects having a total of 748,001 commits, and a total of 711,495 refactoring operations from 111,884 refactoring commits. Additionally, these projects contain on average 935 commits and 19 developers. An overview of the project's statistics is provided in Table~\ref{Table:DATA_Overview}. This table shows the total number of Java projects used in this study (800), the total number of commits across all projects combined (748,001), the total number of refactoring commits and the associated refactoring operations respectively, 111,884 and 711,495. The table also details the number of refactoring operations per code element at different levels of granularity, including method, attribute, class, variable, parameter,  package, and interface, ordered from highest down to the lowest. Additionally, the standard deviation reported in the table shows that these projects are very diverse.
\begin{table*}[h!]
\begin{center}
\caption{Projects overview.}
\label{Table:DATA_Overview}
\begin{adjustbox}{width=1.0\textwidth,center}
\begin{tabular}{lrr}\hline
\toprule
\bfseries Item & \bfseries Count & \bfseries Standard Deviation \\
\midrule
Total of projects & 800 & N/A \\
Total commits & 748,001 & 1233.69 \\
Refactoring commits & 111,884 & 195.48 \\
Refactoring operations & 711,495 & 2402.12 \\
\midrule 
\multicolumn{2}{c}{\textbf{\textit{Considered Projects - Refactored Code Elements}}}\\
\bfseries Code Element & \bfseries \# of Refactorings & \bfseries Standard Deviation  \\
\midrule
Method & 222,785 & 415.55 \\
Attribute & 201,791 & 1854.35 \\
Class & 121,625 & 273.24  \\
Variable & 115,717 & 383.91 \\
Parameter & 48,054 & 127.48\\
Package & 2380 & 8.25 \\
Interface & 1742 & 6.01 \\
\bottomrule
\end{tabular}
\end{adjustbox}
\end{center}
\end{table*}

\subsection{Phase 2: Refactoring Detection}
To extract the entire refactoring history of each project, we used the Refactoring Miner\footnote{\url{https://github.com/tsantalis/RefactoringMiner}} tool introduced by  \citep{tsantalis2018accurate}. We decided to use Refactoring Miner as it has shown promising results in detecting refactorings compared to the state-of-the-art available tools \citep{tsantalis2018accurate} and is suitable for a study that requires a high degree of automation since it can be used through its external API. The Eclipse plug-in refactoring detection tools (e.g., Ref-Finder \citep{kim2010ref}), in contrast, require user interaction to select projects as inputs and trigger the refactoring detection, which is impractical since multiple releases of the same project have to be imported to Eclipse to identify the refactoring history.

\subsection{Phase 3: Commits Classification}
\label{Phase 3: Commits Classification}
After all refactoring operations are collected, we need to classify them. As part of the development workflow, developers associate a message with each commit they make to the project repository. These commit messages are usually written using natural language, and generally convey some information about the commit they represent. In this study, we aim to determine the type of refactoring activity performed by the developer based on the message associated with a refactoring-based commit.  We started by collecting the different motivations that drive developers to refactor their code as reported in the literature \citep{6802406,7833023,Fowler:1999:RID:311424,lanza2007object,Silva:2016:WWR:2950290.2950305,Tsantalis:2013:MES:2555523.2555539,palomba2017exploratory,6112738}. Then, we search for common categories among the reported motivations. The following step involves identifying categories clustering functional requirements, 
 quality attributes and software issues 
under the identified categories. This process resulted in five different categories. 
 Hence, we aim to classify the refactoring commit, into one of five main categories: `\textit{Functional}', `\textit{Bug Fix}', `\textit{Internal Quality Attribute}', `\textit{Code Smell Resolution}', and `\textit{External Quality Attribute}'. Table \ref{Table:ClassificationCategories} provides a description of each category.

In this supervised multi-class classification problem, we followed a multi-staged approach to build our model for commit messages classification. The first stage consists of the model construction. In the second stage, we utilized the built  model to classify the entire dataset of commit messages. An overview of our methodology is depicted in Figure \ref{fig:approach_overview }. In the following subsections, we describe in detail the different steps in each stage.
\begin{table}
\centering
\caption{Classification categories.}
\label{Table:ClassificationCategories}
\begin{adjustbox}{width=0.8\textwidth,center}
\begin{tabular}{@{}ll@{}}
\toprule
\multicolumn{1}{c}{\textbf{Category}} & \multicolumn{1}{c}{\textbf{Description}} \\ \midrule
Functional & Feature implementation, modification or removal \\
Bug Fix & Tagging, debugging, and application of bug fixes \\
Internal QA & \begin{tabular}[c]{@{}l@{}}Restructuring and repackaging the system's code elements \\ to improve its internal design such as coupling and cohesion\end{tabular} \\
Code Smell Resolution & \begin{tabular}[c]{@{}l@{}}Removal of design defects that might violate the fundamentals \\ of software design principles and decrease code quality such \\ as duplicated code and long method\end{tabular} \\
External QA & \begin{tabular}[c]{@{}l@{}}Property or feature that indicates the effectiveness of a system \\ such as testability, understandability, and readability\end{tabular} \\ \bottomrule
\end{tabular}
\end{adjustbox}
\end{table}

\vspace{2mm} \noindent \textbf{Model Construction} 

In the first stage of the experiment, our goal is to build a model from a corpus real world documented refactorings (i.e., commit message) to be utilized in the second stage to classify commit messages. The following subsections detail the different steps in the model construction phase.

\subsubsection{Data Annotation}
In order to construct a machine learning model, a gold set of labeled data is needed to train and test the model. To prepare this gold set, a manual annotation (i.e., labeling) of commit messages needs to be performed by subject experts. To this end, we annotated 1,702 commit messages. This quantity roughly equates to a sample size with a confidence level of 95\% and a confidence interval of 2.  Confidence level and interval are utilized to obtain an accurate and statistically significant sample size from a population \citep{brownlee2018statistical}. 
 The authors of this paper performed the annotation of the commit messages. Provided to each author was a random set of commit messages along with details defining the annotation labels. Each annotator had to label each provided commit message with a label of either `Functional', `Bug Fix', `Internal Quality Attribute', `Code Smell Resolution', and `External Quality Attribute'. To mitigate bias in the annotation process, the annotated commit messages were peer-reviewed by the same group. All decisions made during the review had to be unanimous; discordant commit messages were discarded and replaced. 
In total, we annotated 348 commit messages as `Functional', `Bug Fix', `Internal Quality Attribute', and `Code Smell Resolution', while 310 messages were labeled as `External Quality Attribute'. 

\subsubsection{Text Pre-Processing}
To better support the model in correctly classifying commit messages, we performed a series of text normalization activities. Normalization is a process of transforming non-standard words into a standard and convenient format \citep{jurafsky2019speech}. Similar to \citep{kochhar2014automatic, le2015rclinker}, the activities involved in our pre-processing stage included: (1) expansion of word contractions (e.g., `I\textquotesingle m' $\rightarrow$ `I am'), (2) removal of URLs, single-character words, numbers, punctuation and non-alphabet characters, stop words, and (3) reducing each word to its lemma. The lemmatization process either replaces the suffix of a word with a different one or removes the suffix of a word to get the basic word form (lemma) \citep{lane2019natural}. In our work, the lemmatization process involves sentence separation, part-of-speech identification, and generating dictionary form. We split the commit messages into sentences, since input text could constitute a long chunk of text. The part-of-speech identification helps in filtering words used as features that aid in key-phrase extraction. Lastly, since the word could have multiple dictionary forms, only the most probable form is generated. We opted to use lemmatization over stemming, as the lemma of a word is a valid English word \citep{lane2019natural}. In relation to stopwords, we used the default set of stopwords supplied by NLTK \citep{Bird2002NLTKTN} and also added our own set of custom stop words. To derive the set of custom stop words, we generated and manually analyzed the set of frequently occurring words in our corpus.  Custom stop words include `git', `code', `refactor', `svn', `gitsvnid', `signedoffby', `reviewedon', `testedby', `us', id', `changeid', `lot', `small', `thing', `way'. Additionally, for more effective pre-processing, we tokenized each commit message. Tokenization is the process of dividing the text into its constituent set of words.


\subsubsection{Training/Test Split}
To gauge the accuracy of a machine learning model, the implemented model must be evaluated on a never-seen-before set of observations with known labels. To construct this set of observations, the set of annotated commit messages were divided into two sub-datasets - a training set and a test set. The training set was utilized to construct the model while the test set was utilized to evaluate the classification ability of the model. For our experiment, we performed a shuffled stratified split of the annotated dataset. Our test dataset contained 25\% of the annotated commit messages, while the training dataset contained the remaining 75\% of annotated commit messages. This split results in the training dataset containing a total of 1,276 commit messages, which breaks down to 246 `Functional', 271 `BugFix', 255 `Internal', 276 `CodeSmell', and 228 `External' labeled commit messages. The stratification was performed based on the class (i.e., annotated label) of the commit messages. The use of a random stratified split ensures a better representation of the different types (i.e., labels) of commit messages and helps reduce the variability within the strata \citep{singh2013elements}.

\subsubsection{Feature Extraction}
In order to create a model, we need to provide the classifier with a set of properties or features that are associated with the observations (i.e., commit messages) in our dataset. However, not all features associated with each observation will be useful in improving the prediction abilities of the model. Hence, a feature engineering task is required to determine the set of optimum features \citep{zheng2018feature}. In our study, we constructed our model using the text in the commit message. Hence, the feature for this model is limited to the commit message. We utilized Term Frequency-Inverse Document Frequency (TF-IDF) \citep{manning2008introduction}, commonly used in the literature \citep{lin2013empirical,le2015rclinker},  to convert the textual data into a vector space model that can be passed into the classifier. In our experiments, we evaluate the accuracy of the model by constructing the TF-IDF vectors using different types of N-Grams and feature sizes. The N-Gram technique is a set of \textit{n-word} that occurs in a text set and could be used as a feature to represent that text \citep{kowsari2019text}. In general, the N-Gram term has more semantic than an isolated word. Some of the keywords (e.g., \say{\textit{extract}}) do not provide much information when used on its own. However, when collecting N-Gram from commit message (e.g., \textit{Refactor createOrUpdate method in MongoChannelStore to extract methods and make code more readable}), the keyword \say{extract} clearly indicates that this refactoring commit belongs to \textit{Extract Method} refactoring. In our classification, we use N-Grams since it is very common to enhance the performance of text classification \citep{tan2002use}. Using TF-IDF, we can determine words that are common and rare across the documents (i.e., commit messages) in our dataset; the model utilizes these words. In other words, The value for each N-Gram is proportional to its TF score multiplied by its IDF score. Thus, each preprocessed word in the commit message is assigned a value which is the weight of the word computed using this weighting scheme. TF-IDF gives greater weight (e.g., value) to words which occur frequently in fewer documents rather than words which occur frequently in many documents.

\subsubsection{Model Training}
For our study, we evaluated the accuracy of six machine learning classifiers: Random Forest, Logistic Regression, Multinomial Naive Bayes, K-Nearest Neighbors, Support Vector Classification (C-Support Vector Classification based on LIBSVM \citep{cSupportVector,LIBSVM}), and Decision Tree (CART \citep{CART}). We selected these classifiers since they are widely adopted in several classification problems in software engineering, as reported in Section \ref{sec:RelatedWork}. 
  It is important to note that the library containing the classification algorithms are capable of multiclass classification. As per the Python's SKlearn documentation, Random Forest, K-Nearest Neighbors, Logistic Regression, and Multinomial Naive Bayes are inherently multiclass \citep{inherently}, while SVC utilizes a one-vs-one approach to handle multiclass \citep{SVC}. Moreover, to ensure consistency, we ran each classifier with the same set of test and training data each time we updated the input features. 


\subsubsection{Model Tuning \& Evaluation}
The purpose of this stage in the model construction process is to obtain the optimal set of classifier parameters that provide the highest performance; in other words, the objective of this task is to tune the hyperparameters. For example, for the K-Nearest Neighbors classifier, we tuned the number of neighbors hyperparameter  (i.e., `k') by evaluating the accuracy of the model as we increased the value of `k' from 1 to 50 in increments of one. We tuned at least one hyperparameter associated with each classifier in our list. For numeric-based hyperparameters, we determined the bounds/range for testing through continuously running the classifier with a different range of values to identify the appropriate minimum and maximum value.

We performed our hyperparameter tuning on the training dataset using a combination of 10-fold cross-validation and an exhaustive grid search \citep{dangeti2017statistics}. Our test dataset did not take part in the training process, which provides a more realistic model evaluation. This approach is also known to prevent overfitting that leads to incorrect conclusions. 
 Grid search utilizes a brute force technique to evaluate all combinations of hyperparameters to obtain the best performance. It is used to find the optimal hyperparameters of a model which results in the most accurate predictions. Since our classification is multiclass, we relied on the Micro-F1 score. The combination of hyperparameters that resulted in the highest Micro-F1 score was selected to construct the model. We provide, in Table \ref{Table:Parameters}, the optimal hyperparameter values for the classification algorithms in our study. 

\begin{table}[h!]
\centering
\caption{Optimal parameter values for the classification algorithms.}
\label{Table:Parameters}
\begin{adjustbox}{width=0.40\textwidth,center}
\begin{tabular}{@{}lll@{}}
\toprule
\multicolumn{1}{l}{\textbf{Algorithm}} & \multicolumn{1}{l}{\textbf{Parameter}} & \multicolumn{1}{l}{\textbf{Value}} \\ \midrule
\multirow{4}{*}{Random Forest} & max\_depth & 78 \\
 & n\_estimators & 500 \\
 & criterion & gini \\
 & bootstrap & false \\ \midrule
\multirow{3}{*}{Support Vector Classification} & c & 1.99 \\
 & gamma & scale \\
 & kernel & linear \\ \midrule
\multirow{2}{*}{Decision Tree} & criterion & gini \\
 & max\_depth & 75 \\ \midrule
\multirow{3}{*}{Logistic Regression} & penalty & l1 \\
 & solver & liblinear \\
 & c & 1.0 \\ \midrule
Multinomial Naive Bayes & alpha & 2.63 \\ \midrule
\multirow{2}{*}{K-Nearest Neighbors} & n\_neighbors & 69 \\
 & weights & uniform \\ \bottomrule
\end{tabular}
\end{adjustbox}
\end{table}

\subsubsection{Optimized Model}
In this stage, the optimized model produced by the training phase is utilized to predict the labels of the test dataset. Based on the predictions, we measure the precision and recall for each label as well as the overall F1-score of the model. In Section \ref{sec:Experimental Results}, we detail our classification results. 

\vspace{2mm} \noindent \textbf{Model Classification}

In this stage of our experiment, we utilized the optimized model that we created in the prior stage. However, to be consistent, before classifying each commit message, we performed the same text pre-processing activities, as in the prior stage, on the commit message. The result of this stage is the classification of each refactoring commit into one of the five categories. The output of this classification process was utilized in our experiments in order to answer our research questions.

\subsection{Phase 4: Unit Test File Detection} 
As part of our study, we distinguish between refactorings applied to production and unit test files and perform comparisons against both production-file-based refactorings versus test-file-based refactorings. To identify all test files that were refactored, we followed the same detection approach as \citep{10.5555/3370272.3370293}. In this approach, following JUnit's file naming standards\footnote{\url{https://junit.org/junit4/faq.html\#running\_15}}, we first extracted all refactored Java source files where the filename either starts or ends with the word \say{test}. Next, we utilized JavaParser\footnote{\url{https://javaparser.org/}} to parse each extracted file.  By parsing the files, we were able to eliminate Java files that contained syntax errors and were able to detect if the file contained JUnit-based unit test methods accurately, thereby cutting down on false positives. Finally, to ensure that the files were indeed unit test files, we checked if the files contained unit test methods. As per JUnit specifications, a test method should have a public access modifier, and either has an annotation called @Test (JUnit 4), or the method name should start with \say{test} (JUnit 3).

\subsection{Phase 5: Refactoring Patterns Extraction} 
To identify self-affirmed refactoring patterns, we perform manual analysis similar to our previous work \citep{alomar2019can}. Since commit messages are written in natural language and we need to understand how developers document their refactoring activities, we manually analyzed commit messages by reading through each message to identify self-affirmed refactorings. We then extracted these commit comments to specific patterns (i.e., a keyword or phrase). To avoid redundancy of any kind of patterns, we only considered one phrase if we found different forms of patterns that have the same meaning. For example, if we find patterns such as \say{simplifying the code}, \say{code simplification}, and \say{simplify code}, we add only one of these similar phrases in the list of patterns. This enables having a list of the most insightful and unique patterns. It also helps in making more concise patterns that are usable for readers. We also analyzed the top 100 features, distilled by the classifier, for each category. 

The manual analysis process took approximately 20 days in total. In the first two weeks, the authors had regular meetings to discuss top features, extracted from each category, to understand how each class was represented by its corresponding set of keywords, along with extracting any patterns that are most likely to be descriptive to refactoring, besides being another verification level of the classification accuracy. Moreover, during these meetings, the extraction of textual patterns from commit messages was also performed by the authors. Due to the subjective nature of this process, we opted to report as many keywords as possible for better coverage. When reporting keywords from top features, we kept the majority of keywords, for each category. keywords that were removed were either proper names of code elements (method names, identifiers, etc.), or languages and frameworks. For the identification of patterns from commit messages, the authors kept any keyword that can be either tightly or loosely coupled to refactoring. Such decision mitigates the selection bias, at the expenses of reporting keywords that may or may not be relevant to refactoring documentation. During the last week, two authors have finished analyzing the remaining commit messages. This step resulted in analyzing 59,745 commit messages. Then, we iterated over the set again while excluding the terms identified in our previous work, to identify additional self-affirmed refactoring patterns. We manually read through 21,193 commit messages. Our in-depth inspection resulted in a list of \sarInitialNumber potential self-affirmed refactoring candidates, identified across the considered projects, as illustrated later in Tables~\ref{Table:GeneralPatterns} and~\ref{Table:SpecificPatterns_1}. 

\subsection{Phase 6: Manual Analysis}
To get a more qualitative sense of the classification results, we created five case studies that demonstrate GitHub developers' intentions when refactoring source code. Case study is one of the empirical methods used for studying phenomena in a real-life context \citep{wohlin2012experimentation}. In our study, we performed a combination of manual analysis and quantitative analysis using custom-built scripts. For each case study, we provide the commit message and its corresponding refactoring operations detected by Refactoring Miner. We elaborate in detail these case studies in Section \ref{sec:casestudies}, where we report on our results.

\section{Experimental Results}
\label{sec:Experimental Results}
This section reports and discusses our experimental results and aims to answer the research questions in Section~\ref{sec:Introduction}. 

\textbf{Replication package.} We provide our comprehensive experiments package available in \citep{SAR2020WEB} to further replicate and extend our study. The package includes the selected Java projects, the detailed refactoring and non-refactoring commits and documentation, manual commits classification, the automatic commits classification, and the JUnit file detection.

\subsection{RQ1: To what purposes developers refactor their code?}

To answer this research question, we present the refactoring commit messages classification results explained in Subsection~\ref{Phase 3: Commits Classification}. This section details the classification of 111,884 commit messages containing 711,495 refactoring operations. 
The complete set of scores for all the classifiers including the Precision, Recall, and F-measure scores per class for each machine learning classifier is provided in Table \ref{Table:ClassifierScores_Details}. The best performing model was used to classify the test dataset. Based on our findings, we observed that \textbf{\textit{Random Forest achieved the best F1 score: 87\%}} which is higher than its competitors. Random Forest belongs to the family of ensemble learning machines, and has typically yielded superior predictive performance mainly due to the fact that it aggregates several learners. Hence, we utilized this machine learning algorithm (and its optimal set of hyperparameters) as the optimum model for our study. In order to compare classification algorithms performance, we use the McNemar test \citep{dietterich1998approximate}. We compare the performance of Random Forest against the other five classifiers. As shown in Table \ref{Table:McNemar’sTest}, the McNemar’s test results show that there are statistically significant differences in the performance of the classifiers except for the classifier Support Vector Classification in which the difference is not statistically significant. 

\begin{table}

\centering
\caption{Detailed classification metrics (Precision, Recall, and F-measure) of each classifier.} 
\label{Table:ClassifierScores_Details}
\centering
\begin{adjustbox}{width=1.0\textwidth,center}
\begin{tabular}{lrrrlrrrlrrr}
\hline
\multicolumn{4}{|c|}{\textit{\textbf{Random Forest}}} & \multicolumn{4}{c|}{\textit{\textbf{Support Vector Classification}}} & \multicolumn{4}{c|}{\textit{\textbf{Decision Tree}}} \\ \hline
\multicolumn{1}{c}{\textbf{Category}} & \multicolumn{1}{c}{\textbf{Precision}} & \multicolumn{1}{c}{\textbf{Recall}} & \multicolumn{1}{c|}{\textbf{F1}} & \multicolumn{1}{c}{\textbf{Category}} & \multicolumn{1}{c}{\textbf{Precision}} & \multicolumn{1}{c}{\textbf{Recall}} & \multicolumn{1}{c|}{\textbf{F1}} & \textbf{Category} & \multicolumn{1}{l}{\textbf{Precision}} & \multicolumn{1}{l}{\textbf{Recall}} & \multicolumn{1}{l}{\textbf{F1}} \\ \hline
Bug Fix & 0.83 & 0.79 & \multicolumn{1}{r|}{0.81} & Bug Fix & 0.75 & 0.78 & \multicolumn{1}{r|}{0.77} & Bug Fix & 0.77 & 0.80 & 0.78 \\
Code Smell & 0.93 & 0.95 & \multicolumn{1}{r|}{0.94} & Code Smell & 0.93 & 0.94 & \multicolumn{1}{r|}{0.93} & Code Smell & 0.89 & 0.91 & 0.90 \\
External QA & 0.85 & 0.91 & \multicolumn{1}{r|}{0.88} & External QA & 0.92 & 0.89 & \multicolumn{1}{r|}{0.90} & External QA & 0.77 & 0.90 & 0.83 \\
Functional & 0.81 & 0.91 & \multicolumn{1}{r|}{0.86} & Functional & 0.77 & 0.88 & \multicolumn{1}{r|}{0.82} & Functional & 0.92 & 0.83 & 0.87 \\
Internal QA & 0.95 & 0.81 & \multicolumn{1}{r|}{0.87} & Internal QA & 0.95 & 0.84 & \multicolumn{1}{r|}{0.89} & Internal QA & 0.91 & 0.80 & 0.85 \\ \hline
Average F1 & 0.87 & 0.87 & \multicolumn{1}{r|}{0.87} & Average F1 & 0.87 & 0.86 & \multicolumn{1}{r|}{0.86} & Average F1 & 0.85 & 0.85 & 0.85 \\ \hline
\multicolumn{12}{l}{} \\ \hline
\multicolumn{4}{|c|}{\textit{\textbf{Logistic Regression}}} & \multicolumn{4}{c|}{\textit{\textbf{Multinomial Naive Bayes}}} & \multicolumn{4}{c|}{\textit{\textbf{K-Nearest Neighbors}}} \\ \hline
\multicolumn{1}{c}{\textbf{Category}} & \multicolumn{1}{c}{\textbf{Precision}} & \multicolumn{1}{c}{\textbf{Recall}} & \multicolumn{1}{c|}{\textbf{F1}} & \multicolumn{1}{c}{\textbf{Category}} & \multicolumn{1}{c}{\textbf{Precision}} & \multicolumn{1}{c}{\textbf{Recall}} & \multicolumn{1}{c|}{\textbf{F1}} & \multicolumn{1}{c}{\textbf{Category}} & \multicolumn{1}{c}{\textbf{Precision}} & \multicolumn{1}{c}{\textbf{Recall}} & \multicolumn{1}{c}{\textbf{F1}} \\ \hline
Bug Fix & 0.66 & 0.70 & \multicolumn{1}{r|}{0.68} & Bug Fix & 0.63 & 0.77 & \multicolumn{1}{r|}{0.69} & Bug Fix & 0.62 & 0.71 & 0.66 \\
Code Smell & 0.89 & 0.94 & \multicolumn{1}{r|}{0.91} & Code Smell & 0.82 & 0.94 & \multicolumn{1}{r|}{0.87} & Code Smell & 0.76 & 0.93 & 0.84 \\
External QA & 0.88 & 0.88 & \multicolumn{1}{r|}{0.88} & External QA & 0.97 & 0.71 & \multicolumn{1}{r|}{0.82} & External QA & 0.85 & 0.75 & 0.79 \\
Functional & 0.77 & 0.87 & \multicolumn{1}{r|}{0.82} & Functional & 0.66 & 0.83 & \multicolumn{1}{r|}{0.74} & Functional & 0.68 & 0.73 & 0.71 \\
Internal QA & 0.96 & 0.78 & \multicolumn{1}{r|}{0.86} & Internal QA & 0.99 & 0.67 & \multicolumn{1}{r|}{0.80} & Internal QA & 0.97 & 0.71 & 0.82 \\ \hline
Average F1 & 0.83 & 0.83 & \multicolumn{1}{r|}{0.83} & Average F1 & 0.81 & 0.78 & \multicolumn{1}{r|}{0.78} & Average F1 & 0.78 & 0.77 & 0.76 \\ \hline
\end{tabular}
\end{adjustbox}
\end{table}

\begin{table}

\centering
\caption{McNemar’s test results.} 
\label{Table:McNemar’sTest}
\centering
\begin{tabular}{lrrr}
\hline
\multicolumn{4}{|c|}{\textit{\textbf{Random Forest}}}  \\ \hline
\multicolumn{1}{c}{\textbf{Classifier}} & \multicolumn{1}{c}{\textbf{p-value}} & \multicolumn{1}{c}{\textbf{}} & \multicolumn{1}{c}{\textbf{}}  \\ \hline
Support Vector Classification & 0.1 &  \multicolumn{1}{r}{}  \\
Decision Tree & 0.04 &  \multicolumn{1}{r}{} \\
Logistic Regression & 0.02 &  \multicolumn{1}{r}{}  \\
Multinomial Naive Bayes & 0.01  & \multicolumn{1}{r}{}  \\
K-Nearest Neighbors& 0.01 &  \multicolumn{1}{r}{}  \\ \hline
\end{tabular}
\end{table}

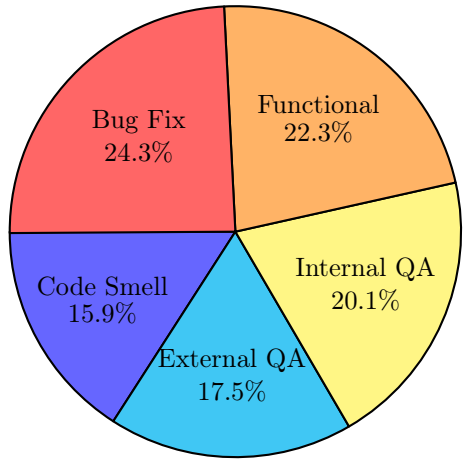
\begin{figure*}[t]
\centering 
\begin{tikzpicture}
\pie[rotate = 180,pos ={0,0},text=inside,outside under=50,no number]{15.9/Code Smell\and15.9\%,17.5/External QA\and17.5\%, 20.1/Internal QA\and20.1\%, 22.3/Functional\and22.3\%,24.3/Bug Fix\and24.3\%}
\end{tikzpicture}
\caption{Percentage of classified commits per category in all projects combined.}
\label{fig:commit_classification}
\end{figure*}

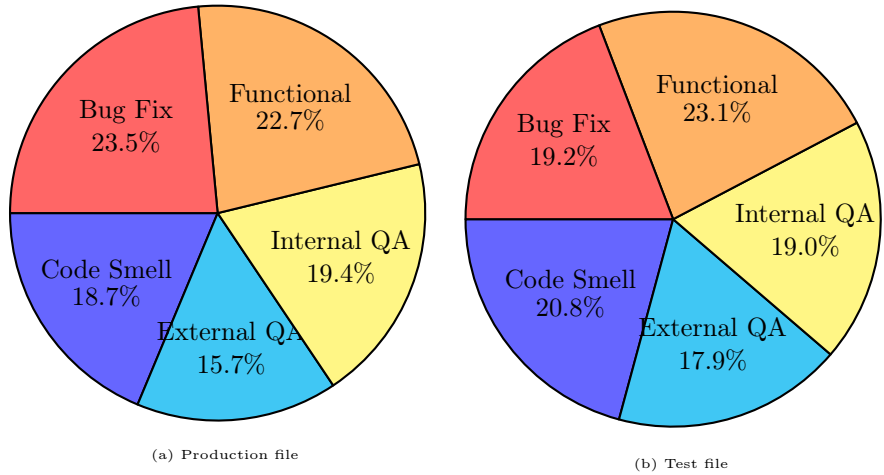
\begin{figure*}[h]
\begin{subfigure}{.5\textwidth}
\centering
\begin{tikzpicture}
\begin{scope}[scale=1.15]
\pie[rotate = 180,radius =2.4,text=inside,outside under=45,no number]{18.7/Code Smell\and18.7\%,15.7/External QA\and15.7\%,  19.4/Internal QA\and19.4\%, 22.7/Functional\and22.7\%,23.5/Bug Fix\and23.5\%}{Production}
\end{scope}
 \end{tikzpicture}
 \caption{Production file}
 \end{subfigure}%
\begin{subfigure}{.5\textwidth}
\centering
\begin{tikzpicture}
\begin{scope}[scale=1.15]
\pie[rotate = 180,radius =2.4,pos ={3,0}, text=inside,outside under=45,no number]{20.8/Code Smell\and20.8\%,17.9/External QA\and17.9\%, 19.0/Internal QA\and19.0\%,23.1/Functional\and23.1\%, 19.2/Bug Fix\and19.2\%}
{Test File}
\end{scope}
 \end{tikzpicture}
 \caption{Test file}
 \end{subfigure}%
 
\caption{Percentage of classified commits per category in production and test files.} 
\label{fig:commit_classification_prodtest}
\end{figure*}

Figure~\ref{fig:commit_classification} shows the categorization of commits, from all projects combined. We observe that all of the categories had almost a uniform distribution of refactoring classes with low variability. For instance, Bug Fix, Functional, Internal Quality Attribute, External Quality Attribute, and Code Smell Resolution had  commit message distribution percentages of 24.3\%, 22.3\%, 20.1\%, 17.5\%, and 15.9\%, respectively. 

The first observation that we can draw from these findings is that developers do not solely refactor their code to fix code smells. 
They instead refactor the code for multiple purposes. Our manual analysis show that developers tend to make design-improvement decisions that include re-modularizing packages by moving classes, reducing class-level coupling, increasing cohesion by moving methods, and renaming elements to increase naming quality in the refactored design. 
Developers also tend to split classes and extract methods for: 1) separation of concerns, 2) helping in easily adding new features, 3) reducing bug propagation, and 4) improving the system's non-functional attributes such as extensibility and maintainability 

Figure ~\ref{fig:commit_classification_prodtest} depicts the distribution of refactoring commits for all production and test files for each refactoring motivation. As can be seen, developers tend to refactor these two types of source files for several refactoring intentions, and they care about refactoring the logic of the application and refactoring the test code that verifies if the application works as expected. Although developers usually handle production and test code differently, the similarity of the patterns shows that they refactor these source files for the same reasons with unnoticeable differences.

\textbf{Production code.} Concerning refactorings applied in the production files, developers perform refactoring for several motivations. For the Bug Fix category, an interpretation for this comes from the nature of the debugging process that includes the disambiguation of identifier naming that may not reflect the appropriate code semantics or that may be infected with lexicon bad smells (i.e., linguistic anti-patterns \citep{abebe2011effect,arnaoudova2013new}). Another debugging practice would be the separation of concerns, which helps in reducing the core complexity of a larger module and reduces its proneness to errors \citep{tsantalis2011identification}. Regarding the Internal Quality Attributes category, developers move code elements for design-level changes \citep{stroggylos2007refactoring,alshayeb2009empirical,bavota2015experimental,mkaouer2015many},  e.g., developers tend to re-modularize classes to make packages more cohesive, and extract methods to reduce coupling between classes. 
 As for the External Quality Attributes category, developers often optimize the code to improve the non-functional quality attributes such as readability, understandability, and maintainability of the production files. For the Code Smell Resolution category, developers eliminate any bad practices and adhere to object-oriented design principles. Finally, for the Functional category, developers implement a new feature or modify the existing ones. 

\textbf{Test code.} With regards to test files, developers perform refactoring to improve the design of the code. An example can be shown by renaming a given code element such as a class, a package or an attribute. Finding better names for code identifiers serves the purpose of increasing the software's comprehensibility. Developers explicitly mention the use of the renaming operations for the purpose of disambiguation the redundancy of methods names and enhancing their usability. Another activity to refactor test files could be moving methods, or pushing code elements across hierarchies, e.g., pushing up attributes. Each of these activities are performed to support several refactoring motivations.

\begin{tcolorbox}
\textit{Summary}. Our study has shown that fixing code smells is not the main driver for developers to refactor their code bases. Indeed, the percentage of commits belonging to this category account for only 15.9 \% of the overall classified refactoring commits, making this class the least among all other categories. Bug Fix is found to be leading with a percentage of 24.3\%, but, the sum of the design-related categories, namely Code Smell, Internal Quality Attribute, and External Quality Attribute, represent the majority with a 53.5\%. As explicitly mentioned by the developers in their commits messages, refactoring is solicited for a wide variety of reasons, going beyond its traditional definition, such as reducing the software's proneness to bugs, easing the addition of functionality, resolving lexical ambiguity, enforcing code styling, and improving the design's testability and reusability.
\end{tcolorbox}


\subsection{Case Studies}
\label{sec:casestudies}

This subsection reveals more details with respect to our classified commits. As we validate our classification results, we have selected an example from each category. For each example, we checkout the corresponding commit to obtain the source code, then two authors manually analyze the code changes. The purpose is not to verify the consistency between the commit message and its corresponding changes, but to capture the context in which refactorings were applied. In each analyzed commit, we report its class, its message, the distribution of its corresponding refactoring, along with our understanding of their usage context.  

\subsubsection{Case Study 1. Refactoring to improve internal quality attributes}



\begin{center}
\includegraphics[width=\columnwidth]{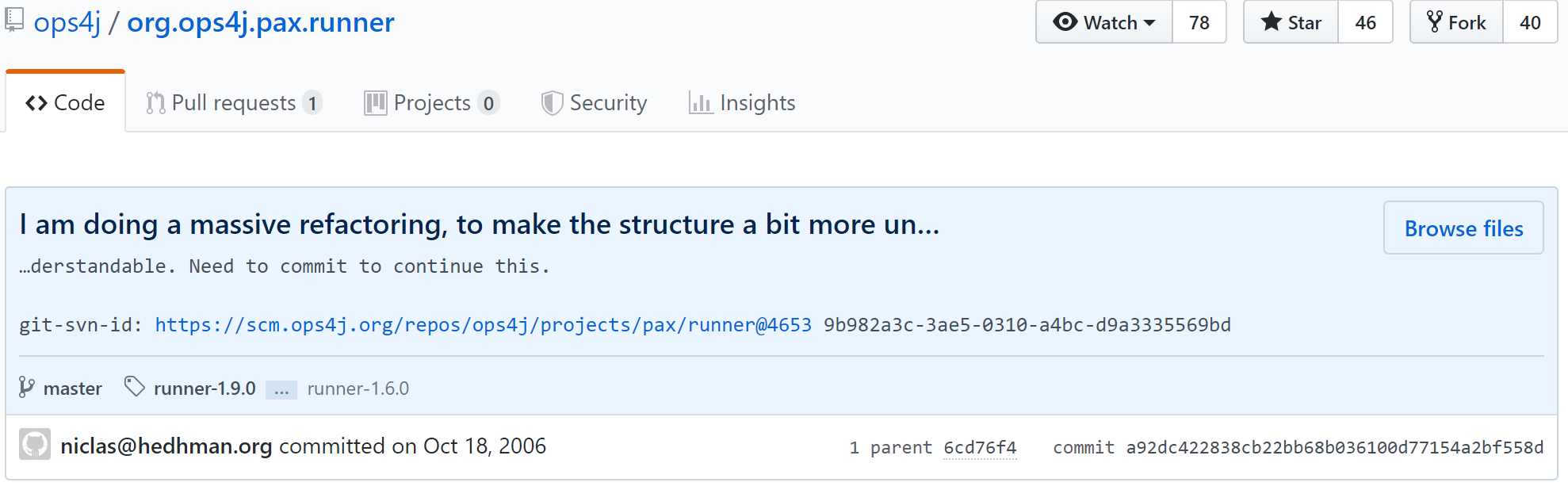}
\captionof{figure}{Commit message stating the restructuring of code to improve its structure.}
\label{fig:Example of Design}

\vspace{0.70cm}

\begin{tikzpicture}
\pie[text=legend]{2.90/Move \& Rename Class, 7.25/Move Source Folder,8.70/Change Package, 81.16/Move Class}
\end{tikzpicture}
\captionof{figure}{Distribution of refactoring operations.}
\label{fig:Refactoring_Operations_Design}
\end{center}

\vspace{0.60cm}

This case study aims to demonstrate one of the five refactoring motivations reported in this study. The commit message mainly discussed two refactoring practices: (1) performing large refactorings, and (2) optimizing the structure of the codebase. It is apparent from this commit that the main intention behind refactoring the code is to improve the design. Specifically, Refactoring Miner detected 69 refactoring operations associated with this commit message. We observe there is consistency between what is documented in the commit message and the actual size of refactoring operations. 

Closer inspection of the nature and type of the 69 refactorings and the corresponding source code shows that the GitHub commit author massively optimized the package structure within existing modularizations. Particularly, as Figure~\ref{fig:Refactoring_Operations_Design} shows, the developer performed four refactoring types, namely, \textit{Move Class, Change Package, Move Source Folder}, and \textit{Move and Rename Class}. A percentage of 80.16\% of these refactorings were \textit{Move Class} refactorings, 8.70\% were \textit{Change Package}, 7.25\% were \textit{Move Source Folder}, and 2.90\% were composite refactorings (\textit{Move and Rename Class}). As pointed out in Refactoring Miner documentation\footnote{\url{https://github.com/tsantalis/RefactoringMiner}}, 
 \textit{Change Package} refactoring involves several package-level refactorings (i.e., Rename, Move, Split, and Merge packages). 

We observe that the developer is optimizing the design by performing repackaging, i.e., extracting packages and moving the classes between these packages, merging packages that have classes strongly related to each other, and renaming packages to reflect the actual behavior of the package. The present observations are significant in at least two major respects: (1) improving the quality of packages structure when optimizing intra-package (i.e., cohesion) and inter-package (i.e., coupling) dependencies and minimizing package cycles and (2) avoiding increasing the size of the large packages and/or merging packages into larger ones. Developer intention to distribute classes over packages, however, might depend on other design factors than package cohesion and coupling. This remodularization activity helps to identify packages containing classes poorly related to each other. 

In order to confirm the main refactoring intention when performing this refactoring, we emailed the GitHub contributor and asked about the main motivation behind performing this massive refactorings in the commit message (Figure~\ref{fig:Example of Design}). The GitHub contributor confirmed that the intention was to improve the design and this motivation is best illustrated in the following response about the commit we examined:

\begin{quote}
  \say{\textit{there are a few reasons for large refactorings: (1) the codebase is becoming increasingly difficult to evolve. Sometimes relatively small conceptual changes can make a huge difference, but requires a lot of changes in many places.}} and \say{\textit{(2) analysis of codebase dependencies, call sequences and so on reveal that the codebase is a mess and needs to be fixed to avoid current or future bugs.}}  
\end{quote}

The most striking observation to emerge from the response was that as a program evolves in size it is vital to design it by splitting it into modules, so that developer does not need to understand all of it to make a small modification. Generally, refactoring to improve the design at different levels of granularity is crucial. This case study sheds light on the importance of refactoring at package-level of granularity  and how it plays a crucial role in the quality and maintainability of the software. In future investigations, it might be possible to extend this work by learning from existing remodularization process and then recommending the right package for a given class taking into account the design quality (e.g., coupling, cohesion, and complexity). Future studies on remodularization topic can develop refactoring tool which can refactor software systems at different levels of granularity.


\subsubsection{Case Study 2. Refactoring to remove code smells}
\vspace{1em}

\begin{center}
\includegraphics[width=\columnwidth]{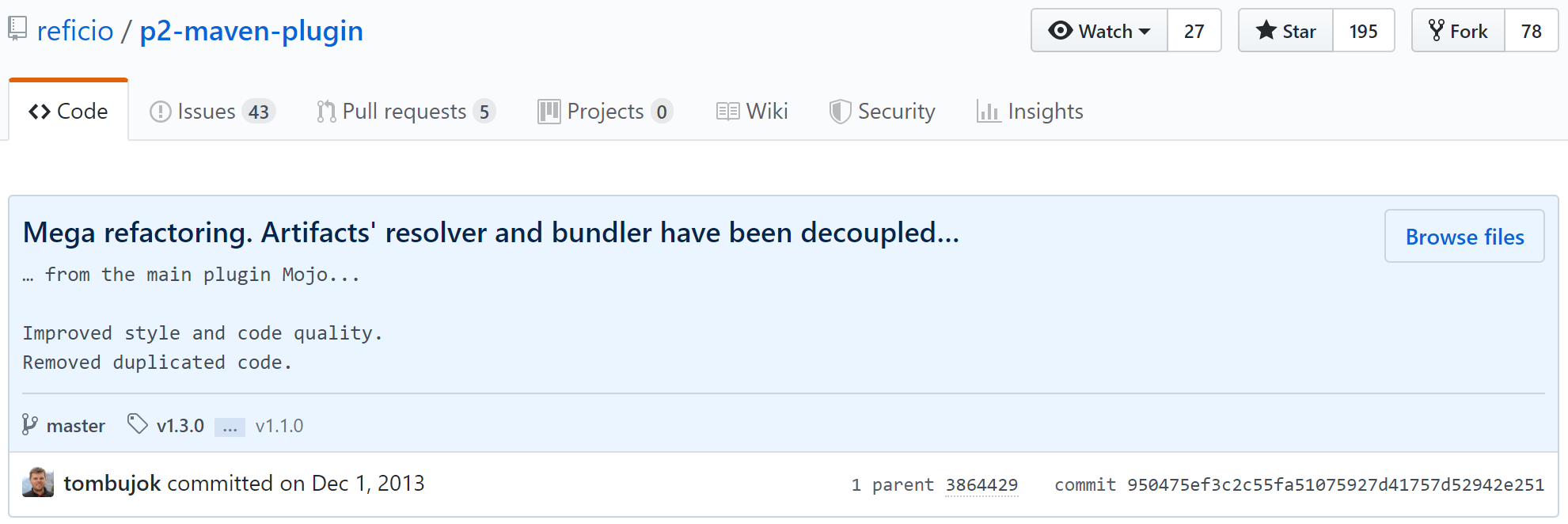}
\captionof{figure}{Commit message stating the removal of duplicate code.}
\label{fig:Example of code smell}

\vspace{0.75cm}

\begin{tikzpicture}
\pie[rotate = 160,outside under=30,text=legend]{2.63/Change Package, 2.63/Extract Method,2.63/Extract Variable,2.63/Rename Variable,5.26/Move Class,5.26/Move \& Rename Class,10.54/Extract Class,13.16/Move Method,21.05/Move Attribute,34.21/Rename Method}
\end{tikzpicture}
\captionof{figure}{Distribution of refactoring operations.} 
\label{fig:Refactoring_Operations_CodeSmell}
\end{center}

\vspace{0.75cm}
This case study illustrates another developer’s perception of refactoring which is mainly about code smell resolution. Figure~\ref{fig:Example of code smell} shows that the developer performed large-scale refactoring to eliminate duplicated code. Generally, code duplication belongs to the \say{Dispensable} code smell category, i.e., code fragments that are unneeded and whose absence would make the code cleaner and more efficient. 

Figure~\ref{fig:Refactoring_Operations_CodeSmell} depicts the 38 refactoring operations performed in which the developer removed duplicated code. 
 The developer performed 10 different types of refactorings associated with the commit message shown in Figure~\ref{fig:Example of code smell}: \textit{Rename Method, Move Attribute, Move Method, Extract Class, Move and Rename Class, Move Class, Rename Variable, Extract Variable, Extract Method}, and \textit{Change Package}. 

From the pie chart, it is clear that the majority of refactorings performed were \textit{Rename Method} and \textit{Move Attribute} with 34.21\% and 21.05\% respectively, followed by \textit{Move Method} with 13.16\% and \textit{Extract Class} refactorings with 10.54\%. Nearly 5\% were \textit{Move Class} and \textit{Move and Rename Class} refactorings and only a small percentage of refactoring commits were \textit{Change Package, Extract Method, Extract Variable}, and \textit{Rename Variable}. 

On further examination of the source code and the corresponding refactorings detected by the tool, we notice that there are a variety of cases in which the code fragments are considered duplicate. One case is when the same code structure is found in more than one place in the same class, and the other one is when the same code expression is written in two different and unrelated classes. The developer treated the former case by using \textit{Extract Method} refactoring followed by the necessarily naming and moving operations and then invoked the code from both places. As for the latter case, the developer solved it by using \textit{Extract Class} refactoring and the corresponding renaming and moving operations for the class and/or attribute that maintained the common functionalities. The developer also performed \textit{Change Package} refactorings when removing code duplication as a complementary step of refactoring, which could indicate motivations outside of those described in the commit message.  

It appears to us that composite refactorings have been performed for resolving this code smell. These activities help eliminate code duplication since merging duplicate code simplifies the design of the code. Additionally, these activities could help improving many code metrics such as the lines of code (LOC), the cyclomatic complexity (CC), and coupling between objects (CBO).

\subsubsection{Case Study 3. Refactoring to improve external quality Attributes}
\vspace{1em}

\begin{center}
\includegraphics[width=\columnwidth]{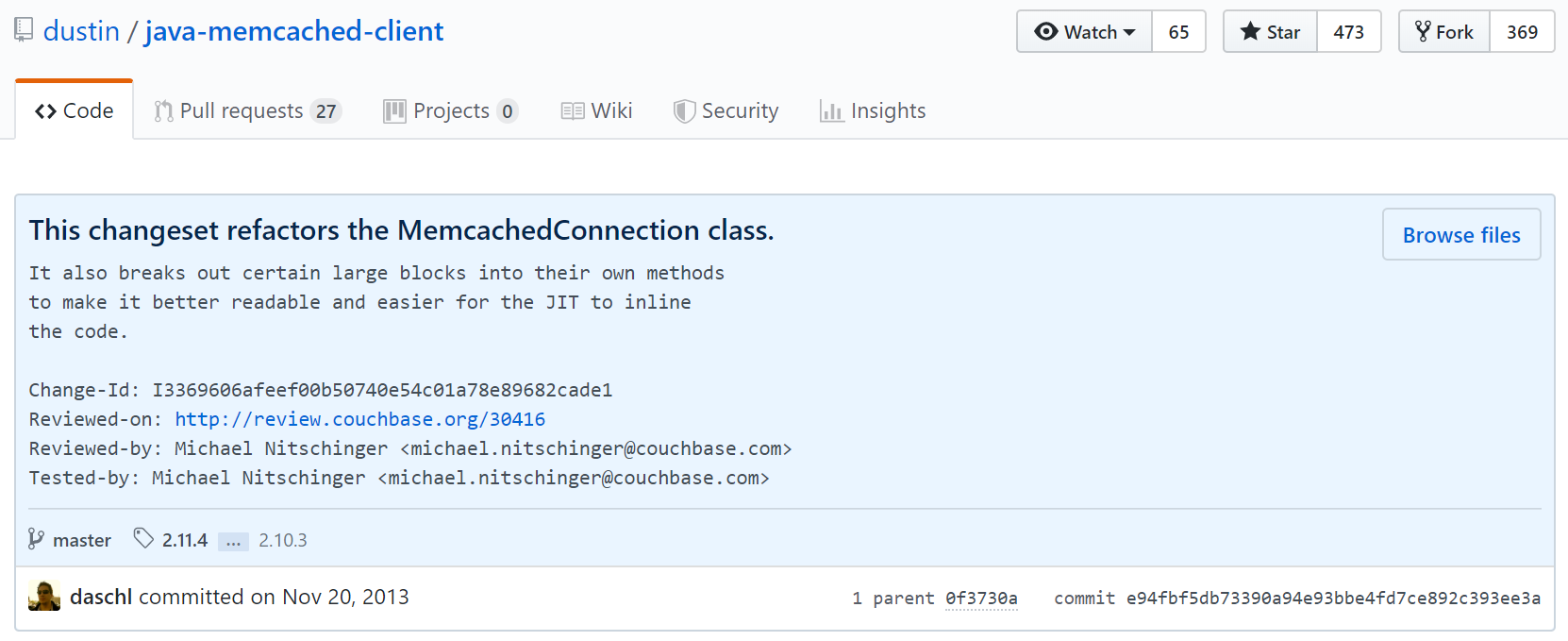}
\captionof{figure}{Commit message stating the refactoring to improve code readability.}
\label{fig:Example of Non-Functional}

\vspace{0.75cm}

\begin{tikzpicture}
\pie[text=legend]{ 6.25/Parameterize Variable, 25/Rename Variable,31.25/Extract Method,37.50/Rename Parameter}
\end{tikzpicture}
\captionof{figure}{Distribution of refactoring operations.}
\label{fig:Refactoring_Operations_nonfunctional}
\end{center}

\vspace{0.75cm}
This case study demonstrates another refactoring intention which is related to improving external quality attributes (i.e., indication of  the enhancement of non-functional attributes such as readability and understandability of the source code). As shown in Figure~\ref{fig:Example of Non-Functional}, the developer stated that the purpose of performing this refactoring is to improve the readability of the source code by breaking large blocks of code into separate methods. 

Figure~\ref{fig:Refactoring_Operations_nonfunctional} illustrates the breakdown of 16 refactoring operations related to readability associated with this commit message. It can be seen that \textit{Rename Parameter} and  \textit{Extract Method} refactorings have the highest refactoring-related commits with 37.50\% and 31.25\%, respectively. \textit{Rename Variable} is the third most performed refactoring with 25\%, in front of \textit{Parameterize Variable} refactorings at 6.25\%. By analyzing the corresponding source code, it is clear that developer decomposed four methods for better readability, namely, \texttt{handleIO(), handleIO(sk SelectionKey), handleReads(sk SelectionKey, qa MemcachedNode)}, and \texttt{attemptReconnects()}. The name could also change for a reason (e.g., when \textit{Extract Method} is applied to a method, the method name and its parameters or variables also update as a result).

To improve code readability, developer used \textit{Extract Method} refactorings as a treatment for this case study to reduce the length of the method body. Additionally, renaming operations were used to improve naming quality in the refactored code and reflect the actual purpose of the parameters and variables. Converting variables to parameters could also make the methods more readable and understandable. To develop a full picture of how to create readable code, future studies will be needed to focus on code readability guidelines or rules for developers. 



\subsubsection{Case Study 4. Refactoring to add feature}
\vspace{1em}

\begin{center}
\includegraphics[width=\columnwidth]{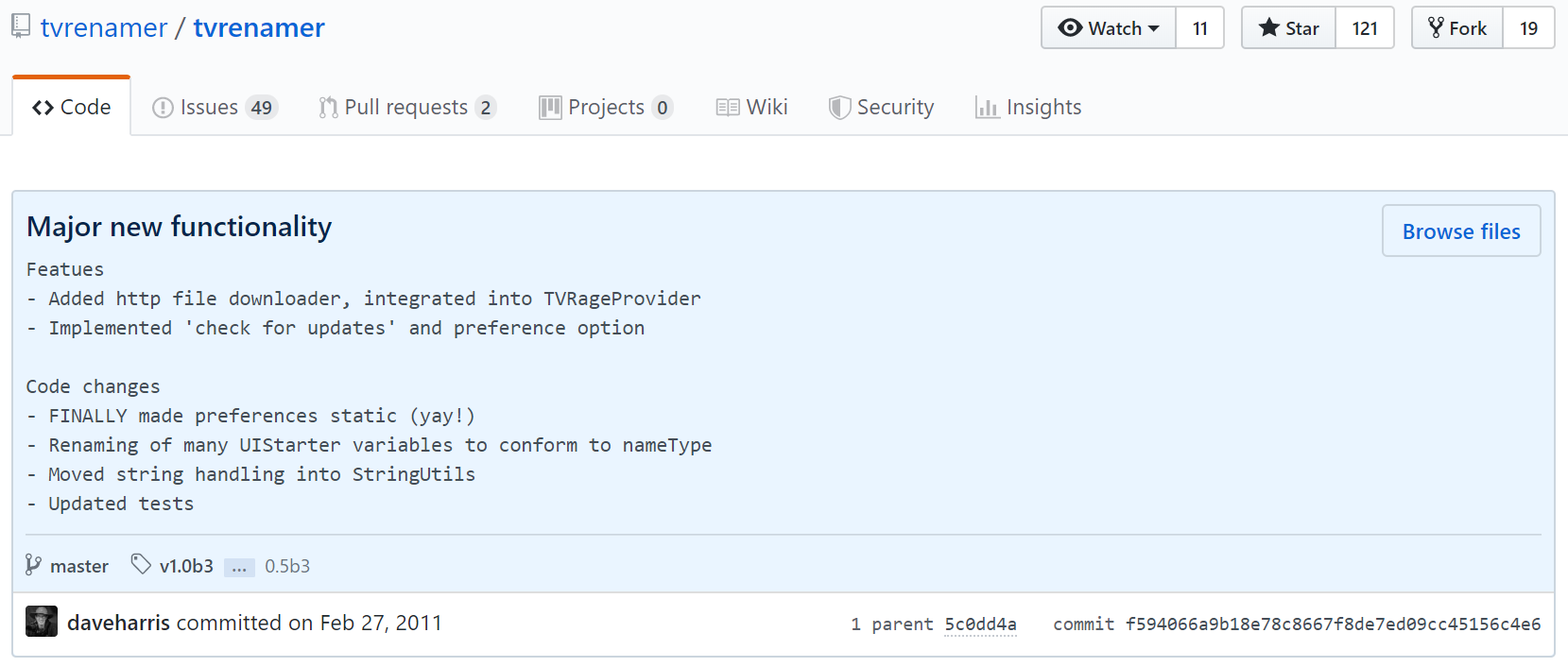}
\captionof{figure}{Commit message stating the addition of a new functionality.}
\label{fig:Example of Feature}

\vspace{0.75cm}

\begin{tikzpicture}
\pie[text=legend]{4.76/Move Method,4.76/Move \& Rename Class,9.52/Rename Method,14.29/Move Class,28.57/Rename Attribute,38.10/Rename Variable}
\end{tikzpicture}
\captionof{figure}{Distribution of refactoring operations.}
\label{fig:Refactoring_Operations_Feature}
\end{center}

\vspace{0.75cm}
This case study discusses another motivation of refactoring that is different than the traditional design improvement motivation. As shown in Figure~\ref{fig:Example of Feature}, developers interleaved refactoring practices with other development-related tasks, i.e., adding feature. Specifically, the developer implemented two new functionalities (i.e., allow the user to download files, and “check for updates” and “preference option” features. Developers also performed other code changes which involved renaming, moving, etc. 

Figure~\ref{fig:Refactoring_Operations_Feature} portrays the 21 refactoring operations performed in which the developer added features and made other related code changes. With regards to the type of refactoring operations used to perform these implementations, the developer mainly performed moving and renaming related operations that are associated with code elements related to that implementation. 
Overall, \textit{Rename Variable} and \textit{Rename Attribute} constitute the main refactoring operations performed accounting for 38.10\% and 28.57\% respectively, followed by \textit{Move Class} with 14.29\% and \textit{Rename Method} with 9.52\% . The percentage of \textit{Move Method} and \textit{Move and Rename Class} refactorings, by contrast, made up a mere 4.76\%. 

Upon exploring the source code, it appears to us that the developer performed moving-related refactorings when adding features (e.g., update checker functionality and activate user preference option) to the system, and renaming-related operations have been performed for several enhancements related to the UI (e.g., renaming buttons, task bar, progress bar, etc). These observations may explain that adding feature is one type of development task that refactorings were interleaved with and the refactoring definition in practice seems to deviate from the rigorous academic definition of refactoring, i.e., refactoring to improve the design of the code. 

\subsubsection{Case Study 5. Refactoring to fix bug}
\vspace{1em}

\begin{center}
\includegraphics[width=\columnwidth]{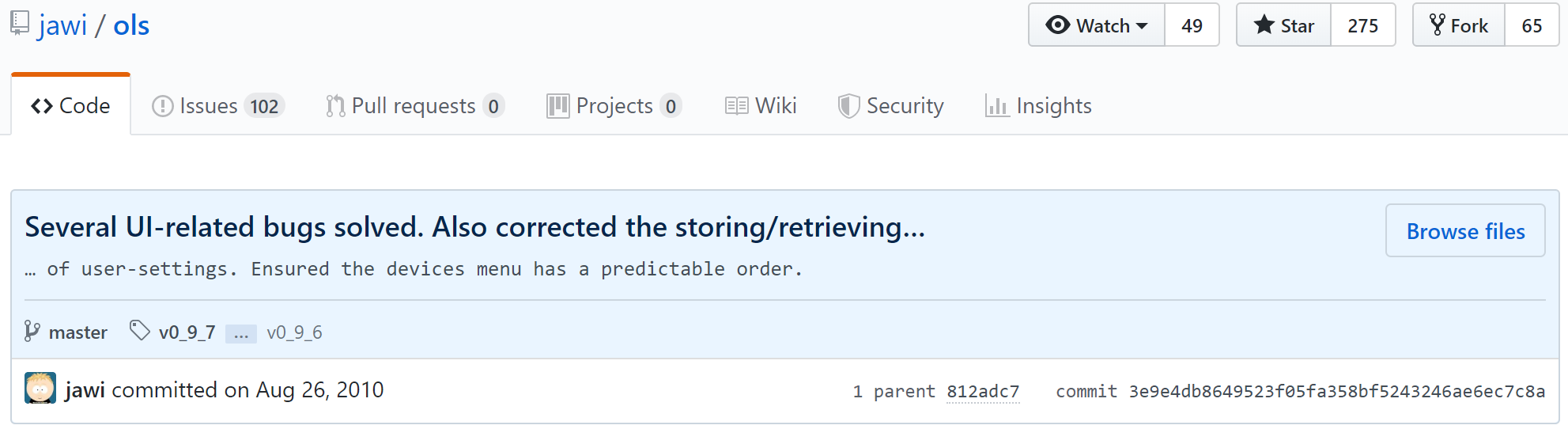}
\captionof{figure}{Commit message stating the correction of user interface related bugs.}
\label{fig:Example of BugFix}

\vspace{0.75cm}

\begin{tikzpicture}
\pie[text=legend]{2.70/Extract Class, 2.70/Rename Method,5.41/Rename Variable,13.51/Extract Method,16.22/Move Attribute, 16.22/Rename Attribute, 18.92/Rename Parameter,24.32/Move Method}
\end{tikzpicture}
\captionof{figure}{Distribution of refactoring operations.}
\label{fig:Refactoring_Operations_BugFix}
\end{center}

\vspace{0.75cm}
This case study presents another refactoring intention, i.e., refactoring to fix bugs that differs from the academic definition of refactoring. It can be seen from the above commit message (Figure~\ref{fig:Example of BugFix}) that several UI-related bugs have been solved while performing refactorings. Similar to the commit in case study 4, the developer interleaved these changes with other types of refactoring. 

The pie chart above shows 7 distinct refactoring operations performed that constituted 37 refactoring instances for bug fixing-related process. The type of refactorings involved in this activity are mainly focused on extracting, moving, and renaming-related operations. 
 
From the graph above we can see that roughly a quarter of refactorings were \textit{Move Method}. \textit{Rename Parameter, Rename Attribute}, and \textit{Move Attribute} constituted almost the same percentage with slight advantage to \textit{Rename Parameter}. \textit{Extract Method} was comprised of 13,51\%, whereas \textit{Rename Variable, Rename Method}, and \textit{Extract Class} combined just constituted under a fifth. 
The present results are significant in at least two major respects: (1) developers flossly refactor the code to reach a specific goal, i.e.,  fix bugs, and  (2) developers did not separate refactoring techniques from bug fixing-related activities. Interleaving these activities may not guarantee behavior preserving transformation as reported by 
\citep{Fowler:1999:RID:311424}. Developers are encouraged to frequently refactor the code to make finding and debugging bugs much easier. Fowler et al. pointed out that developers should stop refactoring if they notice a bug that needs to be fixed since mixing both tasks may lead to changing the behavior of the system. Testing the impact of these changes is a topic beyond the scope of this paper, but it is an interesting research direction that we can take into account in the future. 

\vspace{0.5cm}

Analyzing the distributions of refactoring operations in the case studies, and observing how they vary due to the context of refactoring and due to the difference between production and test files, has raised our curiosity about whether we can observe similar difference if we analyze distributions of refactorings across classification categories. In the next subsection, we define the following research question to investigate the frequency of refactorings, spit by target refactored element (production vs. test) per category.

\subsection{RQ1.1: Do software developers perform different types of refactoring operations on test code and production code between categories?} 

\begin{table}[htbp]
\centering
\caption{Refactoring frequency in production and test files in all projects combined.}
\label{Table:prod_vs_test_v2}
\begin{sideways}
\begin{adjustbox}{width=1.6\textwidth,totalheight=\textwidth,keepaspectratio}
\begin{tabular}{l>{\columncolor[gray]{0.8}}l>{\columncolor[gray]{0.8}}lll>{\columncolor[gray]{0.8}}l>{\columncolor[gray]{0.8}}lll>{\columncolor[gray]{0.8}}l>{\columncolor[gray]{0.8}}l}
\hline
\toprule
\multirow{2}{*}{\textbf{Refactoring}}    & \multicolumn{2}{c}{\textbf{Internal QA}}                                     & \multicolumn{2}{c}{\textbf{Code Smell}}                           & \multicolumn{2}{c}{\textbf{External QA}}                                     & \multicolumn{2}{c}{\textbf{Functional}}                                         & \multicolumn{2}{c}{\textbf{BugFix}}                                          \\ 
                                         & \multicolumn{1}{c}{Prod.} & \multicolumn{1}{c}{Test} & \multicolumn{1}{c}{Prod.} & \multicolumn{1}{c}{Test} & \multicolumn{1}{c}{Prod.} & \multicolumn{1}{c}{Test} & \multicolumn{1}{c}{Prod.} & \multicolumn{1}{c}{Test} & \multicolumn{1}{c}{Prod.} & \multicolumn{1}{c}{Test} \\ 
                                         \midrule
\textbf{Change Package} &  4626 (0.46\%)  & 0 (0.0\%) & 12693 (0.56\%) & 0 (0.0\%) & 3760 (0.57\%) & 0 (0.0\%)  & 5120 (0.50\%)  & 0 (0.0\%)  & 3541 (0.29\%) & 0 (0.0\%)  \\
\textbf{Extract \& Move Method} & 20822 (2.08\%) & 32 (2.50\%) & 8643 (0.38\%) & 30 (2.74\%)  & 8369 (1.27\%) & 35 (2.22\%) & 27327 (2.67\%) & 111 (7.86\%) &  24576 (2.08\%) & 51 (2.84\%) \\
\textbf{Extract Class}    & 13156 (1.31\%)  & 13 (1.01\%) &  6785 (0.30\%) & 18 (1.64\%)  & 6019 (0.91\%) & 68 (4.33\%) & 23625 (2.31\%) & 13 (0.92\%) & 19510 (1.65\%) & 21 (1.16\%) \\
\textbf{Extract Interface} & 1700 (0.17\%) & 0 (0.0\%) & 2522 (0.11\%) &  0 (0.0\%) & 1874 (0.28\%) & 0 (0.0\%) & 2104 (0.20\%) & 0 (0.0\%) & 3658 (0.30\%) & 2 (0.11\%) \\              
\textbf{Extract Method}   & 31357 (3.13\%) & 246 (19.19\%) & 18177 (0.80\%) &  52 (4.75\%) & 22772 (3.47\%) & 148 (9.42\%) & 40286 (3.94\%) & 218 (15.43\%) & 113416 (9.60\%) & 287 (15.98\%) \\              
\textbf{Extract Subclass}  & 1578 (0.16\%) & 0 (0.00\%) & 1109 (0.05\%) & 1 (0.09\%) & 1084 (0.16\%) & 0 (0.0\%) & 1171 (0.11\%) & 0 (0.0\%) & 2391 (0.20\%) & 9 (0.50\%) \\              
\textbf{Extract Superclass}  & 7262 (0.72\%) & 8 (0.62\%) & 7380 (0.32\%)  & 4 (0.36\%)   & 6917 (1.05\%) & 11 (0.70\%) & 10290 (1.00\%) & 13 (0.92\%) & 14939 (1.26\%) & 0 (0.0\%) \\           
\textbf{Extract Variable}    & 9922 (0.99\%) & 30 (2.34\%) & 8893 (0.39\%) & 40 (3.65\%)  & 8233 (1.25\%) & 74 (4.71\%) & 15694 (1.53\%) & 43 (3.04\%) & 59323 (5.02\%) & 71 (3.95\%) \\            
\textbf{Inline Method}  & 6139 (0.61\%) & 22 (1.72\%) & 5100 (0.22\%) &  19 (1.73\%) & 3717 (0.56\%) & 14 (0.89\%) & 7219 (0.70\%) & 11 (0.77\%) & 12662 (1.07\%) & 9 (0.50\%) \\                 
\textbf{Inline Variable} & 3107 (0.31\%) & 4 (0.31\%) & 4290 (0.19\%) &  3 (0.27\%) & 1473 (0.22\%) & 12 (0.76\%) & 2443 (0.23\%) & 13 (0.92\%) & 10570 (0.89\%) & 10 (0.55\%) \\                
\textbf{Move \& Rename Attribute} & 171 (0.02\%) & 1 (0.08\%) & 60 (0.0\%) & 0 (0.0\%)  & 318 (0.04\%) & 0 (0.0\%) & 204 (0.01\%) & 0 (0.0\%) & 120 (0.01\%) & 0 (0.0\%) \\

\textbf{Move \& Rename Class} & 14426 (1.44\%) & 4 (0.31\%) & 12493 (0.55\%) & 14 (1.27\%)   & 14958 (2.27\%) & 5 (0.31\%) & 16054 (1.57\%) & 7 (0.49\%) & 11192 (0.94\%) & 0 (0.0\%) \\ 
\textbf{Move Attribute} & 112542 (11.23\%) & 51 (3.98\%) & 43865 (1.92\%) & 61 (5.57\%)  & 22716 (3.46\%) & 16 (1.01\%) & 45025 (4.41\%) & 33 (2.33\%) & 66400 (5.62\%) & 54 (3.00\%) \\       
\textbf{Move Class} & \textbf{213609 (21.32\%)} & 26 (2.03\%) & \textbf{1202376 (52.73\%)} &  42 (3.83\%) & \textbf{129816 (19.78\%)} & 32 (2.03\%) & \textbf{175675 (17.21\%)} & 12 (0.84\%) & 94787 (8.02\%) & 41 (2.28\%)  \\                      
\textbf{Move Method}  & 61349 (6.12\%) & 120 (9.36\%) & 47268 (2.07\%) &  202 (18.46\%) & 21830 (3.32\%) & 168 (10.70\%) & 101096 (9.90\%)  & 185 (13.10\%) & 87099 (7.37\%) & 89 (4.95\%) \\                   
\textbf{Move Source Folder}   & 6219 (0.62\%) & 0 (0.0\%) & 4636 (0.20\%) &  0 (0.0\%)  & 8087 (1.23\%) & 0 (0.0\%) & 9293 (0.91\%) & 0 (0.0\%) & 5906 (0.50\%) & 0 (0.0\%) \\            
\textbf{Parameterize Variable}   & 2595 (0.26\%) & 12 (0.94\%) & 1623 (0.07\%) & 2 (0.18\%) & 1572 (0.23\%) & 26 (1.65\%) & 3548 (0.34\%) & 6 (0.42\%) & 5474 (0.46\%) & 10 (0.55\%) \\        
\textbf{Pull Up Attribute}  & 8803 (0.88\%) & 52 (4.06\%) & 53171 (2.33\%) &  8 (0.73\%) & 8781 (1.33\%) & 40 (2.54\%) & 15023 (0.34\%) & 26 (1.84\%) & 29810 (2.52\%) & 69 (3.84\%) \\           
\textbf{Pull Up Method}  & 71906 (7.18\%) & 133 (10.37\%) & 26439 (1.17\%) &  39 (3.56\%) & 24539 (3.74\%) & 55 (3.50\%) & 31181 (1.47\%) & 39 (2.76\%) & 81997 (6.94\%) & 204 (11.36\%) \\               
\textbf{Push Down Attribute} & 5186 (0.52\%) & 0 (0.0\%) & 5688 (0.25\%)  & 5 (0.45\%)   & 6476 (0.98\%) & 2 (0.12\%) & 4167 (3.05\%) & 0 (0.0\%) & 6778 (0.57\%) & 0 (0.0\%) \\             
\textbf{Push Down Method} & 14215 (1.42\%) & 1 (0.08\%) & 11874 (0.52\%) & 8 (0.73\%)  & 14689 (2.23\%) & 1 (0.06\%) & 9222 (0.90\%) & (0.0\%) & 14581 (1.23\%) & 0 (0.0\%) \\              
\textbf{Rename Attribute} & 68893 (6.88\%) & 43 (3.35\%) & 229670 (10.07\%) &  20 (1.82\%)  & 112477 (17.14\%) & 67 (4.26\%) & 142835 (14.00\%) & 26 (1.84\%) & 109286 (9.25\%) & 29 (1.61\%) \\               
\textbf{Rename Class} & 28254 (2.82\%) & 16 (1.25\%) & 56894 (2.50\%) & 9 (0.82\%)   & 24241 (3.69\%) & 15 (0.95\%) & 27010 (2.64\%) & 18 (1.27\%) & 37555 (3.17\%) & 10 (0.55\%) \\                  
\textbf{Rename Method} & 90809 (9.06\%)  & \textbf{314 (24.49\%)} & 393385 (17.25\%) &   \textbf{393 (35.92\%}) & 97245 (14.82\%) &  \textbf{461 (29.36\%)} & 120188 (11.77\%) & \textbf{371 (26.27\%}) & 125897 (10.65\%) & \textbf{532 (29.63\%)} \\                   
\textbf{Rename Parameter} & 89514 (8.93\%) & 12 (0.94\%) & 41900 (1.84\%) & 16 (1.46\%) & 41483 (6.32\%)  & 17 (1.08\%) & 72275 (7.08\%) & 14 (0.99\%)  & \textbf{138436 (11.72\%)} & 16 (0.89\%)   \\              
\textbf{Rename Variable}  & 104621 (10.44\%) & 127 (9.91\%)  & 70507 (3.09\%)  & 100 (9.14\%)  & 60788 (9.26\%) & 286 (18.21\%) & 108056 (10.59\%) & 211 (14.94\%) &  95289 (8.06\%) & 249 (13.87\%) \\               
\textbf{Replace Attribute} & 356 (0.04\%) &  5 (0.39\%) & 207 (0.01\%) & 0 (0.0\%)  & 32 (0.004\%) & 0 (0.0\%) & 212 (0.02\%) & 0 (0.0\%) & 102 (0.008\%) & 1 (0.05\%) \\              
\textbf{Replace Variable with Attribute} & 8703 (0.87\%) &  10 (0.78\%) & 2546 (0.11\%) &  8 (0.73\%) & 1813 (0.27\%) & 17 (1.08\%) & 3931 (0.38\%)  & 42 (2.97\%) & 5889 (0.49\%) & 31 (1.72\%) \\
\bottomrule
\end{tabular}
\end{adjustbox}
\end{sideways}
\end{table}

In Table \ref{Table:prod_vs_test_v2}, we show the volume of operations for each refactoring operation applied to the refactored test and production files grouped by the classification category associated with the file. Values in bold indicate the most common applied refactoring operation -- \textit{Move Class} and \textit{Rename Parameter} for production files, and \textit{Rename Method} for test files.




Concerning production file-related refactoring motivations, the topmost refactoring operations performed across all refactoring motivations is \textit{Move Class} refactoring, except for Bug Fix in which \textit{Rename Attribute} is the highest performed refactoring. In the case of internal quality attribute-related motivations, developers performed \textit{Move Class} refactoring to move the relevant classes to the right package if there are many dependencies for the class between two packages. This could eliminate undesired dependencies between modules. Another possibility for the reason to perform such refactoring is to introduce a sub-package and move a group of related classes to a new subpackage. With respect to code smell resolution motivation, developers eliminate a redundant sub-package and nesting level in the package structure when performing \textit{Move Class} refactoring operations. With regards to external quality attribute-related motivation, developers can target improving the understandability of the code by repackaging and moving the classes between these packages. Hence, the structure of the code becomes more understandable. Developers could also maintain code compatibility by moving a class back to its original package to maintain backward compatibility. For feature addition or modification, \textit{Move Class} refactoring is performed when adding new or modifying the implemented features. This could be done by moving the class to appropriate containers or  moving a class to a package that is more functionally or conceptually relevant. Lastly, for bug fixing-related motivations, developers mainly improve parameter and method names; they rename a parameter or method to better represent its purpose and to enforce naming consistency and to conform to the project’s naming conventions. Developers need to change the semantics of the code to improve the readability of the code. For test files-related refactoring motivations, the most frequently applied refactoring is \textit{Rename Method}. This can be explained by the fact that test methods are the fundamental elements in a test suite. Test methods are utilized to test the production source code; hence, the high occurrence of method based refactorings in unit test files.  
The observed difference in the distribution of refactorings in production/test files between our study and the related work \citep{Tsantalis:2013:MES:2555523.2555539} is also due to the size (number of projects) effect of the two groups under comparison.

\begin{tcolorbox}
\textit{Summary}. Our findings are aligned with the previous work \citep{Tsantalis:2013:MES:2555523.2555539}. The distribution of refactoring operations differ between production and test files. Operations undertaking production is significantly larger than operations applied to test files. \textit{Rename Method} and \textit{Move Class} are the most solicited operations for both production and test files. 
         Yet, we could not confirm that developers uniformly apply the same set of refactoring types when refactoring either production or test files.
\end{tcolorbox}

\subsection{RQ2: What patterns do developers use to describe their refactoring activities?}


\begin{table*}[htbp]
\centering
\caption{Patterns detected across all classes. Patterns whose occurrence in refactoring commits is significantly higher than non-refactoring commits (i.e., p$<$ 0.05) are in \textbf{bold}.}
\label{Table:GeneralPatterns}
\begin{sideways}
\small
\begin{adjustbox}{width=1.5\textwidth,totalheight=\textwidth,keepaspectratio}

\begin{tabular}{lllll}
\toprule
\textbf{Patterns}\\
\midrule
(1) Add* & (47) Chang* & (93) \textbf{Cleaned out} & (139) \textbf{CleanUp} & (185) \textbf{CleaningUp}  \\ 
(2) \textbf{Clean* up} & (48) \textbf{Clean-up} & (94) Creat* & (140) \textbf{Decompos*} & (186)  \textbf{Encapsulat*}  \\ 
(3) \textbf{Enhanc*}  & (49) \textbf{Extend*} & (95) \textbf{Extract*} & (141) \textbf{Factor* Out} & (187) Fix*   \\ 
(4) \textbf{Improv*} & (50) \textbf{Inlin*} & (96) \textbf{Introduc*} & (142) Merg* & (188) \textbf{Migrat*}  \\ 
(5) Modif*  & (51) \textbf{Modulariz*} &  (97) Mov* & (143) \textbf{Organiz*} & (189) \textbf{Polish*}  \\ 
(6) \textbf{Pull* Up}  & (52) \textbf{PullUp}  & (98) \textbf{Push Down} & (144) PushDown & (190) \textbf{Repackag*}   \\ 
 (7) \textbf{Re packag*} & (53) \textbf{Re-packag*} & (99) \textbf{Redesign*}  & (145) \textbf{Re-design*} & (191) \textbf{Reduc*}  \\ 
(8) \textbf{Refactor*} & (54) \textbf{Refin*}  & (100) Reformat* & (146) Remov*  & (192) \textbf{Renam*}   \\ 
(9) Reorder* & (55) \textbf{Reorganiz*} & (101) \textbf{Re-organiz*}  & (147) Repackag* & (193) \textbf{Replac*}    \\ 
(10) \textbf{Restructur*} & (56) \textbf{Rework*} & (102) \textbf{Rewrit*}  & (148) \textbf{Re-writ*} & (194) \textbf{Rewrot*}    \\ 
(11) \textbf{Simplif*} & (57) \textbf{Split*} & (103) \textbf{TidyUp} & (149) \textbf{Tid*-up} & (195) \textbf{Tid* Up}    \\ 
         (12) A bit of refactor* & (58) Basic code clean up & (104) Chang* code style & (150) Ease maintenance moving forward & (196) Replace it with \\ 
         (13) \textbf{Big refactor*}  & (59) \textbf{Big cleanup} & (105) Clean* up the code style & (151) Ease of code maintenance & (197) Extracted out code \\ 
         (14) Better factored code & (60) Cleanliness & (106) Code style improv* & (152) \textbf{Easier to maintain} & (198) Reduced code dependency\\ 
         (15) \textbf{Code refactor*} & (61) Clean* up unnecessary code & (107) Code style unification & (153) Simplify future maintenance & (199) Pushed down dependencies  \\ 
         (16) Code has been refactored extensively & (62) Cleanup formatting  &  (108) Fix code style & (154) Improve quality &  (200) \textbf{Simplify the code}\\
         (17) \textbf{Extensive refactor*} & (63) \textbf{Code clean} & (109) Improv* code style &  (155) Improvement of code quality & (201) Less code    \\ 
         (18) Refactoring towards nicer name analysis & (64) Code cleanup  & (110) Minor adjustments to code style & (156) Improved style and code quality &  \textbf{(202) Change package}\\ 
         (19) Heavily refactored code & (65) Code cleanliness & (111) Modifications to code style & (157) Maintain quality & (203) Cosmetic changes \\ 
         (20) \textbf{Heavy refactor*} & (66) \textbf{Code clean up}  & (112) Lots of modifications to code style &  (158) More quality cleanup & (204) Full customization  \\      
         (21) \textbf{Little refactor*} & (67) \textbf{Massive cleanup}  & (113) Makes the code easier to program   & (159) \textbf{Better name} & (205) \textbf{Structure change} \\ 
         (22) \textbf{Lot of refactor*}  & (68) Minor cleaning of the code   & (114) \textbf{Code review} & (160) Chang* name & (206) Module structure change  \\
         (23) \textbf{Major refactor*} &  (69) \textbf{Housekeeping}  & (115) Code rewrite & (161) \textbf{Chang* the name} & (207) Module organization structure change\\ 
         (24) \textbf{Massive refactor*}  & (70) Major rewrite and simplification & (116) Code cosmetic & (162) Chang* the package name & (208) Polishing code\\ 
         (25) \textbf{Huge refactor*} & (71) \textbf{Improv* consistency} & (117) Code revision & (163) \textbf{Chang* method name} & (209) \textbf{Improv* code quality} \\ 
         (26) \textbf{Minor refactor*} & (72) Some fix* and optimization  & (118) Code optimization & (164) Chang* method parameter names for consistency & (210) \textbf{Chang* package structure}\\ 
         (27) \textbf{More refactor*} & (73) Minors fix* and tweak & (119) Code reformatting & (165) Enables condensed naming & (211) Fix quality flaws
\\ 
         (28) Refactor* code & (74) Fix* annoying typo  &  (120) Code organization &  (166) Fix* naming convention & (212) \textbf{Get rid of}\\ 
         (29) Refactor* existing schema & (75) Fix* some formatting & (156) Code rearrangement &  (167) \textbf{Fix nam*} & (213) \textbf{Chang* design}\\ 
         (30) \textbf{Refactor out} & (76) Fix* formatting    & (122) Code formatting & (168) \textbf{Typo in method name} & (214) Improv* naming consistency  \\ 
         (31) \textbf{Small refactor*} & (77) Modifications to make it work better  & (123) Code polishing & (169) Maintain convention & (215) Remov* unused classes \\ 
         (32) \textbf{Some refactor*} & (78) Make it simpler to extend  & (124) \textbf{Code simplification} & (170) Maintain naming consistency & (216) Minor simplification  \\ 
         (33) Tactical refactor* & (79) Fix* Regression & (125) Code adjustment &  (171) Major renam* & (217) Fix* quality issue \\ 
         (34) Moved a lot of stuff  & (80) Remov* the useless  & (126)  \textbf{Code improvement} & (172) \textbf{Name cleanup} & (218) \textbf{Naming improvement}  \\ 
         (35) Fix this tidily  & (81) Remov* unneeded variables & (127) Code style & (173) \textbf{Renam* for consistency} & (219) Packaging improvement\\ 
         (36) \textbf{Further tidying}  & (82) Remov* unneeded code   & (128) Code restructur* & (174) Renam* according to java naming conventions  & (220) \textbf{Structural chang*} \\ 
         (37) Tidied up and tweaked &  (83) Remov* redundant & (129) Code beautifying  & (175) Renam* classes for consistency & (221) Hierarchy clean* \\ 
         (38) Tidied up some code & (84) Remov* dependency  & (130) \textbf{Code tidying} & (176) \textbf{Renam* package} & (222) Hierarchy reduction\\
          (39) \textbf{Restructur* package} & (85) Remov* unused dependencies & (131) Code enhancement  & (177) Resolv* naming inconsistency & (223) Enhanc* architecture \\ 
        (40) \textbf{Restructur* code} &  (86) Remov* unused  & (132) Code reshuffling &  (178) Simpler name & (224) Architecture enhanc*\\ 
         (41) Aggregat* code & (87) Remov* unnecessary else blocks & (133) Code modification  & (179) Us* appropriate variable names & (225) Trim unneeded code\\ 
         (42) Beautif* code  & (88) Remov* needless loop   & (134) Code unification & (180) Us* more consistent variable names & (226) Remov* unneeded code   \\ 
         (43) Tidy code & (89) Maintain consistency & (135) Code quality & (181) Neaten up & (227) More consistent formatting  \\ 
         (44) Beautify* & (90) \textbf{Customiz*} & (136) Make code clearer & (182) Moved more code out of & (228) More easily extended \\
         (45) Moved all integration code to separate package & (91) Improve code clarity  & (137) Code clarity  & (183) Fix bad merge & (229) Makes it more extensible-friendly \\ 
         (46) \textbf{Improve code} & (92) \textbf{Simplify code} & (138) Clean* code & (184) \textbf{Cleanup code} & (230) Clean* up code \\ \hline
       
\end{tabular}
\end{adjustbox}
\end{sideways}
\end{table*}

\begin{table*}
\centering
\caption{Patterns detected by class. Patterns whose occurrence in refactoring commits is significantly higher than non-refactoring commits (i.e., p$<$ 0.05) are in \textbf{bold}.}
\label{Table:SpecificPatterns_1}
\begin{sideways}
\small

\begin{adjustbox}{width=2.0\textwidth,totalheight=\textwidth,keepaspectratio}
\begin{tabular}{lllll}
\toprule
\textbf{BugFix} & \textbf{Code Smell} & \textbf{External} & \textbf{Functional} & \textbf{Internal} \\
\midrule

          \textbf{Minor fixes} &  \textbf{Avoid code duplication} & Reusable structure &  Add* feature &  \textbf{Decoupling}\\ 
         Bug* fix* & \textbf{Avoid duplicate code} & Improv* code reuse &  Add new feature & Enhance loose coupling  \\ 
         Fix* bug* & Avoid redundant method & Add* flexibility  & Added a bunch of features & Reduced coupling \\ 
         Bug hunting & \textbf{Code duplication removed} & Increased flexibility & \textbf{New module}  & Reduce coupling and scope of responsibility  \\
         Correction of bug & \textbf{Delet* duplicate code} &  \textbf{More flexibility} & \textbf{Fix some GUI} & Prevent the tight coupling  \\ 
         Improv* error handling  & Remove unnecessary else blocks & Provide flexibility & Added interesting feature &  Reduced the code size \\ 
           Fix further thread safety issues &  \textbf{Eliminate duplicate code} & A bit more readable & Added more features & Complexity has been reduced\\ 
          Fixed major bug & Fix for duplicate method & \textbf{Better readability} &  Adding features to support & \textbf{Reduce complexity}  \\ 
          Fix numerous bug & Filter duplicate & Better readability and testability  & Adding new features & Reduced greatly the complexity \\ 
         Fix several bug & Joining duplicate code &  Code readability optimization & Addition of a new feature & Removed unneeded complexity \\ 
         Fixed a minor bug & Reduce a ton of code duplication & Easier readability & Feature added & Removes much of the complexity\\ 
         Fixed a tricky bug & \textbf{Reduce code duplication}  & \textbf{Improve readability}  & Implement one of the batch features & Add inheritance   \\ 
         Fix* small bug & Reduced code repetition & \textbf{Increase readability} &  Implementation of feature & Added support to the inheritance    \\ 
         Fixed nasty bug  & \textbf{Refactored duplicate code} & Make it better readable & Implemented the experimental feature & Avoid using inheritance and using composition instead  \\ 
          Fix* some bug*  &  Clear up a small design flaw &  Make it more readable & Introduced possibility to erase features & Better support for specification inheritance \\ 
         Fixed some minor bugs & TemporalField has been refactored  & Readability enhancement & New feature & \textbf{Change* inheritance}  \\ 
          bugfix* & Remove commented out code  & Readability and supportability improvement & Remove the default feature & Extend the generated classes using inheritance \\ 
         Fix* typo* & Removed a lot of code duplication & \textbf{Readability improvements} & Removed incomplete features & Improved support for inheritance \\ 
         Fix* broken & \textbf{Remov* code duplication} & Reformatted for readability & Renamed many features &  Perform deep inheritance  \\ 
         Fix* incorrect & Remove some code duplication & Simplify readability & Renamed some of the features for consistency & Remove inheritance \\ 
         Fix* issue* & Removed the big code duplication & Improve* testability & Small feature addition & Inheritance management  \\ 
        Fix* several issue* & Removed some dead and duplicate code &  Update the performance  & Support of optional feature &   Loosened the module dependency \\
        Fix* concurrency issue* with  & \textbf{Remov* duplicate code} & Add* performance  & Supporting for derived features & Prevents circular inheritance  \\ 
        Fixes several issues & Resolved duplicate code & Scalability improvement  & Added functionality &  Avoid using inheritance \\
         Solved some minor bugs  & Sort out horrendous code duplication  & Better performance &  Added functionality for merge & Add composition  \\ 
          Working on a bug  & Remove duplicated field  & Huge performance improvement  &  Adding new functionality & Composition better than inheritance  \\ 
           \textbf{Get rid of} & Remov* dead code &  \textbf{Improv* performance}  & Adds two new pieces of functionality to & Us* composition  \\ 
         A bit of a simple solution to the issue &  Remove some dead-code &   \textbf{More manageable} & Consolidate common functionality & Separates concerns   \\ 
          A fix to the issue  & Removed all dead code & \textbf{More efficient*}  & Development of this functionality  & Better handling of polymorphism \\ 
          Fix a couple of issue & Removed apparently dead code & Make it reusable for other & Export functionality  & Makes polymorphism easier  \\ 
         Issue management & This is a bit of dead code  & Increase efficiency & Extend functionality of & \textbf{Better encapsulation}   \\ 
        Fix* minor issue & Removed more dead code  & Verify correctness & Extract common functionality & Better encapsulation and less dependencies\\ 
         Correct issue & Fix* code smell & Massive performance improvement &   Functionality added   &  Pushed down dependencies \\ 
         \textbf{Additional fixes}  &  Fix* some code smell & Increase performance  & House common functionality & \textbf{Remov* dependency}    \\ 
         Resolv* problem & Remov* some 'code smells'  & Largely resolved performance issues & Improved functionality  &  Split out each module into   \\ 
         Correct* test failure*   & Update data classes  & Lots of performance improvement  &  \textbf{Move functionality}  &  \\ 
         Fix* all failed test*   & Remove useless class &  Measuring the performance & Moved shared functionality   & \\ 
         Fix* compile failure&  Removed obviously unused/faulty code &  Improv* stability & Feature/code improvements &    \\
          A fix for the errors & Lots of modifications to code style & Usability improvements &  Pulling-up common functionality &  \\ 
         Better error handling & Antipattern bad for performances &  Noticeable performance improvement & Push more functionality &  \\
         Better error message handling & Killed really old comments & Optimizing the speed  &  Re-implemented missing functions & \\ 
         Cleanup error message & Less long methods & Performance boost & Refactored functionality & \\ 
            Error fix* &  Removed some unnecessary fields & \textbf{Performance enhancement} & Refactoring existing functionality & \\ 
           Fixed wrong &  & Performance improvement & Add functionality to &  \\ 
          Fix* error*   &  & Performance much improved & Remov* function* &   \\ 
          Fix* some error* &  & Performance optimization   &  Merging its functionality with   &   \\ 
           Fix small error  &  & Performance speed-up &  Remove* unnecessary function* &   \\
          Fix some errors &  & Refactor performance test  & Reworked functionality &  \\
           Fix compile error & & Renamed performance test  &  Removing obsolete functionality & \\ 
           Fix test error & & Speed up performance & Replicating existing functionality with &  \\ 
           Fixed more compilation errors & & Backward compatible with & Split out the GUI function & \\ 
          Fixed some compile errors & & Fix backward compatibility & Add cosmetic changes &  \\ 
           Fixes all build errors & & Fixing migration compatibility & Add* support &  \\ 
           Fixed Failing tests  & & Fully compatible with &  Implement* new architecture &  \\ 
           \textbf{Handle}   & & Keep backwards compatible &  Update   &  \\ 
          Handling error* & & Maintain compatibility & Additional changes for & \\ 
           Error* fix*  & &  Make it compatible with & UI layout has changed & \\ 
           Tweaking error handling  & & \textbf{More compatible} & GUI: Small changes  &  \\ 
           Various fix* & &  Should be backward-compatible &  New UI layout  &  \\ 
           Fix* problem* & & Retains backward compatibility & \textbf{UI changes} & \\ 
           Got rid of deprecated code & & Stay compatible with & UI enhancements & \\ 
           Delet* deprecated code & & Added some robustness &  \\ 
            Remov* deprecated code & & Improve robustness &\\ 
            & &   \textbf{Improve usability} &   \\ 
             & & Robustness improvement & \\ 
          & &  \textbf{To be more robust}  & &  \\ 
           & & Better understanding &  \\ 
            & & Bit to be easier to understand  & & \\ 
           & & \textbf{Easier to understand} & \\ 
           & & Increases understandability &  \\ 
            & & Adding checks on accessibility  &\\ 
           & & Easier accessible and evolvable &  \\ 
            & &  Make class more extendable  &   \\ 
             & &  Allow extensibility  &  \\ 
          & & For future extensibility &  \\ 
           & & Improved modularity and extensibility & \\ 
            & & Increase testability   & &\\ 
          & & More testable design  & & \\ 
             & & Accuracy improvement & & \\ \hline
\end{tabular}      
\end{adjustbox}
\end{sideways}
\end{table*}

In this research question, we explore the set of \sarInitialNumber potential self-affirmed refactoring candidates, extracted by manual inspection from commits messages and categories top 100 features. We classify these SAR candidates into two tables: Table~\ref{Table:GeneralPatterns} contains generic candidate patterns that were found across categories; Table~\ref{Table:SpecificPatterns_1} contains candidate patterns that are specific to each category.

Upon a closer inspection of these refactoring patterns, we have made several observations: we noticed that developers document refactoring activities at different levels of granularity, e.g., package, class, and method level. Furthermore, we observe that developers state the motivation behind refactoring, and some of these patterns are not restricted only to fixing code smells, as in the original definition of refactoring in Fowler's book, i.e., improving the structure of the code. For instance, developers tend often to improve certain non-functional attributes such as the readability and testability of the source code. Additionally, developers occasionally apply the \say{Don’t Repeat Yourself} principle by removing excessive code duplication. A few patterns indicated that developers refactor the code to improve internal quality attributes such as inheritance, polymorphism, and abstraction. We also noticed the application of a single responsibility principle which is meant to improve the cohesion and coupling of the class when developers explicitly mentioned a few patterns related to dependency removal. 

Further, we observe that developers tend to report the executed refactoring operations by explicitly using terms from Fowler’s taxonomy; terms such as \textit{inline class/method}, \textit{Extract Class/Superclass/Method} or \textit{Push Up Field/Method} and \textit{Push Down Field/Method}. 

The generic nature of some of these patterns was a critical observation that we encountered, i.e., many of these patterns are context specific and can be subject to many interpretations, depending on the meaning the developer is trying to convey. For instance, the pattern \textit{fixed a problem} is descriptive of any anomaly developer encountered and it can be either functional or non-functional. Since in our study, we are interested in textual patterns related to refactoring, we decided to filter this list down by reporting patterns whose frequency in commit messages containing refactoring is significantly higher than in messages of commits without refactoring. The rationale behind this idea is to identify patterns that are repeatedly used in the context of refactoring, and less often in other contexts. Since the patterns were extracted from \commitsNumber messages of commits containing refactoring (we call them refactoring commits), we need to build another corpus of messages from commits that do not contain refactorings (we call them non-refactoring commits). As we plan on comparing the frequency of keywords between the two corpora, i.e., refactoring and non-refactoring commit messages, it is important to adequately choose the non-refactoring messages to ensure fairness. To do so, we follow the following heuristics: we randomly select a statistically significant sample of commits (confidence level of 95\%), 1) chosen from the same set of \projectsNumber projects that issued the refactoring commits; 2) whose authors are from the same authors of the refactoring commits; 3) whose timestamps are in the same interval of refactoring commits timestamps; 4) and finally, the average length of commit messages is approximately close (118 for refactoring commits, and 120 for non-refactoring commits).

\begin{figure}[!ht]
 	\centering
 	\includegraphics[width=1.0\linewidth]{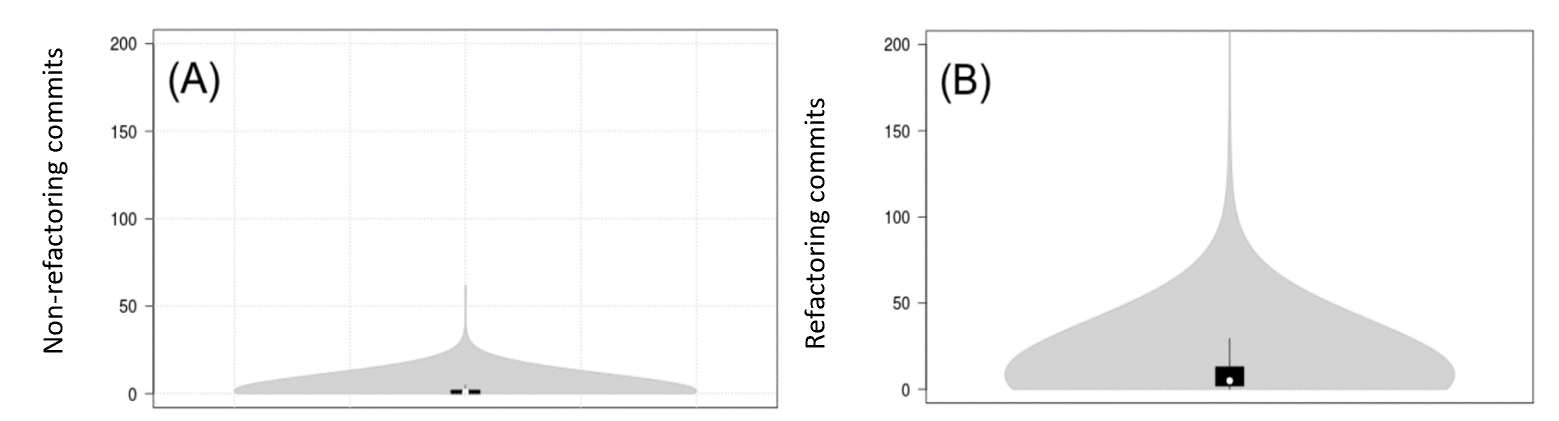}
 	\caption{Violin plots representing the occurrence of \textit{refactor} keyword in (A) non-refactoring corpus vs. (B) refactoring corpus.}
 	\label{Figure:violin-plots-refacoring-occurrences}
\end{figure}

Once the set of non-refactoring commit messages constructed, for each keyword, we calculate its occurrence per project for both corpora. This generates vector of \projectsNumber occurrences per corpus. Each vector dimension contains a positive number representing the keyword occurrence for a project, and zero otherwise. Figure \ref{Figure:violin-plots-refacoring-occurrences} illustrates occurrences violin plots of the keyword \textit{\say{refactor}} in both corpora. While it is observed in Figure \ref{Figure:violin-plots-refacoring-occurrences} that the occurrence of \textit{refactor} in refactoring commits is higher, we need a statistical test to prove it. So, we perform such comparison using the Mann-Whitney U test, a non-parametric test that checks continuous or ordinal data for a significant difference between two independent groups. This applies to our case, since the commits, in the first group, are independent of commits in the second group. We formulate the comparison of each keyword occurrence corpora by defining the alternative hypothesis as follows:

\begin{hypothesis} \label{hyp:RQ6.1}
The occurrence vector of refactoring commits is strictly higher than the occurrence vector of non-refactoring commits.
\end{hypothesis}

And so, the null hypothesis is defined as follows:

\begin{nullhypothesis} \label{hyp:RQ6.2}
The occurrence vector of non-refactoring commits is equal or smaller than the occurrence vector of refactoring commits.
\end{nullhypothesis}

We start with generating occurrence vectors for each keyword, then we perform the statistical test for each pair of vectors. We report our findings in Table ~\ref{Table:GeneralPatterns} and \ref{Table:SpecificPatterns_1}  where keywords in \textbf{bold} are the ones rejecting the null hypothesis (i.e., p$<$ 0.05).

With the analysis of these tables results, we observe the following:

\begin{itemize}
  \item While previous studies have been relying on the detection of refactoring activity in software artifacts using the keyword \say{\textit{refactor*}} \citep{stroggylos2007refactoring,Ratzinger:2008:RRS:1370750.1370759,citeulike:2881658,6112738,6802406}
  , our findings demonstrate that developers use a variety of keywords to describe their refactoring activities. For instance, keywords such as \textit{clean up}, \textit{repackage}, \textit{restructure}, \textit{re-design}, and \textit{modularize} has been used without the mention of the \textit{refactoring} keyword, to imply the existence of refactorings in the committed code. While these keywords are not exclusive to refactoring, and could also be used for general usage, their existence in commits containing refactoring operations has been more significant (i.e., p$<$ 0.05), which qualifies them to be close synonyms to \textit{refactoring}. Table \ref{Table:TopGenericKeywords} enumerates the top-20 keywords, sorted by the percentage of projects they were located in.
  \item The keyword \textit{refactor} was also used in non-refactoring commit messages. This can be explained either by its occasional misuse, like some previous studies found, or by the existence of refactoring operations that were not identified by the tool we are using. Yet, the frequency of misuse of this popular pattern remains significant in the refactoring-related commit messages (i.e., p$<$ 0.05).
  \item We notice that developers document refactoring activities at different levels of granularity, e.g., package, class, and method level. We also observe that developers occasionally state the motivation behind refactoring, which is not restricted only to fixing code smells, as in the original definition of refactoring in the Fowler's book \citep{Fowler:1999:RID:311424}, and so, this supports the rationale behind our classification in the first research question.
  \item Furthermore, our classification has revealed the existence of patterns that are used in specific categories (i.e., motivations). For instance, the traditional code smell category is mainly populated with keywords related to removing duplicate code. Interestingly, all patterns whose existence in refactoring commit messages is statistically significant, were related to duplicate code deletion. Although patterns related to removing code smells exist, e.g., \textit{Clear up a small design flaw} or \textit{fix code smell} or \textit{Antipattern bad for performances}, these patterns occurrence was not large enough to reject the null hypothesis. Nevertheless, Table \ref{Table:TopSpecificKeywords} contains a summary of category-specific patterns that we manually identified. These keywords are found relevant based on how previous studies have been identifying refactoring opportunities (removing code smells, improving structural metrics, optimizing external quality attributes like performance etc.). Note that, in Table \ref{Table:TopSpecificKeywords}, we did not quantify the frequency of these patterns, and we plan on the future to further analyze their popularity, along with the type of refactoring operations that are mostly used with their existence, similarly to previous empirical studies \citep{bavota2013empirical,bavota2015experimental}.
  \item Developers occasionally mention the refactoring operation(s) they perform. The Mann-Whitney U test accepted the alternative hypothesis for all patterns linked to refactoring operations i.e., \textit{Pull Up}, \textit{Push Down}, \textit{Inline}, \textit{Extract}, \textit{Rename}, \textit{Encapsulate}, \textit{Split}, \textit{Extend}, except for the famous \textit{move}. Unlike code smell patterns, \textit{move} does exist in 787 projects (98.37\%, fourth most used keyword, after respectively \textit{Fix}, \textit{Add}, and \textit{Merge}) and it is heavily used by both refactoring and non-refactoring commit messages.
  \item Similarly to move, keywords like \textit{merge}, \textit{reformat}, \textit{remove redundant}, \textit{performance improvement}, \textit{code style}, were popular across many projects, and typically invoked by both refactoring and non-refactoring commits. So, although they do serve in documenting refactoring activities, their generic nature makes them also used in several other contexts. For example, merge is typically used when developers combine classes or methods, as well as describing the resolution of merge conflicts. Similarly, performance improvement is not restricted to non-functional changes, as several performance optimization techniques and genetic improvements are not necessarily linked to refactoring.
\end{itemize}

\begin{table}[]
\begin{center}
\caption{Top generic refactoring patterns.}
\label{Table:TopGenericKeywords}

\begin{adjustbox}{width=0.9\textwidth,center}
\begin{tabular}{llll}

\toprule
\textbf{Patterns} \\
         \midrule 
         Refactor* (89.00\%) 
         & Renam* (83.63\%)
         & Improv* (78.75\%)
         & CleanUp (67.38\%)
         \\
         Replac* (66.88\%)
         & Introduc (53.00\%)
         & Extend (52.63\%)
         & Simplif (52.50\%)
         \\
         Extract (49.00\%)
         & Added support (47.38\%)
         & Split (45.50\%)
         & Reduc* (45.00\%)
         \\
         Chang* name (44.88\%)
         & Migrat (32.88\%)
         & Enhanc (32.63\%)
         & Organiz* (32.25\%)
         \\
         Rework (27.25\%)
         & Rewrit* (27.25\%)
         & Code clean* (25.63\%)
         & Remov* dependency (25.00\%)
         \\
        \bottomrule
\end{tabular} 
\end{adjustbox}
\end{center}
\end{table}
\begin{table*}[h!]
\begin{center}
\caption{Summary of refactoring patterns, clustered by refactoring related categories.
}
\label{Table:TopSpecificKeywords}
\begin{adjustbox}{width=0.8\textwidth,center}
\begin{tabular}{lll}

\toprule
\textbf{Internal} &
\textbf{External} &
\textbf{Code Smell} \\
         \midrule 
         
    Inheritance & Functionality & Duplicate Code \\ 
    Abstraction & Performance & Dead Code  \\ 
    Complexity & Compatibility & Data Class \\ 
    Composition  & Readability & Long Method \\ 
    Coupling & Stability & Switch Statement \\ 
    Encapsulation & Usability & Lazy Class \\
    Design Size  & Flexibility & Too Many Parameters \\
    Polymorphism & Extensibility  & Primitive Obsession \\
    Cohesion  & Efficiency & Feature Envy \\ 
    Messaging & Accuracy & Blob Class \\ 
    Concern Separation & Accessibility & Blob Operation \\
    Dependency & Robustness  & Redundancy   \\ 
    & Testability & Useless class\\ 
    & Correctness & Code style \\
    & Scalability & Antipattern\\
    & Configurability & Design Flaw\\
    & Simplicity & Code Smell \\
    & Reusability & Temporary Field \\
    & Reliability & Old Comment\\
    & Modularity &  \\
    & Maintainability & \\
    & Traceability & \\
    & Interoperability & \\
    & Fault-tolerance & \\
    & Repeatability & \\
    & Understandability & \\
    &  Effectiveness & \\
    & Productivity & \\
    & Modifiability & \\
    & Reproducibility & \\
    & Adaptability & \\
    & Manageability & \\

        \bottomrule
\end{tabular} 
\end{adjustbox}

\end{center}
\end{table*}

\begin{tcolorbox}
\textit{Summary 1}. Developers tend to use a variety of textual patterns to document their refactoring activities, besides '\textit{refactor}', such as '\textit{re-package}', '\textit{redesign}', '\textit{reorganize}', and '\textit{polish}'. 

\textit{Summary 2}. These patterns can be either (1) generic, providing high-level description of the refactoring, e.g., 'clean up unnecessary code', 'ease maintenance moving forward', or (2) specific by explicitly mentioning the rationale behind the refactoring, e.g., 'reduce the code coupling' (internal), 'improving code \textit{readability}' (external), and 'fix \textit{long method}' (code smell).

\textit{Summary 3}. Developers occasionally express their refactoring strategy. We detected several refactoring operations, known from the refactoring catalog, such as \textit{'extract method'}, \textit{'extract class'}, and \textit{'extract interface'}.
\end{tcolorbox}

\begin{figure*}[!ht]
\centering 
\includegraphics[width=.5\textwidth]{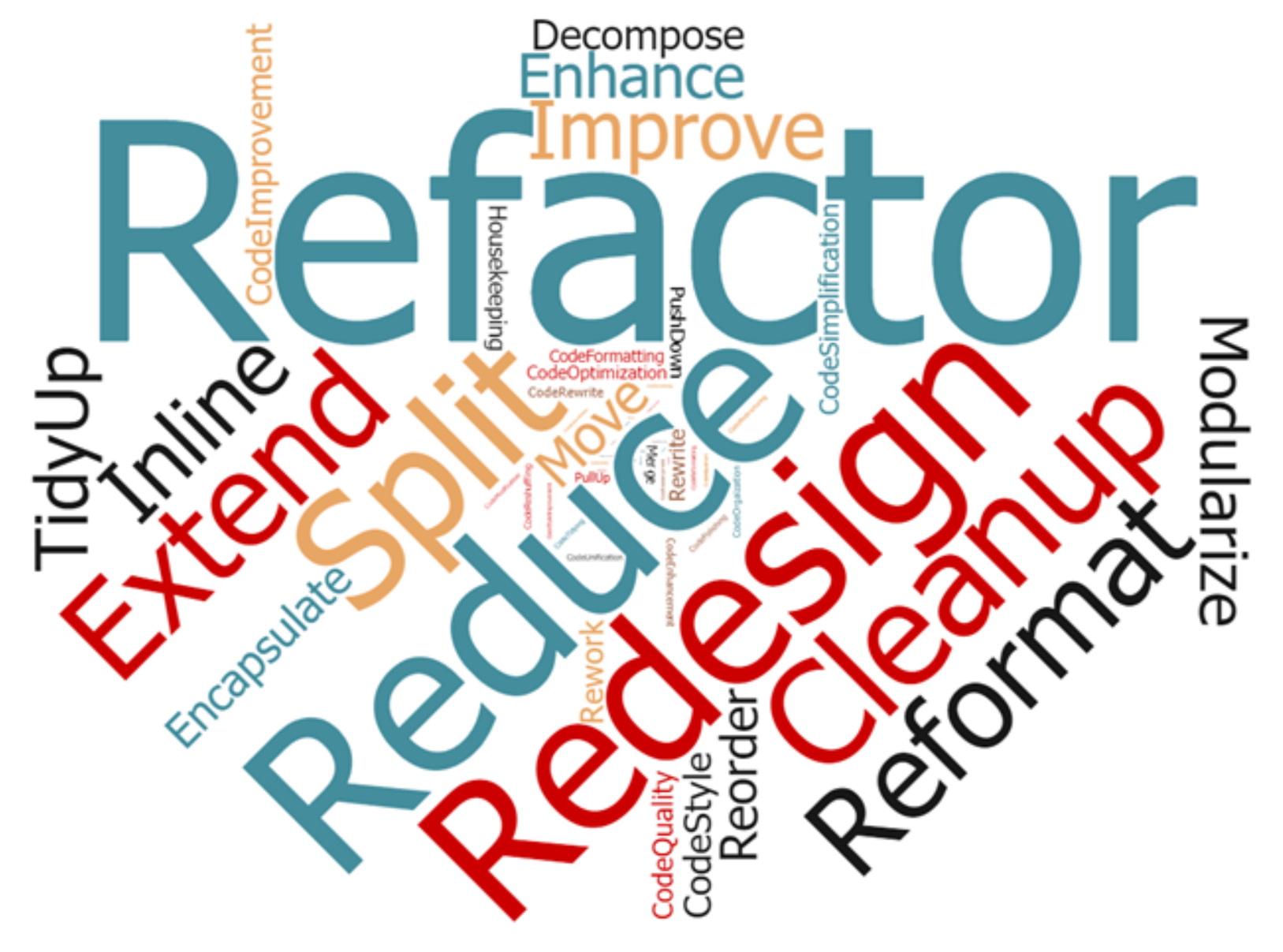}
\caption{Sample of SAR patterns identified in our study.}
\label{fig:wordcloud}
\end{figure*}

The extraction of these patterns raised our curiosity about the extent to which they can represent an alternative to the keyword \textit{refactor}, being the de facto keyword to document refactoring activities. Figure~\ref{fig:wordcloud} reveals examples of these patterns.  In the next subsection, we challenge the hypothesis raised by \citep{murphy2008gathering} about whether developers use a specific pattern, i.e., \textit{\say{refactor}} when describing their refactoring activities. We quantify the messages with the label \textit{\say{refactor}} and without to compare between them.

\subsection{RQ2.1: Do commits containing the label \textit{Refactor} indicate more refactoring activity than those without the label?}

 \citep{murphy2008gathering} proposed several hypotheses related to four methods that gather refactoring data and outlined experiments for testing those hypotheses. One of these methods concerns mining the commit log. \citep{murphy2008gathering} hypothesize that commits labeled with the keyword \textit{\say{refactor}} do not indicate more refactoring instances than unlabeled commits. In an empirical context, we test this hypothesis in two steps. In the first steps, we used the keyword \textit{\say{refactor}}, exactly as dictated by the authors. Thereafter, we quantified the proportion of commits including the searched label across all the considered projects in our benchmark. In the second step, we re-tested the hypothesis using the subset of \numberSAR SAR patterns, whose occurrence in refactoring commits were found to be significant with respect to non-refactoring commits. We counted the percentage of commits containing any of our SAR labels. The result of the two rounds resides in a strict set of commits containing the label \textit{\say{refactor}}, which is included in a larger set containing all patterns, and finally a remaining set of commits which does not contain any patterns. For each of the sets, we count the number of refactoring operations identified in the commits. Then, we break down the set per operation type.   

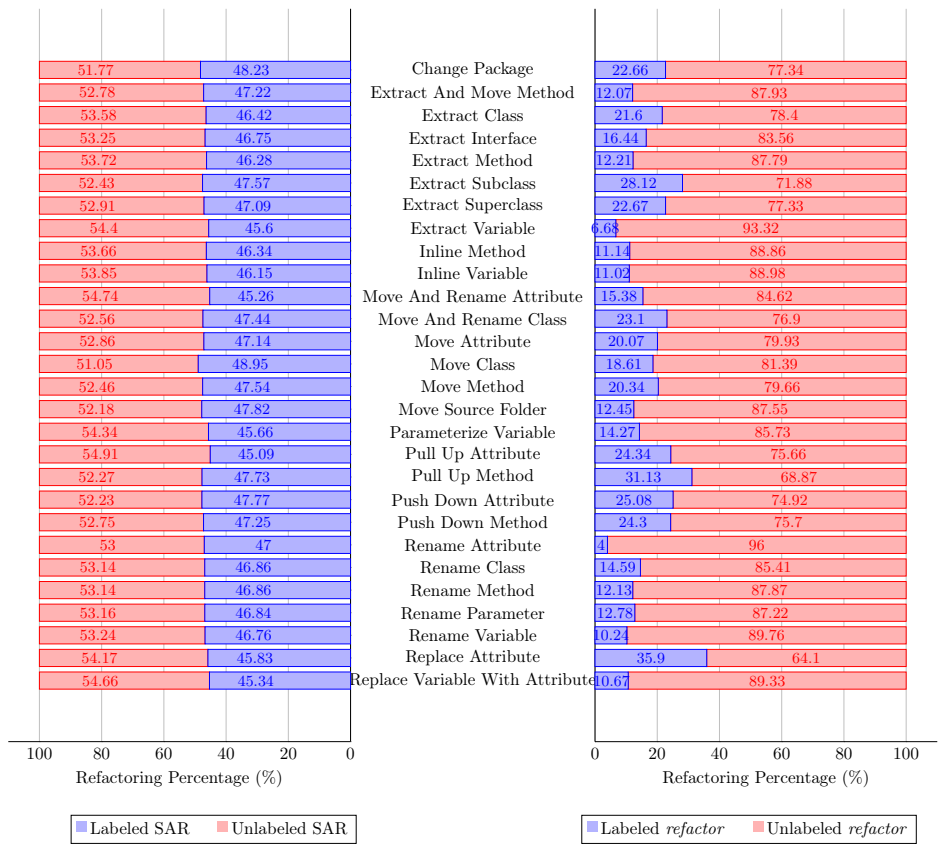
\begin{figure*}[htbp]  
\begin{center}
\begin{tikzpicture}
\begin{scope}[scale=0.65]
\begin{axis}[
name=Unlabeled,
legend pos=outer north east,
scale only axis,
xbar stacked,
 xmin=0, 
xlabel={Refactoring Percentage (\%)},
width=7cm, height= 15cm,
tickwidth         = 0 pt,
tick align=inside,
ytick=data,
grid=major,
ymajorgrids=false,
yticklabels={},
y tick label style={text width=3cm,align=center},
axis x line*=left,
axis y line*=left,
clip=false,
   legend image code/.code={%
   \draw[#1, draw=none,/tikz/.cd,yshift=-0.25em]
        (0cm,1pt) rectangle (6pt,7pt);},
   legend style={
    	at={(0.45,-0.1)},
        anchor=north,
        legend columns=-1,
        /tikz/every even column/.append style={column sep=0.5cm}
        },
       nodes near coords,
    every node near coord/.append style={font=\small},
]
\addplot+[xbar] coordinates {
(10.6741573,1)
(35.8974359,2)
(10.24403823,3)
(12.78092158,4)
(12.12734634,5)
(14.5938174,6)
(03.995472941,7)
(24.30038966,8)
(25.08389262,9)
(31.13196481,10)
(24.33766234,11)
(14.27230047,12)
(12.45033113,13)
(20.3379224,14)
(18.61298608,15)
(20.06552932,16)
(23.10325157,17)
(15.38461538,18)
(11.02003643,19)
(11.14355231,20)
(06.681252104,21)
(22.66584767,22)
(28.11501597,23)
(12.21149811,24)
(16.44444444,25)
(21.6,26)
(12.07233627,27)
(22.65625,28)
};

\addplot+[xbar] coordinates {
(89.3258427,1)
(64.1025641,2)
(89.75596177,3)
(87.21907842,4)
(87.87265366,5)
(85.4061826,6)
(96.00452706,7)
(75.69961034,8)
(74.91610738,9)
(68.86803519,10)
(75.66233766,11)
(85.72769953,12)
(87.54966887,13)
(79.6620776,14)
(81.38701392,15)
(79.93447068,16)
(76.89674843,17)
(84.61538462,18)
(88.97996357,19)
(88.85644769,20)
(93.3187479,21)
(77.33415233,22)
(71.88498403,23)
(87.78850189,24)
(83.55555556,25)
(78.4,26)
(87.92766373,27)
(77.34375,28)
};

     \legend{Labeled \textit{refactor}, Unlabeled \textit{refactor}}
\node[xshift=-2.5cm,align=center] at (axis cs:0,28) {Change Package};
\node[xshift=-2.5cm,align=center] at (axis cs:0,27) {Extract And Move Method};
\node[xshift=-2.5cm,align=center] at (axis cs:0,26) {Extract Class};
\node[xshift=-2.5cm,align=center] at (axis cs:0,25) {Extract Interface};
\node[xshift=-2.5cm,align=center] at (axis cs:0,24) {Extract Method};
\node[xshift=-2.5cm,align=center] at (axis cs:0,23) {Extract Subclass};
\node[xshift=-2.5cm,align=center] at (axis cs:0,22) {Extract Superclass};
\node[xshift=-2.5cm,align=center] at (axis cs:0,21) {Extract Variable};
\node[xshift=-2.5cm,align=center] at (axis cs:0,20) {Inline Method};
\node[xshift=-2.5cm,align=center] at (axis cs:0,19) {Inline Variable};
\node[xshift=-2.5cm,align=center] at (axis cs:0,18) {Move And Rename Attribute};
\node[xshift=-2.5cm,align=center] at (axis cs:0,17) {Move And Rename Class};
\node[xshift=-2.5cm,align=center] at (axis cs:0,16) {Move Attribute};
\node[xshift=-2.5cm,align=center] at (axis cs:0,15) {Move Class};
\node[xshift=-2.5cm,align=center] at (axis cs:0,14) {Move Method};
\node[xshift=-2.5cm,align=center] at (axis cs:0,13) {Move Source Folder};
\node[xshift=-2.5cm,align=center] at (axis cs:0,12) {Parameterize Variable};
\node[xshift=-2.5cm,align=center] at (axis cs:0,11) {Pull Up Attribute};
\node[xshift=-2.5cm,align=center] at (axis cs:0,10) {Pull Up Method};
\node[xshift=-2.5cm,align=center] at (axis cs:0,9) {Push Down Attribute};
\node[xshift=-2.5cm,align=center] at (axis cs:0,8) {Push Down Method};
\node[xshift=-2.5cm,align=center] at (axis cs:0,7) {Rename Attribute};
\node[xshift=-2.5cm,align=center] at (axis cs:0,6) {Rename Class};
\node[xshift=-2.5cm,align=center] at (axis cs:0,5) {Rename Method};
\node[xshift=-2.5cm,align=center] at (axis cs:0,4) {Rename Parameter};
\node[xshift=-2.5cm,align=center] at (axis cs:0,3) {Rename Variable}; 
\node[xshift=-2.5cm,align=center] at (axis cs:0,2) {Replace Attribute};
\node[xshift=-2.5cm,align=center] at (axis cs:0,1) {Replace Variable With Attribute};
\end{axis}
\begin{axis}[ 
name=Labeled,
legend pos=outer north east,
at={(Unlabeled.north west)},anchor=north east, xshift=-5cm,
scale only axis,
xbar stacked,
xmin=0,
xlabel={Refactoring Percentage (\%)},
tickwidth         = 0 pt,
tick align=inside,
ytick=data,
grid=major,
ymajorgrids=false,
yticklabels={},
width=7cm, height= 15cm,
 x dir=reverse,
 every node near coord/.style={
               anchor=east,
            },
axis x line*=left,
axis y line*=right,
 legend image code/.code={%
   \draw[#1, draw=none,/tikz/.cd,yshift=-0.25em]
       (0cm,1pt) rectangle (6pt,7pt);},
   legend style={
    	at={(0.6,-0.1)},
        anchor=north,
        legend columns=-1,
        /tikz/every even column/.append style={column sep=0.5cm}
        },
           nodes near coords,
        every node near coord/.append style={font=\small},
]

\addplot+[xbar] coordinates {
(45.34288639,1)
(45.83333333,2)
(46.75658317,3)
(46.84371106,4)
(46.85834434,5)
(46.85768863,6)
(47.0018358,7)
(47.25336323,8)
(47.76511832,9)
(47.73465759,10)
(45.08629297,11)
(45.66326531,12)
(47.82308224,13)
(47.53775443,14)
(48.94947092,15)
(47.14243058,16)
(47.43628186,17)
(45.26315789,18)
(46.15007357,19)
(46.34464752,20)
(45.59604468,21)
(47.09132272,22)
(47.57118928,23)
(46.28245107,24)
(46.74556213,25)
(46.41956631,26)
(47.2204308,27)
(48.23053589,28)
};

\addplot+[xbar] coordinates {
(54.65711361,1)
(54.16666667,2)
(53.24341683,3)
(53.15628894,4)
(53.14165566,5)
(53.14231137,6)
(52.9981642,7)
(52.74663677,8)
(52.23488168,9)
(52.26534241,10)
(54.91370703,11)
(54.33673469,12)
(52.17691776,13)
(52.46224557,14)
(51.05052908,15)
(52.85756942,16)
(52.56371814,17)
(54.73684211,18)
(53.84992643,19)
(53.65535248,20)
(54.40395532,21)
(52.90867728,22)
(52.42881072,23)
(53.71754893,24)
(53.25443787,25)
(53.58043369,26)
(52.7795692,27)
(51.76946411,28)
};
  
      \legend{Labeled SAR, Unlabeled SAR}
\end{axis}
\end{scope}
\end{tikzpicture}

\caption{Distribution of refactoring operations for commits labeled and unlabeled SAR (left side) and commits labeled and unlabeled \textit{refactor} (right side).} 
\label{fig:Refactoring Operation}
\end{center}
\end{figure*} 

In order to compare the number of refactorings identified for each set, i.e., labeled and unlabeled commits with the keyword \textit{“refactor”}, along with labeled and unlabeled commits with SAR patterns. We used the Wilcoxon test, as suggested by  \citep{murphy2008gathering} for the purpose of testing the hypothesis. We then applied the non-parametric Wilcoxon rank-sum test to estimate the significance of differences between the numbers of the sets. The choice of Wilcoxon rank-sum test is motivated by the independence of sets from each other (the occurrence of \textit{refactor} is independent of the occurrence of the remaining patterns).

Figure~\ref{fig:Refactoring Operation} shows the distribution of refactorings in labeled and unlabeled commits with SAR patterns (group 1 on the left) and labeled and unlabeled commits with the keyword refactor (group 2 on the right). The first observation we can draw is that \textit{Replace Attribute} stands as most labeled refactoring with a percentage of 35.9\% for group 2, while the difference between operations percentages, in group 1, is not significant, with \textit{Move Class} having the highest percentage of 48.95\%. Another observation is that \textit{Pull Up Attribute} turns out to be the most unlabeled refactoring with a score of 54.91\% for group 1, whereas \textit{Rename Attribute} tends to be the most unlabeled refactoring for group 2. This result is consistent with one of the previous studies stating that renames are rarely labeled, as they detected explicit documentation of renames in less than 1\% of their dataset \citep{arnaoudova2014repent}. 

By comparing the different commits that are labeled and unlabeled with SAR patterns, we observe a significant number of labeled refactoring commits for each refactoring operation supported by the tool Refactoring Miner (p-value = 0.0005). This implies that there is a strong trend of developers in using these phrases in refactoring commits. The results for commits labeled and unlabeled \textit{\say{refactor}} with a p-value = 0.0005 engender an opposite observation, which corroborates the expected outcome of Murphy-Hill et al.'s hypothesis. Thus, the use of \textit{\say{refactor}} is not a great indication of refactoring activities. The difference between the two tests indicates the usefulness of the list of SAR patterns that we identified.  

It is to note that we did not perform any correspondence between the mentioned patterns and the corresponding refactoring operation(s). In other terms, if an operation is explicitly mentioned in a commit message, we have not checked whether it was among the applied refactoring at the source code level. We opted for such verification to be outside of the scope of the current study, while it would be an interesting direction we can consider in our future investigations.

\begin{tcolorbox}

\textit{Summary.} In consistency with the previous findings of \citep{murphy2008gathering}, our findings confirm that developers do not exclusively rely on the pattern \textit{\say{refactor}} to describe refactoring activities. However, we found that developers do document their refactoring activities in commit messages with a variety of patterns that we identified in this study.
\end{tcolorbox}

\section{Discussions and Implications}
\label{sec:Discussion}

In this section, we want to further discuss our findings and outline their implications on future research directions in refactoring.

\textbf{Developer's Motivation behind Refactoring.} One of main findings show that developers are not only driven by design improvement and code smell removal when taking decisions about refactoring. According to our RQ1 findings, fixing bugs, and feature implementation play a major role in triggering various refactoring activities. Traditional refactoring tools are still leading their refactoring effort based on how it is needed to cope with design antipatterns, which is acceptable to the extent where it is indeed the developer's intention, otherwise, they have not been designed or tested in different circumstances. So, an interesting future direction is to study how we can augment existing refactoring tools to better frame the developer's perception of refactoring, and then their corresponding objectives to achieve (reducing coupling, improve code readability, renaming to remove ambiguity etc.). This will automatically induce the search for more adequate refactoring operations, to achieve each objective.

\textbf{Refactoring Support.} Classifying refactoring commits by message is an important activity because it allows us to contextualize these refactoring activities with information about the development activities that led to them. This contextualization is critical and will augment our ability to study the reasoning behind decisions to apply different types of refactoring. This will lead to better support for informing developers of when to apply a refactoring and what refactoring to apply. For example, recent studies try to understand how the development context which motivated a rename refactoring affects the way the words in a name changes when the refactoring is applied \citep{peruma2018empirical, peruma2019contextualizing} for the purpose of modeling, more formally, how names evolve given a development context. Without approaches such as the one proposed in this work, these studies will be missing critical data. In particular, our findings show that renames are dominant, for test files, across all categories (Table~\ref{Table:prod_vs_test_v2}). This indicates that renames occur in many, many different development contexts and, with our tool, studies such as these could be extended to study how names change \textit{given each individual context} instead of assuming they are indistinguishable. This extends to other work as well; there is a critical need for assisting developers in determining when to apply a refactoring; what refactoring to apply; and in some cases how to apply the refactoring \citep{peruma2018empirical, peruma2019contextualizing, arnaoudova2013new,arnaoudova2014repent,liuAutomatedRefactoring}.

Additionally, there is a demonstrated need to further automate refactoring support. Prior research by \citep{6802406} has investigated the way developers interact with IDEs when applying refactorings. \citep{10.1007/978-3-642-39038-8_23, 4602672} have shown that refactorings are frequently applied manually instead of automatically. This indicates that current support for refactoring is not enough; the benefit of automated application is outweighed by the cost, which other researchers have highlighted \citep{newman2018study, Li:2012:LMR:2328876.2328881}. Finally, we theorize that it will be beneficial to study how refactorings are applied to solve different types of problems (i.e., in this case, different maintenance tasks). This is supported by research that isolates certain types of code or code changes, such as isolating test from production code \citep{Tufano:2016:EIN:2970276.2970340}. Like this example, future research must understand the context surrounding refactorings by identifying the reasoning (i.e., development context) behind refactoring operations. The results from this work directly impact research in this area by providing a methodology to categorize refactoring commit messages and providing an exploratory discussion of the motivation behind different types of refactorings. We plan to explore this question in greater detail in future research.

\textbf{Refactoring Documentation.} One of the main purposes of the automatic detection of refactoring is to better understand how developers cope with their software decay by extracting any refactoring strategies that can be associated with removing code smells \citep{tsantalis2008jdeodorant,bavota2013empirical}, or improving the design structural measurements \citep{mkaouer2014recommendation,bavota2014recommending}. However, these techniques only analyze the changes at the source code level, and provide the operations performed, without associating it with any textual description, which may infer the rationale behind the refactoring application. Our proposal, of textual patterns, is the first step towards complementing the existing effort in detecting refactorings, by augmenting it with any description that was intended to describe the refactoring activity. As previously shown in Tables \ref{Table:GeneralPatterns}, \ref{Table:SpecificPatterns_1}, and~\ref{Table:TopSpecificKeywords} developers tend to add a high-level description of their refactoring activity, and occasionally mention their intention behind refactoring (remove duplicate code, improve readability, etc.), along with mentioning the refactoring operations they apply (type migration, inline methods, etc.). This paper proposes, combined with the detection of refactoring operations, a solid background for future empirical investigations. For instance, previous studies have analyzed the impact of refactoring operations on structural metrics \citep{bavota2015experimental,cedrim2016does,palomba2017exploratory}. One of the main limitations of these studies is the absence of any context related to the application of refactorings, i.e., it is not clear whether developers did apply these refactoring with the intention of improving design metrics. Therefore, it is important to consider commits whose commit messages specifically express the refactoring for the purpose of optimizing structural metrics, such as coupling, and complexity, and so, many empirical studies can be revisited with a more adequate dataset. 

Furthermore, our study provides software practitioners with a catalog of common refactoring documentation patterns (cf., Tables \ref{Table:GeneralPatterns}, \ref{Table:SpecificPatterns_1}, and \ref{Table:TopSpecificKeywords}) which would represent concrete examples of common ways to document refactoring activities in commit messages. This catalog of SAR patterns can encourage developers to follow best documentation patterns and to further extend these patterns to improve refactoring changes documentation in particular and code changes in general. Indeed, reliable and accurate documentation is always of crucial importance in any software project. The presence of documentation for low level changes such as refactoring operations and commit changes helps to keep track of all aspects of software development and it improves on the quality of the end product. Its main focuses are learning and knowledge transfer to other developers.

Another important research direction that requires further attention concerns the documentation of refactoring. It has been known that there is a general shortage of refactoring documentation, as developers typically focus on describing their functional updates and bug patches. Also, there is no consensus about how refactoring should be documented, which makes it subjective and developer specific. Moreover, the fine-grained description of refactoring can be time consuming, as typical description should contain indication about the operations performed, refactored code elements, and a hint about the intention behind the refactoring. In addition, the developer specification can be ambiguous as it reflects the developer's understanding of what has been improved in the source code, which can be different in reality, as the developer may not necessarily adequately estimate the refactoring impact on the quality improvement. Therefore, our model can help to build a corpus of refactoring descriptions, and so many studies can better analyze the typical syntax used by developers in order to develop better natural language models to improve it, and potentially automate it, just like existing studies related to other types of code changes \citep{buse2010automatically,linares2015changescribe,liu2018neural}. 

\textbf{Refactoring and Developer's Experience.} While refactoring is being applied by various developers \citep{AlOmarIWoR2020}, it would be interesting to evaluate their refactoring practices. We would like to capture and better understand the code refactoring best practices and learn from these developers so that we can recommend them for other developers. Previous work \citep{alomar2019can} performed an exploratory study on how developers document their refactoring activities in commit messages, this activity is called Self-Affirmed Refactoring (SAR). They found that developers tend to use a variety of textual patterns to document their refactoring activities, such as \textit{\say{refactor}}, \textit{\say{move}} and \textit{\say{extract}}. In follow-up work,  \citep{alomar2019impact} identified which quality models are more in-line with the developer's vision of quality optimization when they explicitly mention in the commit messages that they refactor to improve these quality attributes. Since we noticed that various developers are responsible for performing refactorings, one potential research direction is to investigate which developers are responsible for the introduction of SARs in order to examine whether experience plays a role in the introduction of SARs or not. Another potential research direction is to study if developer experience is one of the factors that might contribute to the significant improvement of the quality metrics that are aligned with developer description in the commit message. In other words, we would like to evaluate the top contributors refactoring practice against all the rest of refactoring contributors by assessing their contributions on the main internal quality attributes improvement (e.g., cohesion, coupling, and complexity). 

\textbf{Refactoring Automation.} There have been various studies targeting the automation of refactoring \citep{harman2007pareto,simons2015search,mkaouer2015many,lin2016interactive,mkaouer2016use}. They mainly rely on the correspondence between the impact of refactoring on the source code to guide the generation of code changes that will potentially improve it. Therefore, existing studies heavily rely on structural measurements to guide the search for these code changes, and so, improving quality attributes and removing anti-patterns were the main drivers for automated refactoring. Clearly, the challenge facing such approaches is applicability. Performing large-scale code changes, impacting various components in the source code, may be catchy for its quality, but it also drastically disturbs the existing software design. Although developers are in favor for optimizing the quality of their software, they still want to recognize their own design. 

\section{Threats to Validity}
\label{sec:Threats}

We identify, in this section, potential threats to the validity of our approach and our experiments. 

\textbf{Internal Validity.} In this paper, we analyzed only the 28 refactoring operations detected by Refactoring Miner, which can be viewed as a validity threat because the tool did not consider all refactoring types mentioned by  \citep{Fowler:1999:RID:311424}. However, in a previous study,  \citep{6112738}  reported that these types are amongst the most common refactoring types. Moreover, we did not perform a manual validation of refactoring types detected by Refactoring Miner to assess its accuracy, so our study is mainly threatened by the accuracy of the detection tool. Yet, \citep{tsantalis2018accurate} report that Refactoring Miner has a precision of 98\% and a recall of 87\% which significantly outperforms the previous state-of-the-art tools, which gives us confidence in using the tool. 

Further, the set of commit messages used in this study may represent a threat to validity, because not all of the messages it may indicate refactoring activities. To mitigate this risk, we manually inspected a subset of change messages and ensured that projects selected are well-commented and use meaningful commit messages. 
 Additionally, since extracting refactoring patterns heavily depends on the content of commit messages, our results may be impacted by the quantity and quality of commits in a software project. To alleviate this threat, we examined multiple projects. Moreover, our manual analysis is a time consuming and an error prone task, which we tried to mitigate by focusing mainly on commits known to contain refactorings. Also, since our keywords largely overlap with keywords used in previous studies, this raised our confidence about the found set but does not guarantee that we did not miss any patterns.

Another threat relates to the detection of JUnit test files. The task of associating a unit test file with its production file was an automated process (performed based on filename/string matching associations). If developers deviate from JUnit guidelines on file naming, false positives may be triggered. However, our manual verification of random associations and the extensiveness of our dataset acts as a means of countering this risk.

\textbf{External Validity.} The first threat is that the analysis was restricted to only open source, Java-based, Git-based repositories. However, we were still able to analyze 800 projects that are highly varied in size, contributors, number of commits and refactorings. Another threat concerns the generalization of the identified recurring patterns in the refactoring commits. Our choice of patterns may have an impact on our findings and may not generalize to other open source or commercial projects since the identified refactoring patterns may be different for another set of projects (e.g., outside the Java developers community or projects that have a low number of or no commit messages). Consequently, we cannot claim that the results of refactoring motivation (see Figure~\ref{fig:commit_classification}) can be generalized to other programming languages in which different refactoring tools have been used, projects with a significantly larger number of commits, and different software systems where the need for improving the design might be less important.

\textbf{Construct validity.} The classification of refactorings heavily relies on commit messages. Even when projects are well-commented, they might not contain SAR, as developers might not document refactoring activities in the commit messages. We mitigate this risk by choosing projects that are appropriate for our analysis. Another potential threat relates to manual classification. Since the manual classification of training commit messages is a human intensive task and it is subject to personal bias, we mitigate manual classification related errors by discarding short and ambiguous commits from our dataset and replacing them with other commits. Another important limitation concerns the size of the dataset used for training and evaluation. The size of the used dataset was determined similarly to previous commit classification studies, but we are not certain that this number is optimal for our problem. It is better to use a systematic technique for choosing the size of the evaluation set. 

To mitigate the impact of different commit message styles and auto-generated messages, we diversified the set of projects to extract commits from. We also randomly sampled from the two commits clusters, those containing detected refactorings and those without. An additional threat to validity relates to the construction of our set of refactoring patterns. One pattern could be used as an umbrella term for lots of different types of activity (e.g., “Cleaning” might mean totally different things to different developers). However, we mitigate this threat by focusing mainly on commits known to contain refactorings.  Further,  recent studies \citep{YAN2016296,Kirinuki:2014:HYC:2597008.2597798} indicate that commit comments could capture more than one type of classification (i.e., mixed maintenance activity). In this work, we only consider single-labeled classification, but this is an interesting direction that we can take into account in our future work.

\textbf{Conclusion Validity.} The refactoring documentation research question has been provided along with the corresponding hypotheses in order to aid in drawing a conclusion. In this context, statistical tests have been used to test the significance of the results gained. Specifically, we applied the Wilcoxon test and the Mann-Whitney U test, widely used non-parametric tests, to test whether refactoring patterns are significant or not, and to test the occurrence of \textit{refactor} in refactoring commits and non-refactoring commits, respectively. These tests make no assumption that the data is normally distributed. Refactoring motivation categories and the way we grouped the refactoring concepts described in previous papers and established relations between them pose a threat to the conclusion validity of our study. If some information was not described in the literature, it may affect our conclusions.

\section{Conclusion}
\label{sec:Conclusion}
In this paper, we performed a large-scale empirical study to explore the motivation driving refactorings, the documentation of refactoring activities, and the proportion of refactoring operations performed on production and test code. In summary, the main conclusions are: (1) our study shows that code smell resolution is not the only driver for developers to factor out their code. Refactoring activity is also driven by changes in requirements, correction of errors, structural design optimization and nonfunctional quality attributes enhancement. Developers are using wide variety of refactoring operations to refactor production and test files, and (2) a wide variety of textual patterns is used to document refactoring activities in the commit messages. These patterns could demonstrate developer perception of refactoring or report a specific refactoring operation name following Fowler's names. 

As future work, we aim to investigate the effect of refactoring on both change and fault-proneness in large-scale open source systems. Specifically, we would like to investigate commit-labeled refactoring to determine if certain refactoring motivations lead to decreased change and fault-prone classes. Further, since a commit message could potentially belong to multiple categories (e.g., improve the design and fix a bug), future research could usefully explore how to automatically classify commits into this kind of hybrid categories. Another potentially interesting future direction will be to conduct additional studies using other refactoring detection tools to analyze open source and industrial software projects and compare findings. Since we observed that feature requests and fix bugs are also refactoring motivators for developers, researchers are encouraged to adopt a maintenance-related refactoring beside design-related refactoring when building a refactoring tool in the future.

{\footnotesize\bibliography{references.bib}}

\begin{thebibliography}{101}
\expandafter\ifx\csname natexlab\endcsname\relax\def\natexlab#1{#1}\fi
\providecommand{\url}[1]{\texttt{#1}}
\providecommand{\href}[2]{#2}
\providecommand{\path}[1]{#1}
\providecommand{\DOIprefix}{doi:}
\providecommand{\ArXivprefix}{arXiv:}
\providecommand{\URLprefix}{URL: }
\providecommand{\Pubmedprefix}{pmid:}
\providecommand{\doi}[1]{\href{http://dx.doi.org/#1}{\path{#1}}}
\providecommand{\Pubmed}[1]{\href{pmid:#1}{\path{#1}}}
\providecommand{\bibinfo}[2]{#2}
\ifx\xfnm\relax \def\xfnm[#1]{\unskip,\space#1}\fi
\bibitem[{Abebe et~al.(2011)Abebe, Haiduc, Tonella \& Marcus}]{abebe2011effect}
\bibinfo{author}{Abebe, S.~L.}, \bibinfo{author}{Haiduc, S.},
  \bibinfo{author}{Tonella, P.}, \& \bibinfo{author}{Marcus, A.}
  (\bibinfo{year}{2011}).
\newblock \bibinfo{title}{The effect of lexicon bad smells on concept location
  in source code}.
\newblock In {\it \bibinfo{booktitle}{Source Code Analysis and Manipulation
  (SCAM), 2011 11th IEEE International Working Conference on}\/} (pp.
  \bibinfo{pages}{125--134}).
\newblock \bibinfo{organization}{Ieee}.
\bibitem[{AlDallal \& Abdin(2017)}]{7833023}
\bibinfo{author}{AlDallal, J.}, \& \bibinfo{author}{Abdin, A.}
  (\bibinfo{year}{2017}).
\newblock \bibinfo{title}{Empirical evaluation of the impact of object-oriented
  code refactoring on quality attributes: A systematic literature review}.
\newblock {\it \bibinfo{journal}{IEEE Transactions on Software Engineering}\/},
   {\it \bibinfo{volume}{PP}\/}, \bibinfo{pages}{1--1}.
  \DOIprefix\doi{10.1109/TSE.2017.2658573}.
\bibitem[{Alkadhi et~al.(2018)Alkadhi, Nonnenmacher, Guzman \&
  Bruegge}]{alkadhi2018developers}
\bibinfo{author}{Alkadhi, R.}, \bibinfo{author}{Nonnenmacher, M.},
  \bibinfo{author}{Guzman, E.}, \& \bibinfo{author}{Bruegge, B.}
  (\bibinfo{year}{2018}).
\newblock \bibinfo{title}{How do developers discuss rationale?}
\newblock In {\it \bibinfo{booktitle}{2018 IEEE 25th International Conference
  on Software Analysis, Evolution and Reengineering (SANER)}\/} (pp.
  \bibinfo{pages}{357--369}).
\newblock \bibinfo{organization}{IEEE}.
\bibitem[{AlOmar(2020 (last accessed October 20, 2020))}]{SAR2020WEB}
\bibinfo{author}{AlOmar, E.~A.} (\bibinfo{year}{2020 (last accessed October 20,
  2020)}).
\newblock {\it \bibinfo{title}{self-affirmed-refactoring repository}\/}.
\newblock \URLprefix
  \url{https://smilevo.github.io/self-affirmed-refactoring/}.
\bibitem[{AlOmar et~al.(2019{\natexlab{a}})AlOmar, Mkaouer \&
  Ouni}]{alomar2019can}
\bibinfo{author}{AlOmar, E.~A.}, \bibinfo{author}{Mkaouer, M.~W.}, \&
  \bibinfo{author}{Ouni, A.} (\bibinfo{year}{2019}{\natexlab{a}}).
\newblock \bibinfo{title}{Can refactoring be self-affirmed? an exploratory
  study on how developers document their refactoring activities in commit
  messages}.
\newblock In {\it \bibinfo{booktitle}{Proceedings of the 3nd International
  Workshop on Refactoring-accepted. IEEE}\/}.
\bibitem[{AlOmar et~al.(2020{\natexlab{a}})AlOmar, Mkaouer \&
  Ouni}]{alomar2020toward}
\bibinfo{author}{AlOmar, E.~A.}, \bibinfo{author}{Mkaouer, M.~W.}, \&
  \bibinfo{author}{Ouni, A.} (\bibinfo{year}{2020}{\natexlab{a}}).
\newblock \bibinfo{title}{Toward the automatic classification of self-affirmed
  refactoring}.
\newblock {\it \bibinfo{journal}{Journal of Systems and Software}\/},  (p.
  \bibinfo{pages}{110821}).
\bibitem[{AlOmar et~al.(2019{\natexlab{b}})AlOmar, Mkaouer, Ouni \&
  Kessentini}]{alomar2019impact}
\bibinfo{author}{AlOmar, E.~A.}, \bibinfo{author}{Mkaouer, M.~W.},
  \bibinfo{author}{Ouni, A.}, \& \bibinfo{author}{Kessentini, M.}
  (\bibinfo{year}{2019}{\natexlab{b}}).
\newblock \bibinfo{title}{On the impact of refactoring on the relationship
  between quality attributes and design metrics}.
\newblock In {\it \bibinfo{booktitle}{2019 ACM/IEEE International Symposium on
  Empirical Software Engineering and Measurement (ESEM)}\/} (pp.
  \bibinfo{pages}{1--11}).
\newblock \bibinfo{organization}{IEEE}.
\bibitem[{AlOmar et~al.(2020{\natexlab{b}})AlOmar, Peruma, Newman, Mkaouer \&
  Ouni}]{AlOmarIWoR2020}
\bibinfo{author}{AlOmar, E.~A.}, \bibinfo{author}{Peruma, A.},
  \bibinfo{author}{Newman, C.~D.}, \bibinfo{author}{Mkaouer, M.~W.}, \&
  \bibinfo{author}{Ouni, A.} (\bibinfo{year}{2020}{\natexlab{b}}).
\newblock \bibinfo{title}{On the relationship between developer experience and
  refactoring: An exploratory study and preliminary results}.
\newblock In {\it \bibinfo{booktitle}{Proceedings of the 4th International
  Workshop on Refactoring}\/} IWoR 2020.
\newblock \bibinfo{address}{New York, NY, USA}: \bibinfo{publisher}{Association
  for Computing Machinery}.
\bibitem[{AlOmar et~al.(2020{\natexlab{c}})AlOmar, Rodriguez, ~, Wang, Adepoju,
  Lopez, Newman, Ouni \& Mkaouer}]{alomar2020how}
\bibinfo{author}{AlOmar, E.~A.}, \bibinfo{author}{Rodriguez, P.~T.},
  \bibinfo{author}{~, J., Bowman}, \bibinfo{author}{Wang, T.},
  \bibinfo{author}{Adepoju, B.}, \bibinfo{author}{Lopez, K.},
  \bibinfo{author}{Newman, C.~D.}, \bibinfo{author}{Ouni, A.}, \&
  \bibinfo{author}{Mkaouer, M.~W.} (\bibinfo{year}{2020}{\natexlab{c}}).
\newblock \bibinfo{title}{How do developers refactor code to improve code
  reusability?}
\newblock In {\it \bibinfo{booktitle}{International Conference on Software and
  Systems Reuse}\/}.
\newblock \bibinfo{organization}{Springer}.
\bibitem[{Alshayeb(2009)}]{alshayeb2009empirical}
\bibinfo{author}{Alshayeb, M.} (\bibinfo{year}{2009}).
\newblock \bibinfo{title}{Empirical investigation of refactoring effect on
  software quality}.
\newblock {\it \bibinfo{journal}{Information and software technology}\/},  {\it
  \bibinfo{volume}{51}\/}, \bibinfo{pages}{1319--1326}.
\bibitem[{Amor et~al.(2006)Amor, Robles, Gonzalez-Barahona, Navarro~Gsyc,
  Carlos \& Madrid}]{article}
\bibinfo{author}{Amor, J.}, \bibinfo{author}{Robles, G.},
  \bibinfo{author}{Gonzalez-Barahona, J.}, \bibinfo{author}{Navarro~Gsyc, A.},
  \bibinfo{author}{Carlos, J.}, \& \bibinfo{author}{Madrid, S.}
  (\bibinfo{year}{2006}).
\newblock \bibinfo{title}{Discriminating development activities in versioning
  systems: A case study}.
\bibitem[{Arnaoudova et~al.(2013)Arnaoudova, Di~Penta, Antoniol \&
  Gueheneuc}]{arnaoudova2013new}
\bibinfo{author}{Arnaoudova, V.}, \bibinfo{author}{Di~Penta, M.},
  \bibinfo{author}{Antoniol, G.}, \& \bibinfo{author}{Gueheneuc, Y.-G.}
  (\bibinfo{year}{2013}).
\newblock \bibinfo{title}{A new family of software anti-patterns: Linguistic
  anti-patterns}.
\newblock In {\it \bibinfo{booktitle}{Software Maintenance and Reengineering
  (CSMR), 2013 17th European Conference on}\/} (pp. \bibinfo{pages}{187--196}).
\newblock \bibinfo{organization}{IEEE}.
\bibitem[{Arnaoudova et~al.(2014)Arnaoudova, Eshkevari, Di~Penta, Oliveto,
  Antoniol \& Gueheneuc}]{arnaoudova2014repent}
\bibinfo{author}{Arnaoudova, V.}, \bibinfo{author}{Eshkevari, L.~M.},
  \bibinfo{author}{Di~Penta, M.}, \bibinfo{author}{Oliveto, R.},
  \bibinfo{author}{Antoniol, G.}, \& \bibinfo{author}{Gueheneuc, Y.-G.}
  (\bibinfo{year}{2014}).
\newblock \bibinfo{title}{Repent: Analyzing the nature of identifier
  renamings}.
\newblock {\it \bibinfo{journal}{IEEE Transactions on Software Engineering}\/},
   {\it \bibinfo{volume}{40}\/}, \bibinfo{pages}{502--532}.
\bibitem[{Barry et~al.(1981)}]{barry1981software}
\bibinfo{author}{Barry, B.} et~al. (\bibinfo{year}{1981}).
\newblock \bibinfo{title}{Software engineering economics}.
\newblock {\it \bibinfo{journal}{New York}\/},  {\it \bibinfo{volume}{197}\/}.
\bibitem[{Bavota et~al.(2015)Bavota, De~Lucia, Di~Penta, Oliveto \&
  Palomba}]{bavota2015experimental}
\bibinfo{author}{Bavota, G.}, \bibinfo{author}{De~Lucia, A.},
  \bibinfo{author}{Di~Penta, M.}, \bibinfo{author}{Oliveto, R.}, \&
  \bibinfo{author}{Palomba, F.} (\bibinfo{year}{2015}).
\newblock \bibinfo{title}{An experimental investigation on the innate
  relationship between quality and refactoring}.
\newblock {\it \bibinfo{journal}{Journal of Systems and Software}\/},  {\it
  \bibinfo{volume}{107}\/}, \bibinfo{pages}{1--14}.
\bibitem[{Bavota et~al.(2013)Bavota, Dit, Oliveto, Di~Penta, Poshyvanyk \&
  De~Lucia}]{bavota2013empirical}
\bibinfo{author}{Bavota, G.}, \bibinfo{author}{Dit, B.},
  \bibinfo{author}{Oliveto, R.}, \bibinfo{author}{Di~Penta, M.},
  \bibinfo{author}{Poshyvanyk, D.}, \& \bibinfo{author}{De~Lucia, A.}
  (\bibinfo{year}{2013}).
\newblock \bibinfo{title}{An empirical study on the developers' perception of
  software coupling}.
\newblock In {\it \bibinfo{booktitle}{Proceedings of the 2013 International
  Conference on Software Engineering}\/} (pp. \bibinfo{pages}{692--701}).
\newblock \bibinfo{organization}{IEEE Press}.
\bibitem[{Bavota et~al.(2014)Bavota, Panichella, Tsantalis, Di~Penta, Oliveto
  \& Canfora}]{bavota2014recommending}
\bibinfo{author}{Bavota, G.}, \bibinfo{author}{Panichella, S.},
  \bibinfo{author}{Tsantalis, N.}, \bibinfo{author}{Di~Penta, M.},
  \bibinfo{author}{Oliveto, R.}, \& \bibinfo{author}{Canfora, G.}
  (\bibinfo{year}{2014}).
\newblock \bibinfo{title}{Recommending refactorings based on team
  co-maintenance patterns}.
\newblock In {\it \bibinfo{booktitle}{Proceedings of the 29th ACM/IEEE
  international conference on Automated software engineering}\/} (pp.
  \bibinfo{pages}{337--342}).
\newblock \bibinfo{organization}{ACM}.
\bibitem[{Bird(2002)}]{Bird2002NLTKTN}
\bibinfo{author}{Bird, S.} (\bibinfo{year}{2002}).
\newblock \bibinfo{title}{Nltk: The natural language toolkit}.
\newblock {\it \bibinfo{journal}{ArXiv}\/},  {\it
  \bibinfo{volume}{cs.CL/0205028}\/}.
\bibitem[{Boehm(2002)}]{Boehm:2002:SEE:944331.944370}
\bibinfo{author}{Boehm, B.~W.} (\bibinfo{year}{2002}).
\newblock \bibinfo{title}{Software pioneers}.
\newblock chapter \bibinfo{chapter}{Software Engineering Economics}. (pp.
  \bibinfo{pages}{641--686}).
\newblock \bibinfo{address}{Berlin, Heidelberg}:
  \bibinfo{publisher}{Springer-Verlag}.
\newblock \URLprefix \url{http://dl.acm.org/citation.cfm?id=944331.944370}.
\bibitem[{Breiman(2017)}]{CART}
\bibinfo{author}{Breiman, L.} (\bibinfo{year}{2017}).
\newblock {\it \bibinfo{title}{Classification and Regression Trees}\/}.
\newblock \bibinfo{publisher}{CRC Press}.
\bibitem[{Brownlee(2018)}]{brownlee2018statistical}
\bibinfo{author}{Brownlee, J.} (\bibinfo{year}{2018}).
\newblock {\it \bibinfo{title}{Statistical Methods for Machine Learning:
  Discover how to Transform Data into Knowledge with Python}\/}.
\newblock \bibinfo{publisher}{Machine Learning Mastery}.
\newblock \URLprefix \url{https://books.google.com/books?id=386nDwAAQBAJ}.
\bibitem[{Buse \& Weimer(2010)}]{buse2010automatically}
\bibinfo{author}{Buse, R.~P.}, \& \bibinfo{author}{Weimer, W.}
  (\bibinfo{year}{2010}).
\newblock \bibinfo{title}{Automatically documenting program changes.}
\newblock In {\it \bibinfo{booktitle}{ASE}\/} (pp. \bibinfo{pages}{33--42}).
\newblock volume~\bibinfo{volume}{10}.
\bibitem[{Cedrim et~al.(2016)Cedrim, Sousa, Garcia \& Gheyi}]{cedrim2016does}
\bibinfo{author}{Cedrim, D.}, \bibinfo{author}{Sousa, L.},
  \bibinfo{author}{Garcia, A.}, \& \bibinfo{author}{Gheyi, R.}
  (\bibinfo{year}{2016}).
\newblock \bibinfo{title}{Does refactoring improve software structural quality?
  a longitudinal study of 25 projects}.
\newblock In {\it \bibinfo{booktitle}{Proceedings of the 30th Brazilian
  Symposium on Software Engineering}\/} (pp. \bibinfo{pages}{73--82}).
\newblock \bibinfo{organization}{ACM}.
\bibitem[{Chang \& Lin(2011)}]{LIBSVM}
\bibinfo{author}{Chang, C.-C.}, \& \bibinfo{author}{Lin, C.-J.}
  (\bibinfo{year}{2011}).
\newblock \bibinfo{title}{Libsvm: A library for support vector machines}, .
\newblock {\it \bibinfo{volume}{2}\/}. \URLprefix
  \url{https://doi.org/10.1145/1961189.1961199}.
  \DOIprefix\doi{10.1145/1961189.1961199}.
\bibitem[{Dangeti(2017)}]{dangeti2017statistics}
\bibinfo{author}{Dangeti, P.} (\bibinfo{year}{2017}).
\newblock {\it \bibinfo{title}{Statistics for Machine Learning}\/}.
\newblock \bibinfo{publisher}{Packt Publishing}.
\bibitem[{Deng et~al.(2012)Deng, Tian \& Zhang}]{cSupportVector}
\bibinfo{author}{Deng, N.}, \bibinfo{author}{Tian, Y.}, \&
  \bibinfo{author}{Zhang, C.} (\bibinfo{year}{2012}).
\newblock {\it \bibinfo{title}{Support Vector Machines: Optimization Based
  Theory, Algorithms, and Extensions}\/}.
\newblock Chapman \& Hall/CRC Data Mining and Knowledge Discovery Series.
\newblock \bibinfo{publisher}{Taylor \& Francis}.
\bibitem[{Dietterich(1998)}]{dietterich1998approximate}
\bibinfo{author}{Dietterich, T.~G.} (\bibinfo{year}{1998}).
\newblock \bibinfo{title}{Approximate statistical tests for comparing
  supervised classification learning algorithms}.
\newblock {\it \bibinfo{journal}{Neural computation}\/},  {\it
  \bibinfo{volume}{10}\/}, \bibinfo{pages}{1895--1923}.
\bibitem[{Dig et~al.(2006)Dig, Comertoglu, Marinov \& Johnson}]{Dig2006}
\bibinfo{author}{Dig, D.}, \bibinfo{author}{Comertoglu, C.},
  \bibinfo{author}{Marinov, D.}, \& \bibinfo{author}{Johnson, R.}
  (\bibinfo{year}{2006}).
\newblock \bibinfo{title}{Automated detection of refactorings in evolving
  components}.
\newblock In \bibinfo{editor}{D.~Thomas} (Ed.), {\it \bibinfo{booktitle}{ECOOP
  2006 -- Object-Oriented Programming: 20th European Conference, Nantes,
  France, July 3-7, 2006. Proceedings}\/} (pp. \bibinfo{pages}{404--428}).
\newblock \bibinfo{address}{Berlin, Heidelberg}: \bibinfo{publisher}{Springer
  Berlin Heidelberg}.
\newblock \URLprefix \url{https://doi.org/10.1007/11785477_24}.
  \DOIprefix\doi{10.1007/11785477_24}.
\bibitem[{Erlikh(2000)}]{erlikh2000leveraging}
\bibinfo{author}{Erlikh, L.} (\bibinfo{year}{2000}).
\newblock \bibinfo{title}{Leveraging legacy system dollars for e-business}.
\newblock {\it \bibinfo{journal}{IT professional}\/},  {\it
  \bibinfo{volume}{2}\/}, \bibinfo{pages}{17--23}.
\bibitem[{Fowler et~al.(1999)Fowler, Beck, Brant, Opdyke \&
  Roberts}]{Fowler:1999:RID:311424}
\bibinfo{author}{Fowler, M.}, \bibinfo{author}{Beck, K.},
  \bibinfo{author}{Brant, J.}, \bibinfo{author}{Opdyke, W.}, \&
  \bibinfo{author}{Roberts, d.} (\bibinfo{year}{1999}).
\newblock {\it \bibinfo{title}{Refactoring: Improving the Design of Existing
  Code}\/}.
\newblock \bibinfo{address}{Boston, MA, USA}:
  \bibinfo{publisher}{Addison-Wesley Longman Publishing Co., Inc.}
\newblock \URLprefix \url{http://dl.acm.org/citation.cfm?id=311424}.
\bibitem[{Harman \& Tratt(2007)}]{harman2007pareto}
\bibinfo{author}{Harman, M.}, \& \bibinfo{author}{Tratt, L.}
  (\bibinfo{year}{2007}).
\newblock \bibinfo{title}{Pareto optimal search based refactoring at the design
  level}.
\newblock In {\it \bibinfo{booktitle}{Proceedings of the 9th annual conference
  on Genetic and evolutionary computation}\/} (pp.
  \bibinfo{pages}{1106--1113}).
\newblock \bibinfo{organization}{ACM}.
\bibitem[{Hattori \& Lanza(2008)}]{4686322}
\bibinfo{author}{Hattori, L.~P.}, \& \bibinfo{author}{Lanza, M.}
  (\bibinfo{year}{2008}).
\newblock \bibinfo{title}{On the nature of commits}.
\newblock In {\it \bibinfo{booktitle}{2008 23rd IEEE/ACM International
  Conference on Automated Software Engineering - Workshops}\/} (pp.
  \bibinfo{pages}{63--71}).
\newblock \DOIprefix\doi{10.1109/ASEW.2008.4686322}.
\bibitem[{Hindle et~al.(2011)Hindle, Ernst, Godfrey \&
  Mylopoulos}]{Hindle:2011:ATN:1985441.1985466}
\bibinfo{author}{Hindle, A.}, \bibinfo{author}{Ernst, N.~A.},
  \bibinfo{author}{Godfrey, M.~W.}, \& \bibinfo{author}{Mylopoulos, J.}
  (\bibinfo{year}{2011}).
\newblock \bibinfo{title}{Automated topic naming to support cross-project
  analysis of software maintenance activities}.
\newblock In {\it \bibinfo{booktitle}{Proceedings of the 8th Working Conference
  on Mining Software Repositories}\/} MSR '11 (pp. \bibinfo{pages}{163--172}).
\newblock \bibinfo{address}{New York, NY, USA}: \bibinfo{publisher}{ACM}.
\newblock \URLprefix \url{http://doi.acm.org/10.1145/1985441.1985466}.
  \DOIprefix\doi{10.1145/1985441.1985466}.
\bibitem[{Hindle et~al.(2009)Hindle, German, Godfrey \& Holt}]{5090025}
\bibinfo{author}{Hindle, A.}, \bibinfo{author}{German, D.~M.},
  \bibinfo{author}{Godfrey, M.~W.}, \& \bibinfo{author}{Holt, R.~C.}
  (\bibinfo{year}{2009}).
\newblock \bibinfo{title}{Automatic classication of large changes into
  maintenance categories}.
\newblock In {\it \bibinfo{booktitle}{2009 IEEE 17th International Conference
  on Program Comprehension}\/} (pp. \bibinfo{pages}{30--39}).
\newblock \DOIprefix\doi{10.1109/ICPC.2009.5090025}.
\bibitem[{Hindle et~al.(2008)Hindle, German \&
  Holt}]{Hindle:2008:LCT:1370750.1370773}
\bibinfo{author}{Hindle, A.}, \bibinfo{author}{German, D.~M.}, \&
  \bibinfo{author}{Holt, R.} (\bibinfo{year}{2008}).
\newblock \bibinfo{title}{What do large commits tell us?: A taxonomical study
  of large commits}.
\newblock In {\it \bibinfo{booktitle}{Proceedings of the 2008 International
  Working Conference on Mining Software Repositories}\/} MSR '08 (pp.
  \bibinfo{pages}{99--108}).
\newblock \bibinfo{address}{New York, NY, USA}: \bibinfo{publisher}{ACM}.
\newblock \URLprefix \url{http://doi.acm.org/10.1145/1370750.1370773}.
  \DOIprefix\doi{10.1145/1370750.1370773}.
\bibitem[{H{\"o}nel et~al.(2019)H{\"o}nel, Ericsson, L{\"o}we \&
  Wingkvist}]{honel2019importance}
\bibinfo{author}{H{\"o}nel, S.}, \bibinfo{author}{Ericsson, M.},
  \bibinfo{author}{L{\"o}we, W.}, \& \bibinfo{author}{Wingkvist, A.}
  (\bibinfo{year}{2019}).
\newblock \bibinfo{title}{Importance and aptitude of source code density for
  commit classification into maintenance activities}.
\newblock In {\it \bibinfo{booktitle}{The 19th IEEE International Conference on
  Software Quality, Reliability, and Security}\/}.
\bibitem[{Jurafsky \& Martin(2019)}]{jurafsky2019speech}
\bibinfo{author}{Jurafsky, D.}, \& \bibinfo{author}{Martin, J.~H.}
  (\bibinfo{year}{2019}).
\newblock \bibinfo{title}{Speech and language processing: An introduction to
  natural language processing, computational linguistics, and speech
  recognition}.
\newblock {\it \bibinfo{journal}{Prentic e Hall}\/}, .
\bibitem[{Kim et~al.(2010)Kim, Gee, Loh \& Rachatasumrit}]{kim2010ref}
\bibinfo{author}{Kim, M.}, \bibinfo{author}{Gee, M.}, \bibinfo{author}{Loh,
  A.}, \& \bibinfo{author}{Rachatasumrit, N.} (\bibinfo{year}{2010}).
\newblock \bibinfo{title}{Ref-finder: a refactoring reconstruction tool based
  on logic query templates}.
\newblock In {\it \bibinfo{booktitle}{Proceedings of the eighteenth ACM SIGSOFT
  international symposium on Foundations of software engineering}\/} (pp.
  \bibinfo{pages}{371--372}).
\newblock \bibinfo{organization}{ACM}.
\bibitem[{Kim et~al.(2014)Kim, Zimmermann \& Nagappan}]{6802406}
\bibinfo{author}{Kim, M.}, \bibinfo{author}{Zimmermann, T.}, \&
  \bibinfo{author}{Nagappan, N.} (\bibinfo{year}{2014}).
\newblock \bibinfo{title}{An empirical study of refactoring challenges and
  benefits at microsoft}.
\newblock {\it \bibinfo{journal}{IEEE Transactions on Software Engineering}\/},
   {\it \bibinfo{volume}{40}\/}, \bibinfo{pages}{633--649}.
  \DOIprefix\doi{10.1109/TSE.2014.2318734}.
\bibitem[{Kirinuki et~al.(2014)Kirinuki, Higo, Hotta \&
  Kusumoto}]{Kirinuki:2014:HYC:2597008.2597798}
\bibinfo{author}{Kirinuki, H.}, \bibinfo{author}{Higo, Y.},
  \bibinfo{author}{Hotta, K.}, \& \bibinfo{author}{Kusumoto, S.}
  (\bibinfo{year}{2014}).
\newblock \bibinfo{title}{Hey! are you committing tangled changes?}
\newblock In {\it \bibinfo{booktitle}{Proceedings of the 22Nd International
  Conference on Program Comprehension}\/} ICPC 2014 (pp.
  \bibinfo{pages}{262--265}).
\newblock \bibinfo{address}{New York, NY, USA}: \bibinfo{publisher}{ACM}.
\newblock \URLprefix \url{http://doi.acm.org/10.1145/2597008.2597798}.
  \DOIprefix\doi{10.1145/2597008.2597798}.
\bibitem[{Kochhar et~al.(2014)Kochhar, Thung \& Lo}]{kochhar2014automatic}
\bibinfo{author}{Kochhar, P.~S.}, \bibinfo{author}{Thung, F.}, \&
  \bibinfo{author}{Lo, D.} (\bibinfo{year}{2014}).
\newblock \bibinfo{title}{Automatic fine-grained issue report
  reclassification}.
\newblock In {\it \bibinfo{booktitle}{Engineering of Complex Computer Systems
  (ICECCS), 2014 19th International Conference on}\/} (pp.
  \bibinfo{pages}{126--135}).
\newblock \bibinfo{organization}{IEEE}.
\bibitem[{Kowsari et~al.(2019)Kowsari, Jafari~Meimandi, Heidarysafa, Mendu,
  Barnes \& Brown}]{kowsari2019text}
\bibinfo{author}{Kowsari, K.}, \bibinfo{author}{Jafari~Meimandi, K.},
  \bibinfo{author}{Heidarysafa, M.}, \bibinfo{author}{Mendu, S.},
  \bibinfo{author}{Barnes, L.}, \& \bibinfo{author}{Brown, D.}
  (\bibinfo{year}{2019}).
\newblock \bibinfo{title}{Text classification algorithms: A survey}.
\newblock {\it \bibinfo{journal}{Information}\/},  {\it
  \bibinfo{volume}{10}\/}, \bibinfo{pages}{150}.
\bibitem[{Lane et~al.(2019)Lane, Hapke \& Howard}]{lane2019natural}
\bibinfo{author}{Lane, H.}, \bibinfo{author}{Hapke, H.}, \&
  \bibinfo{author}{Howard, C.} (\bibinfo{year}{2019}).
\newblock {\it \bibinfo{title}{Natural Language Processing in Action:
  Understanding, Analyzing, and Generating Text with Python}\/}.
\newblock \bibinfo{publisher}{Manning Publications Company}.
\bibitem[{Lanza \& Marinescu(2007)}]{lanza2007object}
\bibinfo{author}{Lanza, M.}, \& \bibinfo{author}{Marinescu, R.}
  (\bibinfo{year}{2007}).
\newblock {\it \bibinfo{title}{Object-oriented metrics in practice: using
  software metrics to characterize, evaluate, and improve the design of
  object-oriented systems}\/}.
\newblock \bibinfo{publisher}{Springer Science \& Business Media}.
\bibitem[{Le et~al.(2015)Le, Linares-V{\'a}squez, Lo \&
  Poshyvanyk}]{le2015rclinker}
\bibinfo{author}{Le, T.-D.~B.}, \bibinfo{author}{Linares-V{\'a}squez, M.},
  \bibinfo{author}{Lo, D.}, \& \bibinfo{author}{Poshyvanyk, D.}
  (\bibinfo{year}{2015}).
\newblock \bibinfo{title}{Rclinker: Automated linking of issue reports and
  commits leveraging rich contextual information}.
\newblock In {\it \bibinfo{booktitle}{2015 IEEE 23rd International Conference
  on Program Comprehension}\/} (pp. \bibinfo{pages}{36--47}).
\newblock \bibinfo{organization}{IEEE}.
\bibitem[{Levin \& Yehudai(2017)}]{Levin:2017:BAC:3127005.3127016}
\bibinfo{author}{Levin, S.}, \& \bibinfo{author}{Yehudai, A.}
  (\bibinfo{year}{2017}).
\newblock \bibinfo{title}{Boosting automatic commit classification into
  maintenance activities by utilizing source code changes}.
\newblock In {\it \bibinfo{booktitle}{Proceedings of the 13th International
  Conference on Predictive Models and Data Analytics in Software
  Engineering}\/} PROMISE (pp. \bibinfo{pages}{97--106}).
\newblock \bibinfo{address}{New York, NY, USA}: \bibinfo{publisher}{ACM}.
\newblock \URLprefix \url{http://doi.acm.org/10.1145/3127005.3127016}.
  \DOIprefix\doi{10.1145/3127005.3127016}.
\bibitem[{Li \& Thompson(2012)}]{Li:2012:LMR:2328876.2328881}
\bibinfo{author}{Li, H.}, \& \bibinfo{author}{Thompson, S.}
  (\bibinfo{year}{2012}).
\newblock \bibinfo{title}{Let's make refactoring tools user-extensible!}
\newblock In {\it \bibinfo{booktitle}{Proceedings of the Fifth Workshop on
  Refactoring Tools}\/} WRT '12 (pp. \bibinfo{pages}{32--39}).
\newblock \bibinfo{address}{New York, NY, USA}: \bibinfo{publisher}{ACM}.
\newblock \URLprefix \url{http://doi.acm.org/10.1145/2328876.2328881}.
  \DOIprefix\doi{10.1145/2328876.2328881}.
\bibitem[{Lin et~al.(2013)Lin, Ma \& Chen}]{lin2013empirical}
\bibinfo{author}{Lin, S.}, \bibinfo{author}{Ma, Y.}, \& \bibinfo{author}{Chen,
  J.} (\bibinfo{year}{2013}).
\newblock \bibinfo{title}{Empirical evidence on developer's commit activity for
  open-source software projects.}
\newblock In {\it \bibinfo{booktitle}{SEKE}\/} (pp. \bibinfo{pages}{455--460}).
\newblock volume~\bibinfo{volume}{13}.
\bibitem[{Lin et~al.(2016)Lin, Peng, Cai, Dig, Zheng \&
  Zhao}]{lin2016interactive}
\bibinfo{author}{Lin, Y.}, \bibinfo{author}{Peng, X.}, \bibinfo{author}{Cai,
  Y.}, \bibinfo{author}{Dig, D.}, \bibinfo{author}{Zheng, D.}, \&
  \bibinfo{author}{Zhao, W.} (\bibinfo{year}{2016}).
\newblock \bibinfo{title}{Interactive and guided architectural refactoring with
  search-based recommendation}.
\newblock In {\it \bibinfo{booktitle}{Proceedings of the 2016 24th ACM SIGSOFT
  International Symposium on Foundations of Software Engineering}\/} (pp.
  \bibinfo{pages}{535--546}).
\newblock \bibinfo{organization}{ACM}.
\bibitem[{Linares-V{\'a}squez et~al.(2015)Linares-V{\'a}squez, Cort{\'e}s-Coy,
  Aponte \& Poshyvanyk}]{linares2015changescribe}
\bibinfo{author}{Linares-V{\'a}squez, M.}, \bibinfo{author}{Cort{\'e}s-Coy,
  L.~F.}, \bibinfo{author}{Aponte, J.}, \& \bibinfo{author}{Poshyvanyk, D.}
  (\bibinfo{year}{2015}).
\newblock \bibinfo{title}{Changescribe: A tool for automatically generating
  commit messages}.
\newblock In {\it \bibinfo{booktitle}{2015 IEEE/ACM 37th IEEE International
  Conference on Software Engineering}\/} (pp. \bibinfo{pages}{709--712}).
\newblock \bibinfo{organization}{IEEE} volume~\bibinfo{volume}{2}.
\bibitem[{{Liu} et~al.(2013){Liu}, {Guo} \& {Shao}}]{liuAutomatedRefactoring}
\bibinfo{author}{{Liu}, H.}, \bibinfo{author}{{Guo}, X.}, \&
  \bibinfo{author}{{Shao}, W.} (\bibinfo{year}{2013}).
\newblock \bibinfo{title}{Monitor-based instant software refactoring}.
\newblock {\it \bibinfo{journal}{IEEE Transactions on Software Engineering}\/},
   {\it \bibinfo{volume}{39}\/}, \bibinfo{pages}{1112--1126}.
  \DOIprefix\doi{10.1109/TSE.2013.4}.
\bibitem[{Liu et~al.(2018)Liu, Xia, Hassan, Lo, Xing \& Wang}]{liu2018neural}
\bibinfo{author}{Liu, Z.}, \bibinfo{author}{Xia, X.}, \bibinfo{author}{Hassan,
  A.~E.}, \bibinfo{author}{Lo, D.}, \bibinfo{author}{Xing, Z.}, \&
  \bibinfo{author}{Wang, X.} (\bibinfo{year}{2018}).
\newblock \bibinfo{title}{Neural-machine-translation-based commit message
  generation: how far are we?}
\newblock In {\it \bibinfo{booktitle}{Proceedings of the 33rd ACM/IEEE
  International Conference on Automated Software Engineering}\/} (pp.
  \bibinfo{pages}{373--384}).
\newblock \bibinfo{organization}{ACM}.
\bibitem[{Manning et~al.(2008)Manning, Raghavan \&
  Sch{\"u}tze}]{manning2008introduction}
\bibinfo{author}{Manning, C.}, \bibinfo{author}{Raghavan, P.}, \&
  \bibinfo{author}{Sch{\"u}tze, H.} (\bibinfo{year}{2008}).
\newblock {\it \bibinfo{title}{Introduction to Information Retrieval}\/}.
\newblock \bibinfo{publisher}{Cambridge University Press}.
\bibitem[{Mauczka et~al.(2015)Mauczka, Brosch, Schanes \& Grechenig}]{7180125}
\bibinfo{author}{Mauczka, A.}, \bibinfo{author}{Brosch, F.},
  \bibinfo{author}{Schanes, C.}, \& \bibinfo{author}{Grechenig, T.}
  (\bibinfo{year}{2015}).
\newblock \bibinfo{title}{Dataset of developer-labeled commit messages}.
\newblock In {\it \bibinfo{booktitle}{2015 IEEE/ACM 12th Working Conference on
  Mining Software Repositories}\/} (pp. \bibinfo{pages}{490--493}).
\newblock \DOIprefix\doi{10.1109/MSR.2015.71}.
\bibitem[{Mauczka et~al.(2012)Mauczka, Huber, Schanes, Schramm, Bernhart \&
  Grechenig}]{Mauczka2012}
\bibinfo{author}{Mauczka, A.}, \bibinfo{author}{Huber, M.},
  \bibinfo{author}{Schanes, C.}, \bibinfo{author}{Schramm, W.},
  \bibinfo{author}{Bernhart, M.}, \& \bibinfo{author}{Grechenig, T.}
  (\bibinfo{year}{2012}).
\newblock \bibinfo{title}{Tracing your maintenance work -- a cross-project
  validation of an automated classification dictionary for commit messages}.
\newblock In \bibinfo{editor}{J.~de~Lara}, \& \bibinfo{editor}{A.~Zisman}
  (Eds.), {\it \bibinfo{booktitle}{Fundamental Approaches to Software
  Engineering: 15th International Conference, FASE 2012, Held as Part of the
  European Joint Conferences on Theory and Practice of Software, ETAPS 2012,
  Tallinn, Estonia, March 24 - April 1, 2012. Proceedings}\/} (pp.
  \bibinfo{pages}{301--315}).
\newblock \bibinfo{address}{Berlin, Heidelberg}: \bibinfo{publisher}{Springer
  Berlin Heidelberg}.
\newblock \URLprefix \url{https://doi.org/10.1007/978-3-642-28872-2_21}.
  \DOIprefix\doi{10.1007/978-3-642-28872-2_21}.
\bibitem[{McBurney et~al.(2017)McBurney, Jiang, Kessentini, Kraft, Armaly,
  Mkaouer \& McMillan}]{mcburney2017towards}
\bibinfo{author}{McBurney, P.~W.}, \bibinfo{author}{Jiang, S.},
  \bibinfo{author}{Kessentini, M.}, \bibinfo{author}{Kraft, N.~A.},
  \bibinfo{author}{Armaly, A.}, \bibinfo{author}{Mkaouer, M.~W.}, \&
  \bibinfo{author}{McMillan, C.} (\bibinfo{year}{2017}).
\newblock \bibinfo{title}{Towards prioritizing documentation effort}.
\newblock {\it \bibinfo{journal}{IEEE Transactions on Software Engineering}\/},
   {\it \bibinfo{volume}{44}\/}, \bibinfo{pages}{897--913}.
\bibitem[{Mkaouer et~al.(2016)Mkaouer, Kessentini, Bechikh, Cinn{\'e}ide \&
  Deb}]{mkaouer2016use}
\bibinfo{author}{Mkaouer, M.~W.}, \bibinfo{author}{Kessentini, M.},
  \bibinfo{author}{Bechikh, S.}, \bibinfo{author}{Cinn{\'e}ide, M.~{\'O}.}, \&
  \bibinfo{author}{Deb, K.} (\bibinfo{year}{2016}).
\newblock \bibinfo{title}{On the use of many quality attributes for software
  refactoring: a many-objective search-based software engineering approach}.
\newblock {\it \bibinfo{journal}{Empirical Software Engineering}\/},  {\it
  \bibinfo{volume}{21}\/}, \bibinfo{pages}{2503--2545}.
\bibitem[{Mkaouer et~al.(2014)Mkaouer, Kessentini, Bechikh, Deb \&
  {\'O}~Cinn{\'e}ide}]{mkaouer2014recommendation}
\bibinfo{author}{Mkaouer, M.~W.}, \bibinfo{author}{Kessentini, M.},
  \bibinfo{author}{Bechikh, S.}, \bibinfo{author}{Deb, K.}, \&
  \bibinfo{author}{{\'O}~Cinn{\'e}ide, M.} (\bibinfo{year}{2014}).
\newblock \bibinfo{title}{Recommendation system for software refactoring using
  innovization and interactive dynamic optimization}.
\newblock In {\it \bibinfo{booktitle}{Proceedings of the 29th ACM/IEEE
  international conference on Automated software engineering}\/} (pp.
  \bibinfo{pages}{331--336}).
\newblock \bibinfo{organization}{ACM}.
\bibitem[{Mkaouer et~al.(2015)Mkaouer, Kessentini, Shaout, Koligheu, Bechikh,
  Deb \& Ouni}]{mkaouer2015many}
\bibinfo{author}{Mkaouer, W.}, \bibinfo{author}{Kessentini, M.},
  \bibinfo{author}{Shaout, A.}, \bibinfo{author}{Koligheu, P.},
  \bibinfo{author}{Bechikh, S.}, \bibinfo{author}{Deb, K.}, \&
  \bibinfo{author}{Ouni, A.} (\bibinfo{year}{2015}).
\newblock \bibinfo{title}{Many-objective software remodularization using
  nsga-iii}.
\newblock {\it \bibinfo{journal}{ACM Transactions on Software Engineering and
  Methodology (TOSEM)}\/},  {\it \bibinfo{volume}{24}\/}, \bibinfo{pages}{17}.
\bibitem[{Moser et~al.(2007)Moser, Abrahamsson, Pedrycz, Sillitti \&
  Succi}]{moser2008case}
\bibinfo{author}{Moser, R.}, \bibinfo{author}{Abrahamsson, P.},
  \bibinfo{author}{Pedrycz, W.}, \bibinfo{author}{Sillitti, A.}, \&
  \bibinfo{author}{Succi, G.} (\bibinfo{year}{2007}).
\newblock \bibinfo{title}{A case study on the impact of refactoring on quality
  and productivity in an agile team}.
\newblock In {\it \bibinfo{booktitle}{Balancing Agility and Formalism in
  Software Engineering}\/} (pp. \bibinfo{pages}{252--266}).
\newblock \bibinfo{publisher}{Springer}.
\bibitem[{Moser et~al.(2006)Moser, Sillitti, Abrahamsson \&
  Succi}]{moser2006does}
\bibinfo{author}{Moser, R.}, \bibinfo{author}{Sillitti, A.},
  \bibinfo{author}{Abrahamsson, P.}, \& \bibinfo{author}{Succi, G.}
  (\bibinfo{year}{2006}).
\newblock \bibinfo{title}{Does refactoring improve reusability?}
\newblock In {\it \bibinfo{booktitle}{International Conference on Software
  Reuse}\/} (pp. \bibinfo{pages}{287--297}).
\newblock \bibinfo{organization}{Springer}.
\bibitem[{Munaiah et~al.(2017)Munaiah, Kroh, Cabrey \&
  Nagappan}]{munaiah2017curating}
\bibinfo{author}{Munaiah, N.}, \bibinfo{author}{Kroh, S.},
  \bibinfo{author}{Cabrey, C.}, \& \bibinfo{author}{Nagappan, M.}
  (\bibinfo{year}{2017}).
\newblock \bibinfo{title}{Curating github for engineered software projects}.
\newblock {\it \bibinfo{journal}{Empirical Software Engineering}\/},  {\it
  \bibinfo{volume}{22}\/}, \bibinfo{pages}{3219--3253}.
\bibitem[{Murphy-Hill \& Black(2008)}]{4602672}
\bibinfo{author}{Murphy-Hill, E.}, \& \bibinfo{author}{Black, A.~P.}
  (\bibinfo{year}{2008}).
\newblock \bibinfo{title}{Refactoring tools: Fitness for purpose}.
\newblock {\it \bibinfo{journal}{IEEE Software}\/},  {\it
  \bibinfo{volume}{25}\/}, \bibinfo{pages}{38--44}.
  \DOIprefix\doi{10.1109/MS.2008.123}.
\bibitem[{Murphy-Hill et~al.(2008)Murphy-Hill, Black, Dig \&
  Parnin}]{murphy2008gathering}
\bibinfo{author}{Murphy-Hill, E.}, \bibinfo{author}{Black, A.~P.},
  \bibinfo{author}{Dig, D.}, \& \bibinfo{author}{Parnin, C.}
  (\bibinfo{year}{2008}).
\newblock \bibinfo{title}{Gathering refactoring data: a comparison of four
  methods}.
\newblock In {\it \bibinfo{booktitle}{Proceedings of the 2nd Workshop on
  Refactoring Tools}\/} (p.~\bibinfo{pages}{7}).
\newblock \bibinfo{organization}{ACM}.
\bibitem[{Murphy-Hill et~al.(2012)Murphy-Hill, Parnin \& Black}]{6112738}
\bibinfo{author}{Murphy-Hill, E.}, \bibinfo{author}{Parnin, C.}, \&
  \bibinfo{author}{Black, A.~P.} (\bibinfo{year}{2012}).
\newblock \bibinfo{title}{How we refactor, and how we know it}.
\newblock {\it \bibinfo{journal}{IEEE Transactions on Software Engineering}\/},
   {\it \bibinfo{volume}{38}\/}, \bibinfo{pages}{5--18}.
  \DOIprefix\doi{10.1109/TSE.2011.41}.
\bibitem[{Negara et~al.(2013)Negara, Chen, Vakilian, Johnson \&
  Dig}]{10.1007/978-3-642-39038-8_23}
\bibinfo{author}{Negara, S.}, \bibinfo{author}{Chen, N.},
  \bibinfo{author}{Vakilian, M.}, \bibinfo{author}{Johnson, R.~E.}, \&
  \bibinfo{author}{Dig, D.} (\bibinfo{year}{2013}).
\newblock \bibinfo{title}{A comparative study of manual and automated
  refactorings}.
\newblock In \bibinfo{editor}{G.~Castagna} (Ed.), {\it
  \bibinfo{booktitle}{ECOOP 2013 -- Object-Oriented Programming}\/} (pp.
  \bibinfo{pages}{552--576}).
\newblock \bibinfo{address}{Berlin, Heidelberg}: \bibinfo{publisher}{Springer
  Berlin Heidelberg}.
\bibitem[{Newman et~al.(2018)Newman, Mkaouer, Collard \&
  Maletic}]{newman2018study}
\bibinfo{author}{Newman, C.~D.}, \bibinfo{author}{Mkaouer, M.~W.},
  \bibinfo{author}{Collard, M.~L.}, \& \bibinfo{author}{Maletic, J.~I.}
  (\bibinfo{year}{2018}).
\newblock \bibinfo{title}{A study on developer perception of transformation
  languages for refactoring}.
\newblock In {\it \bibinfo{booktitle}{Proceedings of the 2nd International
  Workshop on Refactoring}\/} (pp. \bibinfo{pages}{34--41}).
\newblock \bibinfo{organization}{ACM}.
\bibitem[{Paix{\~a}o et~al.(2020)Paix{\~a}o, Uch{\^o}a, Bibiano, Oliveira,
  Garcia, Krinke \& Arvonio}]{paixao2020behind}
\bibinfo{author}{Paix{\~a}o, M.}, \bibinfo{author}{Uch{\^o}a, A.},
  \bibinfo{author}{Bibiano, A.~C.}, \bibinfo{author}{Oliveira, D.},
  \bibinfo{author}{Garcia, A.}, \bibinfo{author}{Krinke, J.}, \&
  \bibinfo{author}{Arvonio, E.} (\bibinfo{year}{2020}).
\newblock \bibinfo{title}{Behind the intents: An in-depth empirical study on
  software refactoring in modern code review}.
\newblock {\it \bibinfo{journal}{17th MSR}\/}, .
\bibitem[{Palomba et~al.(2017)Palomba, Zaidman, Oliveto \&
  De~Lucia}]{palomba2017exploratory}
\bibinfo{author}{Palomba, F.}, \bibinfo{author}{Zaidman, A.},
  \bibinfo{author}{Oliveto, R.}, \& \bibinfo{author}{De~Lucia, A.}
  (\bibinfo{year}{2017}).
\newblock \bibinfo{title}{An exploratory study on the relationship between
  changes and refactoring}.
\newblock In {\it \bibinfo{booktitle}{2017 IEEE/ACM 25th International
  Conference on Program Comprehension (ICPC)}\/} (pp.
  \bibinfo{pages}{176--185}).
\newblock \bibinfo{organization}{IEEE}.
\bibitem[{Pantiuchina et~al.(2020)Pantiuchina, Zampetti, Scalabrino,
  Piantadosi, Oliveto, Bavota \& Di~Penta}]{pantiuchina2018developers}
\bibinfo{author}{Pantiuchina, J.}, \bibinfo{author}{Zampetti, F.},
  \bibinfo{author}{Scalabrino, S.}, \bibinfo{author}{Piantadosi, V.},
  \bibinfo{author}{Oliveto, R.}, \bibinfo{author}{Bavota, G.}, \&
  \bibinfo{author}{Di~Penta, M.} (\bibinfo{year}{2020}).
\newblock \bibinfo{title}{Why developers refactor source code: A mining-based
  study}, .
\bibitem[{Peruma et~al.(2019{\natexlab{a}})Peruma, Almalki, Newman, Mkaouer,
  Ouni \& Palomba}]{10.5555/3370272.3370293}
\bibinfo{author}{Peruma, A.}, \bibinfo{author}{Almalki, K.},
  \bibinfo{author}{Newman, C.~D.}, \bibinfo{author}{Mkaouer, M.~W.},
  \bibinfo{author}{Ouni, A.}, \& \bibinfo{author}{Palomba, F.}
  (\bibinfo{year}{2019}{\natexlab{a}}).
\newblock \bibinfo{title}{On the distribution of test smells in open source
  android applications: An exploratory study}.
\newblock In {\it \bibinfo{booktitle}{Proceedings of the 29th Annual
  International Conference on Computer Science and Software Engineering}\/}
  CASCON ’19 (p. \bibinfo{pages}{193–202}).
\newblock \bibinfo{address}{USA}: \bibinfo{publisher}{IBM Corp.}
\bibitem[{Peruma et~al.(2018)Peruma, Mkaouer, Decker \&
  Newman}]{peruma2018empirical}
\bibinfo{author}{Peruma, A.}, \bibinfo{author}{Mkaouer, M.~W.},
  \bibinfo{author}{Decker, M.~J.}, \& \bibinfo{author}{Newman, C.~D.}
  (\bibinfo{year}{2018}).
\newblock \bibinfo{title}{An empirical investigation of how and why developers
  rename identifiers}.
\newblock In {\it \bibinfo{booktitle}{Proceedings of the 2nd International
  Workshop on Refactoring}\/} (pp. \bibinfo{pages}{26--33}).
\newblock \bibinfo{organization}{ACM}.
\bibitem[{Peruma et~al.(2019{\natexlab{b}})Peruma, Mkaouer, Decker \&
  Newman}]{peruma2019contextualizing}
\bibinfo{author}{Peruma, A.}, \bibinfo{author}{Mkaouer, M.~W.},
  \bibinfo{author}{Decker, M.~J.}, \& \bibinfo{author}{Newman, C.~D.}
  (\bibinfo{year}{2019}{\natexlab{b}}).
\newblock \bibinfo{title}{Contextualizing rename decisions using refactorings
  and commit messages}.
\newblock In {\it \bibinfo{booktitle}{Proceedings of the 19th IEEE
  International Working Conference on Source Code Analysis and Manipulation,
  IEEE}\/}.
\bibitem[{Potdar \& Shihab(2014)}]{potdar2014exploratory}
\bibinfo{author}{Potdar, A.}, \& \bibinfo{author}{Shihab, E.}
  (\bibinfo{year}{2014}).
\newblock \bibinfo{title}{An exploratory study on self-admitted technical
  debt}.
\newblock In {\it \bibinfo{booktitle}{Software Maintenance and Evolution
  (ICSME), 2014 IEEE International Conference on}\/} (pp.
  \bibinfo{pages}{91--100}).
\newblock \bibinfo{organization}{IEEE}.
\bibitem[{Ratzinger(2007)}]{citeulike:2881658}
\bibinfo{author}{Ratzinger, J.} (\bibinfo{year}{2007}).
\newblock {\it \bibinfo{title}{{sPACE: Software Project Assessment in the
  Course of Evolution}}\/}.
\newblock Ph.D. thesis.
\newblock \URLprefix
  \url{http://www.infosys.tuwien.ac.at/Staff/ratzinger/publications/ratzinger\_phd-thesis\_space.pdf}.
\bibitem[{Ratzinger et~al.(2008)Ratzinger, Sigmund \&
  Gall}]{Ratzinger:2008:RRS:1370750.1370759}
\bibinfo{author}{Ratzinger, J.}, \bibinfo{author}{Sigmund, T.}, \&
  \bibinfo{author}{Gall, H.~C.} (\bibinfo{year}{2008}).
\newblock \bibinfo{title}{On the relation of refactorings and software defect
  prediction}.
\newblock In {\it \bibinfo{booktitle}{Proceedings of the 2008 International
  Working Conference on Mining Software Repositories}\/} MSR '08 (pp.
  \bibinfo{pages}{35--38}).
\newblock \bibinfo{address}{New York, NY, USA}: \bibinfo{publisher}{ACM}.
\newblock \URLprefix \url{http://doi.acm.org/10.1145/1370750.1370759}.
  \DOIprefix\doi{10.1145/1370750.1370759}.
\bibitem[{Silva et~al.(2016)Silva, Tsantalis \&
  Valente}]{Silva:2016:WWR:2950290.2950305}
\bibinfo{author}{Silva, D.}, \bibinfo{author}{Tsantalis, N.}, \&
  \bibinfo{author}{Valente, M.~T.} (\bibinfo{year}{2016}).
\newblock \bibinfo{title}{Why we refactor? confessions of github contributors}.
\newblock In {\it \bibinfo{booktitle}{Proceedings of the 2016 24th ACM SIGSOFT
  International Symposium on Foundations of Software Engineering}\/} FSE 2016
  (pp. \bibinfo{pages}{858--870}).
\newblock \bibinfo{address}{New York, NY, USA}: \bibinfo{publisher}{ACM}.
\newblock \URLprefix \url{http://doi.acm.org/10.1145/2950290.2950305}.
  \DOIprefix\doi{10.1145/2950290.2950305}.
\bibitem[{Simons et~al.(2015)Simons, Singer \& White}]{simons2015search}
\bibinfo{author}{Simons, C.}, \bibinfo{author}{Singer, J.}, \&
  \bibinfo{author}{White, D.~R.} (\bibinfo{year}{2015}).
\newblock \bibinfo{title}{Search-based refactoring: Metrics are not enough}.
\newblock In {\it \bibinfo{booktitle}{International Symposium on Search Based
  Software Engineering}\/} (pp. \bibinfo{pages}{47--61}).
\newblock \bibinfo{organization}{Springer}.
\bibitem[{Singh \& Mangat(2013)}]{singh2013elements}
\bibinfo{author}{Singh, R.}, \& \bibinfo{author}{Mangat, N.}
  (\bibinfo{year}{2013}).
\newblock {\it \bibinfo{title}{Elements of Survey Sampling}\/}.
\newblock Texts in the Mathematical Sciences.
\newblock \bibinfo{publisher}{Springer Netherlands}.
\bibitem[{SKlearn(2007{\natexlab{a}})}]{inherently}
\bibinfo{author}{SKlearn} (\bibinfo{year}{2007}{\natexlab{a}}).
\newblock \bibinfo{title}{1.12. multiclass and multilabel algorithms —
  scikit-learn 0.23.2 documentation}.
\newblock
  \bibinfo{howpublished}{\url{https://scikit-learn.org/stable/modules/multiclass.html}}.
\bibitem[{SKlearn(2007{\natexlab{b}})}]{SVC}
\bibinfo{author}{SKlearn} (\bibinfo{year}{2007}{\natexlab{b}}).
\newblock \bibinfo{title}{sklearn.svm.svc — scikit-learn 0.23.2
  documentation}.
\newblock
  \bibinfo{howpublished}{\url{https://scikit-learn.org/stable/modules/generated/sklearn.svm.SVC.html\#sklear.svm.SVC}}.
\bibitem[{Soares et~al.(2009)Soares, Cavalcanti, Gheyi, Massoni, Serey \&
  Corn{\'e}lio}]{Soares2009safetytool}
\bibinfo{author}{Soares, G.}, \bibinfo{author}{Cavalcanti, D.},
  \bibinfo{author}{Gheyi, R.}, \bibinfo{author}{Massoni, T.},
  \bibinfo{author}{Serey, D.}, \& \bibinfo{author}{Corn{\'e}lio, M.}
  (\bibinfo{year}{2009}).
\newblock \bibinfo{title}{Saferefactor-tool for checking refactoring safety}.
\newblock {\it \bibinfo{journal}{Tools Session at SBES}\/},  (pp.
  \bibinfo{pages}{49--54}).
\bibitem[{Soares et~al.(2013)Soares, Gheyi, Murphy-Hill \&
  Johnson}]{soares2013comparing}
\bibinfo{author}{Soares, G.}, \bibinfo{author}{Gheyi, R.},
  \bibinfo{author}{Murphy-Hill, E.}, \& \bibinfo{author}{Johnson, B.}
  (\bibinfo{year}{2013}).
\newblock \bibinfo{title}{Comparing approaches to analyze refactoring activity
  on software repositories}.
\newblock {\it \bibinfo{journal}{Journal of Systems and Software}\/},  {\it
  \bibinfo{volume}{86}\/}, \bibinfo{pages}{1006--1022}.
\bibitem[{Stroggylos \& Spinellis(2007)}]{stroggylos2007refactoring}
\bibinfo{author}{Stroggylos, K.}, \& \bibinfo{author}{Spinellis, D.}
  (\bibinfo{year}{2007}).
\newblock \bibinfo{title}{Refactoring--does it improve software quality?}
\newblock In {\it \bibinfo{booktitle}{Software Quality, 2007. WoSQ'07: ICSE
  Workshops 2007. Fifth International Workshop on}\/} (pp.
  \bibinfo{pages}{10--10}).
\newblock \bibinfo{organization}{IEEE}.
\bibitem[{Swanson(1976)}]{Swanson:1976:DM:800253.807723}
\bibinfo{author}{Swanson, E.~B.} (\bibinfo{year}{1976}).
\newblock \bibinfo{title}{The dimensions of maintenance}.
\newblock In {\it \bibinfo{booktitle}{Proceedings of the 2Nd International
  Conference on Software Engineering}\/} ICSE '76 (pp.
  \bibinfo{pages}{492--497}).
\newblock \bibinfo{address}{Los Alamitos, CA, USA}: \bibinfo{publisher}{IEEE
  Computer Society Press}.
\newblock \URLprefix \url{http://dl.acm.org/citation.cfm?id=800253.807723}.
\bibitem[{Sz{\H{o}}ke et~al.(2017)Sz{\H{o}}ke, Antal, Nagy, Ferenc \&
  Gyim{\'o}thy}]{szHoke2017empirical}
\bibinfo{author}{Sz{\H{o}}ke, G.}, \bibinfo{author}{Antal, G.},
  \bibinfo{author}{Nagy, C.}, \bibinfo{author}{Ferenc, R.}, \&
  \bibinfo{author}{Gyim{\'o}thy, T.} (\bibinfo{year}{2017}).
\newblock \bibinfo{title}{Empirical study on refactoring large-scale industrial
  systems and its effects on maintainability}.
\newblock {\it \bibinfo{journal}{Journal of Systems and Software}\/},  {\it
  \bibinfo{volume}{129}\/}, \bibinfo{pages}{107--126}.
\bibitem[{Sz{\H{o}}ke et~al.(2014)Sz{\H{o}}ke, Nagy, Ferenc \&
  Gyim{\'o}thy}]{szHoke2014case}
\bibinfo{author}{Sz{\H{o}}ke, G.}, \bibinfo{author}{Nagy, C.},
  \bibinfo{author}{Ferenc, R.}, \& \bibinfo{author}{Gyim{\'o}thy, T.}
  (\bibinfo{year}{2014}).
\newblock \bibinfo{title}{A case study of refactoring large-scale industrial
  systems to efficiently improve source code quality}.
\newblock In {\it \bibinfo{booktitle}{International Conference on Computational
  Science and Its Applications}\/} (pp. \bibinfo{pages}{524--540}).
\newblock \bibinfo{organization}{Springer}.
\bibitem[{Tan et~al.(1999)}]{tan1999text}
\bibinfo{author}{Tan, A.-H.} et~al. (\bibinfo{year}{1999}).
\newblock \bibinfo{title}{Text mining: The state of the art and the
  challenges}.
\newblock In {\it \bibinfo{booktitle}{Proceedings of the PAKDD 1999 Workshop on
  Knowledge Disocovery from Advanced Databases}\/} (pp.
  \bibinfo{pages}{65--70}).
\newblock \bibinfo{organization}{sn} volume~\bibinfo{volume}{8}.
\bibitem[{Tan et~al.(2002)Tan, Wang \& Lee}]{tan2002use}
\bibinfo{author}{Tan, C.-M.}, \bibinfo{author}{Wang, Y.-F.}, \&
  \bibinfo{author}{Lee, C.-D.} (\bibinfo{year}{2002}).
\newblock \bibinfo{title}{The use of bigrams to enhance text categorization}.
\newblock {\it \bibinfo{journal}{Information processing \& management}\/},
  {\it \bibinfo{volume}{38}\/}, \bibinfo{pages}{529--546}.
\bibitem[{Tsantalis et~al.(2008)Tsantalis, Chaikalis \&
  Chatzigeorgiou}]{tsantalis2008jdeodorant}
\bibinfo{author}{Tsantalis, N.}, \bibinfo{author}{Chaikalis, T.}, \&
  \bibinfo{author}{Chatzigeorgiou, A.} (\bibinfo{year}{2008}).
\newblock \bibinfo{title}{Jdeodorant: Identification and removal of
  type-checking bad smells}.
\newblock In {\it \bibinfo{booktitle}{2008 12th European Conference on Software
  Maintenance and Reengineering}\/} (pp. \bibinfo{pages}{329--331}).
\newblock \bibinfo{organization}{IEEE}.
\bibitem[{Tsantalis \& Chatzigeorgiou(2011)}]{tsantalis2011identification}
\bibinfo{author}{Tsantalis, N.}, \& \bibinfo{author}{Chatzigeorgiou, A.}
  (\bibinfo{year}{2011}).
\newblock \bibinfo{title}{Identification of extract method refactoring
  opportunities for the decomposition of methods}.
\newblock {\it \bibinfo{journal}{Journal of Systems and Software}\/},  {\it
  \bibinfo{volume}{84}\/}, \bibinfo{pages}{1757--1782}.
\bibitem[{Tsantalis et~al.(2013)Tsantalis, Guana, Stroulia \&
  Hindle}]{Tsantalis:2013:MES:2555523.2555539}
\bibinfo{author}{Tsantalis, N.}, \bibinfo{author}{Guana, V.},
  \bibinfo{author}{Stroulia, E.}, \& \bibinfo{author}{Hindle, A.}
  (\bibinfo{year}{2013}).
\newblock \bibinfo{title}{A multidimensional empirical study on refactoring
  activity}.
\newblock In {\it \bibinfo{booktitle}{Proceedings of the 2013 Conference of the
  Center for Advanced Studies on Collaborative Research}\/} CASCON '13 (pp.
  \bibinfo{pages}{132--146}).
\newblock \bibinfo{address}{Riverton, NJ, USA}: \bibinfo{publisher}{IBM Corp.}
\newblock \URLprefix \url{http://dl.acm.org/citation.cfm?id=2555523.2555539}.
\bibitem[{Tsantalis et~al.(2018)Tsantalis, Mansouri, Eshkevari, Mazinanian \&
  Dig}]{tsantalis2018accurate}
\bibinfo{author}{Tsantalis, N.}, \bibinfo{author}{Mansouri, M.},
  \bibinfo{author}{Eshkevari, L.~M.}, \bibinfo{author}{Mazinanian, D.}, \&
  \bibinfo{author}{Dig, D.} (\bibinfo{year}{2018}).
\newblock \bibinfo{title}{Accurate and efficient refactoring detection in
  commit history}.
\newblock In {\it \bibinfo{booktitle}{Proceedings of the 40th International
  Conference on Software Engineering}\/} (pp. \bibinfo{pages}{483--494}).
\newblock \bibinfo{organization}{ACM}.
\bibitem[{Tufano et~al.(2016)Tufano, Palomba, Bavota, Di~Penta, Oliveto,
  De~Lucia \& Poshyvanyk}]{Tufano:2016:EIN:2970276.2970340}
\bibinfo{author}{Tufano, M.}, \bibinfo{author}{Palomba, F.},
  \bibinfo{author}{Bavota, G.}, \bibinfo{author}{Di~Penta, M.},
  \bibinfo{author}{Oliveto, R.}, \bibinfo{author}{De~Lucia, A.}, \&
  \bibinfo{author}{Poshyvanyk, D.} (\bibinfo{year}{2016}).
\newblock \bibinfo{title}{An empirical investigation into the nature of test
  smells}.
\newblock In {\it \bibinfo{booktitle}{Proceedings of the 31st IEEE/ACM
  International Conference on Automated Software Engineering}\/} ASE 2016 (pp.
  \bibinfo{pages}{4--15}).
\newblock \bibinfo{address}{New York, NY, USA}: \bibinfo{publisher}{ACM}.
\newblock \DOIprefix\doi{10.1145/2970276.2970340}.
\bibitem[{Vassallo et~al.(2019)Vassallo, Grano, Palomba, Gall \&
  Bacchelli}]{vassallo2019large}
\bibinfo{author}{Vassallo, C.}, \bibinfo{author}{Grano, G.},
  \bibinfo{author}{Palomba, F.}, \bibinfo{author}{Gall, H.~C.}, \&
  \bibinfo{author}{Bacchelli, A.} (\bibinfo{year}{2019}).
\newblock \bibinfo{title}{A large-scale empirical exploration on refactoring
  activities in open source software projects}.
\newblock {\it \bibinfo{journal}{Science of Computer Programming}\/},  {\it
  \bibinfo{volume}{180}\/}, \bibinfo{pages}{1--15}.
\bibitem[{Wang(2009)}]{5306290}
\bibinfo{author}{Wang, Y.} (\bibinfo{year}{2009}).
\newblock \bibinfo{title}{What motivate software engineers to refactor source
  code? evidences from professional developers}.
\newblock In {\it \bibinfo{booktitle}{2009 IEEE International Conference on
  Software Maintenance}\/} (pp. \bibinfo{pages}{413--416}).
\newblock \DOIprefix\doi{10.1109/ICSM.2009.5306290}.
\bibitem[{Wohlin et~al.(2012)Wohlin, Runeson, H{\"o}st, Ohlsson, Regnell \&
  Wessl{\'e}n}]{wohlin2012experimentation}
\bibinfo{author}{Wohlin, C.}, \bibinfo{author}{Runeson, P.},
  \bibinfo{author}{H{\"o}st, M.}, \bibinfo{author}{Ohlsson, M.~C.},
  \bibinfo{author}{Regnell, B.}, \& \bibinfo{author}{Wessl{\'e}n, A.}
  (\bibinfo{year}{2012}).
\newblock {\it \bibinfo{title}{Experimentation in software engineering}\/}.
\newblock \bibinfo{publisher}{Springer Science \& Business Media}.
\bibitem[{Xia et~al.(2016)Xia, Lo, Wang \& Yang}]{xia2016collective}
\bibinfo{author}{Xia, X.}, \bibinfo{author}{Lo, D.}, \bibinfo{author}{Wang,
  X.}, \& \bibinfo{author}{Yang, X.} (\bibinfo{year}{2016}).
\newblock \bibinfo{title}{Collective personalized change classification with
  multiobjective search}.
\newblock {\it \bibinfo{journal}{IEEE Transactions on Reliability}\/},  {\it
  \bibinfo{volume}{65}\/}, \bibinfo{pages}{1810--1829}.
\bibitem[{Yan et~al.(2016)Yan, Fu, Zhang, Yang, Xu \& Kymer}]{YAN2016296}
\bibinfo{author}{Yan, M.}, \bibinfo{author}{Fu, Y.}, \bibinfo{author}{Zhang,
  X.}, \bibinfo{author}{Yang, D.}, \bibinfo{author}{Xu, L.}, \&
  \bibinfo{author}{Kymer, J.~D.} (\bibinfo{year}{2016}).
\newblock \bibinfo{title}{Automatically classifying software changes via
  discriminative topic model: Supporting multi-category and cross-project}.
\newblock (pp. \bibinfo{pages}{296 -- 308}).
\newblock volume \bibinfo{volume}{113}.
\newblock \URLprefix
  \url{http://www.sciencedirect.com/science/article/pii/S016412121500285X}.
  \DOIprefix\doi{https://doi.org/10.1016/j.jss.2015.12.019}.
\bibitem[{Zhang et~al.(2018)Zhang, Li, Li \& Liang}]{zhangpreliminary18}
\bibinfo{author}{Zhang, D.}, \bibinfo{author}{Li, B.}, \bibinfo{author}{Li,
  Z.}, \& \bibinfo{author}{Liang, P.} (\bibinfo{year}{2018}).
\newblock \bibinfo{title}{A preliminary investigation of self-admitted
  refactorings in open source software}.
\newblock \DOIprefix\doi{10.18293/SEKE2018-081}.
\bibitem[{Zheng \& Casari(2018)}]{zheng2018feature}
\bibinfo{author}{Zheng, A.}, \& \bibinfo{author}{Casari, A.}
  (\bibinfo{year}{2018}).
\newblock {\it \bibinfo{title}{Feature Engineering for Machine Learning:
  Principles and Techniques for Data Scientists}\/}.
\newblock \bibinfo{publisher}{O'Reilly Media}.

\end{thebibliography}

\end{document}